\documentclass[epj]{svjour}
\usepackage{graphics}
\usepackage{graphicx,floatflt,amssymb,rotate,epsfig}
\usepackage{mathrsfs}
\usepackage{amssymb}
\usepackage{amsmath}
\usepackage{color}
\usepackage{graphicx}
\usepackage{subfig}
\usepackage{multirow}
\usepackage{mathtools}
\usepackage{cite}
\usepackage{float}
\usepackage{pstricks}
\usepackage[toc,page]{appendix}
\usepackage{pifont}
\usepackage{braket}
\usepackage{bbold}
\usepackage[toc,page]{appendix}
\usepackage%[colorlinks=true,linkcolor=red,urlcolor=blue,citecolor=blue]
{hyperref}
\usepackage{pdflscape}
\usepackage{tablefootnote}
\usepackage{afterpage}
\usepackage{capt-of}% or use the larger `caption` package

\usepackage{relsize}
\def\Babar{{\mbox{\slshape B\kern-0.1em{\smaller A}\kern-0.1em B\kern-0.1em{\smaller A\kern-0.2em R}}}}

\newcommand{\ba}{\begin{array}}
	\newcommand{\ea}{\end{array}}
\def\beq{\begin{equation}}
\def\eeq{\end{equation}}
\def\bea{\begin{eqnarray}}
\def\eea{\end{eqnarray}}
\def\nn{\nonumber}

\def\roughly#1{\mathrel{\raise.3ex\hbox
		{$#1$\kern-.75em\lower1ex\hbox{$\sim$}}}}

\def\sla#1{\raise.15ex\hbox{$/$}\kern-.57em #1}% Feynman slash

\def\bd{B_d^0}

\def\order{\lower 1.8ex \hbox{\LARGE\~{}}}

\def\bdtau{B\to D^{(\ast)}\tau\nu_{\tau}}
\def\rdast{{{\cal B}{(B\to D^{(\ast)}\tau\nu_{\tau})}}/{{\cal B}{(B\to D^{(\ast)}\ell\nu_{\ell})}}}

\def\bdell{B\to D^{(\ast)}\ell\nu_{\ell}}
\def\bd0tau{B\to D \tau\nu_{\tau}}

\def\srdast{\mathcal{R}_{D^{(*)}}}
\def\be {\begin{equation}}
\def\ee {\end{equation}}

\usepackage[normalem]{ulem}

\definecolor{darkgreen}{cmyk}{1,0,1,0.4}

\definecolor{pink}{cmyk}{0.4,1,0.3,0}

\def\com2#1{\textcolor{red}{\it{#1}}}
\usepackage{enumitem}

\begin{document}
\title{$b \to c \tau \nu_{\tau}$ Decays: A Catalogue to  \\Compare, Constrain, and Correlate New Physics Effects}
\subtitle{}
\author{Srimoy Bhattacharya\inst{1}, Soumitra Nandi \inst{1} \and Sunando Kumar Patra\inst{2}% etc
}                     % Do not remove
\offprints{}          % Insert a name or remove this line
\institute{Indian Institute of Technology, North Guwahati, Guwahati 781039, Assam, India \and Department of Physics, Indian Institute of Technology, Kanpur-208016, India.}
\date{Received: date / Revised version: date}
\abstract{In this article, we have predicted the standard model (SM) values of the asymmetric and angular observables in $\bdtau$ decays, using the results of the new up-to-date analysis in $B\to D^{(*)}\ell\nu_{\ell}$. We have also revisited the SM prediction of the inclusive ratio $\mathcal{R}_{X_c}$, and have given its values in different schemes of the charm quark mass. This is the first analysis which includes all the known corrections in the SM. In addition, we have analysed the $b\to c\tau\nu_\tau$ decay modes in a model-independent framework of effective field theory beyond the standard model. Considering all the possible combinations of the effective operators in $b \to c \tau\nu_{\tau}$ decays and using the Akaike Information Criterion, we find out the scenarios which can best explain the available data on these channels. In the selected scenarios, best-fit values and correlations of the new parameters are extracted. Using these results, predictions are made on various observables in the exclusive and inclusive semitaunic $b \to c $ decays. The graphical correlations between these observables are shown, which are found to be useful in discriminating various new physics scenarios.
%
%\PACS{
%      {PACS-key}{discribing text of that key}   \and
%      {PACS-key}{discribing text of that key}
%     } % end of PACS codes
} %end of abstract

\titlerunning{$b \to c \tau \nu_{\tau}$ Decays: A Catalogue to \ldots NP Effects}
\authorrunning{S. Bhattacharya, S. Nandi \& S. K. Patra}
\maketitle
%
%%%%%%%%%%%%%%%%%%%%%%%%%%%%%%%%%%%%%%%%%%%%%%%%%	
	\section{Introduction}\label{intro}
%%%%%%%%%%%%%%%%%%%%%%%%%%%%%%%%%%%%%%%%%%%%%%%%%

	The semileptonic $\bdell$ ($\ell = \mu$ or $e$) decays play important role in the extraction of $|V_{cb}|$ as well as the extraction of the form factors associated with the $b\to c$ decays. These form factors are the major inputs used in the predictions of $\mathcal{R}_{D^{(*)}} = \rdast$, where $\ell = \mu$ or $e$; for recent updates see \cite{Bigi:2016mdz,Bernlochner:2017jka,Bigi:2017njr,Grinstein:2017nlq,Bigi:2017jbd,Jaiswal:2017rve,Bernlochner:2017xyx}. There are additional form factors which can not be extracted directly from the experimental data and one thus needs to rely on the heavy quark effective theory (HQET) inputs. On the other hand, the lattice simulations can predict these form factors at zero and non-zero recoils \cite{Lattice:2015rga,Na:2015kha}. The standard model (SM) predictions of $\mathcal{R}_{D^{(*)}}$ which are used in the literature are given by \cite{Bigi:2016mdz,Fajfer:2012vx}
\begin{equation}\label{eq:expRDRDst}
	\mathcal{R}_{D} = 0.299 \pm 0.003,\ \ \ \mathcal{R}_{D^{*}} = 0.252 \pm 0.003.  
\end{equation}
	The prediction of $\mathcal{R}_{D}$ includes the up-to-date lattice inputs, whereas the prediction of $\mathcal{R}_{D^{*}}$ relies heavily on HQET inputs. Also, the current lattice results suggest that the HQET values of form factors at zero recoil are not in complete agreement with those from lattice \cite{Bigi:2016mdz,Bigi:2017njr}. This discrepancy could be due to the missing higher order ($\alpha^2_s$ and $\alpha_s\Lambda_{QCD}/m_b$) corrections in the HQET relations of the form factors. The inclusion of lattice inputs increases the value of $\mathcal{R}_{D^{*}}$ \cite{Bernlochner:2017jka,Bigi:2017jbd,Jaiswal:2017rve}. Moreover, the data allow the unknown corrections in the ratios of the HQET form factors to be as large as 20\% \cite{Jaiswal:2017rve}. With all these inputs and using the Caprini-Lellouch-Neubert (CLN) \cite{Caprini:1997mu} parametrisation of the form factors, the SM prediction is given by $\mathcal{R}_{D^{*}} = 0.259 \pm 0.006$ \cite{Jaiswal:2017rve}. On the experimental side, the current world averages are given by  \cite{average}
\begin{equation}
	\mathcal{R}_{D} = 0.407 \pm 0.046,\ \ \  \mathcal{R}_{D^{*}} =0.304 \pm 0.015\,.
\end{equation}
	We note that the deviations in both the observables are a little less then 2.6$\sigma$. Though these deviations can be explained by a variety of new physics models, we will follow a model independent analysis, like the one done in ref. \cite{Bhattacharya:2016zcw}. The other model independent analyses, which are relatively new, can be seen in references. \cite{Alonso:2016oyd,Choudhury:2016ulr,Celis:2016azn,Ivanov:2017mrj,Akeroyd:2017mhr,Buttazzo:2017ixm,Gonzalez-Alonso:2017iyc,Azatov:2018knx,Altmannshofer:2017poe}. There are a few other observables, which could be constructed from the $\bdtau$ decays and which are potentially sensitive to the new physics (NP) beyond the SM. For an update, please see \cite{Bhattacharya:2015ida} and references therein. Among these, the $\tau$ polarisation asymmetry has been measured by Belle \cite{Hirose:2016wfn}, and it is consistent with the SM predictions (see section \ref{sec:obscons} for detail). 

	Recently, LHCb has published their result on another $b\to c\tau\nu_{\tau}$ decay mode, where the observable and its measured value are given by \cite{Aaij:2017tyk} 
\begin{equation}
	\mathcal{R}_{J/\psi} = \frac{{\cal B}{(B_c\to J/\psi \tau\nu_{\tau})}}{{\cal B}{(B_c\to J/\psi \ell\nu_{\ell})}} = 0.71 \pm 0.25 ,
\end{equation}
	$\ell = \mu$ or $e$. Though the uncertainties are large, this measurement is 2$\sigma$ above the corresponding SM prediction, which lies in between 0.25 and 0.29 \cite{Ivanov:2005fd,Hernandez:2006gt,Ivanov:2006ni,Wen-Fei:2013uea}. We note that in both $\srdast$ and $\mathcal{R}_{J/\psi}$, the measured values are above the SM predictions and therefore, the NP should contribute constructively to both the decay modes, in order to explain the observed discrepancies. There are a few other $b\to c\tau\nu_{\tau}$ decay modes like $B_c\to\tau\nu_{\tau}$, $\Lambda_b \to \Lambda_c\tau\nu_{\tau}$ and the inclusive decay $B\to X_c\tau\nu_{\tau}$ which are potentially sensitive to the new interactions and the NP affecting $\srdast$ and $\mathcal{R}_{J/\psi}$ should also have an impact on these decay modes.
	
	Therefore, the correlation studies of the various observables associated with these decay modes will be an important probe for an indirect detection of NP. On the other hand, the precise measurements of some of these observables will be useful to constrain the new physics parameters associated with a model. This motivates one to predict the values of all the relevant observables for some specific model, which can then be further checked for consistency with the future measurements. 

	In this article, we have done a model independent analysis of the NP affecting the $b\to c\tau\nu_{\tau}$ decay modes. The operator basis is exactly the same as that given in our earlier works \cite{Bhattacharya:2015ida,Bhattacharya:2016zcw}, which consists of scalar (S), vector (V), and tensor (T) type of operators. We did not consider the scenarios with right handed neutrinos. We have considered all possible combinations of these operators and categorised them as independent models. There are several models capable of describing the observed data and one is thus confronted with the problem of model selection. We use the Akaike information criterion (AIC) to find out the best possible model(s) for the existing data. A model selection criterion is a formula that allows one to compare models; for details, see \cite{akaike}. Alternative related approaches to the model selection are the bootstrap method and cross-validation. Cross-validation works poorly with small sample sizes, as it is in our case, and parametric bootstrap variants of AIC have recently been proposed \cite{SHANG}, which we have not used in the present analysis. Using the AIC, we have first selected the models best suited for explaining the existing data. Then, with the best fitted values of the model parameters, we have predicted the values of various observables associated with the above mentioned decay modes. We have studied the correlations amongst the observables in detail as well. 
%%%%%%%%%%%%%%%%%%%%%%%%%%%%%%%%%%%%%%%%%%%%%%%%%%%%%%%%%%%%%%%%%%%
	\section{New Physics and the observables }\label{sec:obscons}
%	\subsection{Theoretical status}
	\subsection{New operators}
%%%%%%%%%%%%%%%%%%%%%%%%%%%%%%%%%%%%%%%%%%%%%%%%%%%%%%%%%%%%%%%%%%%

	In this subsection, we will discuss the complete operator basis in $b\to c\tau\nu_{\tau}$ decays.
	The most general effective Hamiltonian describing these transitions is given by
	\begin{align}
	{\cal H}_{eff} &= \frac{4 G_F}{\sqrt{2}} V_{cb} \left.[( \delta_{\ell\tau} + C_{V_1}^{\ell}) {\cal O}_{V_1}^{\ell} + 
	C_{V_2}^{\ell} {\cal O}_{V_2}^{\ell} + C_{S_1}^{\ell} {\cal O}_{S_1}^{\ell} \right. \nn \\ & \left. + C_{S_2}^{\ell} {\cal O}_{S_2}^{\ell}
	+ C_{T}^{\ell} {\cal O}_{T}^{\ell}\right.]\,,
	\label{eq1}
	\end{align}
	where  $C^\ell_W (W=V_1,V_2,S_1,S_2,T)$ are the Wilson coefficients corresponding to the following four-fermi operators:
	
	\begin{align}
	{\cal O}_{V_1}^{\ell} &= ({\bar c}_L \gamma^\mu b_L)({\bar \tau}_L \gamma_\mu \nu_{\ell L})\nn,\\
	{\cal O}_{V_2}^{\ell} &= ({\bar c}_R \gamma^\mu b_R)({\bar \tau}_L \gamma_\mu \nu_{\ell L}) \nn, \\
	{\cal O}_{S_1}^{\ell} &= ({\bar c}_L  b_R)({\bar \tau}_R \nu_{\ell L})\nn,\\
	{\cal O}_{S_2}^{\ell} &= ({\bar c}_R b_L)({\bar \tau}_R \nu_{\ell L}) \nn, \\
	{\cal O}_{T}^{\ell}   &= ({\bar c}_R \sigma^{\mu\nu} b_L)({\bar \tau}_R \sigma_{\mu\nu} \nu_{\ell L})\,.
	\label{eq2}
	\end{align}
	Here, we have considered only the left handed neutrinos. 

%%%%%%%%%%%%%%%%%%%%%%%%%%%%%%%%%%%%%%%%%%%%%%%%%%%%%%%%%%%
	\subsection{Observables in $b\to c\tau\nu_{\tau}$ decays }
%%%%%%%%%%%%%%%%%%%%%%%%%%%%%%%%%%%%%%%%%%%%%%%%%%%%%%%%%%%
We will define various observables used in our analysis in this subsection.  

%%%%%%%%%%%%%%%%%%%%%%%%%%%%%%%%%%%%%
	\subsubsection{$B\to D^{(*)}\tau\nu_{\tau}$}
%%%%%%%%%%%%%%%%%%%%%%%%%%%%%%%%%%%%%

	Following ref. \cite{Bhattacharya:2016zcw} and references therein, we can write the differential decay rates for $B\rightarrow D^{(*)} \tau \nu_\tau$ 
	with the Hamiltonian in eq. \ref{eq1}% (see appendix. \ref{sec:app2})
	. The $q^2$-distribution of the decay rate of the decays $\bdell$ are obtained %from equations (\ref{dgambd}) and (\ref{dgambdst})
	by setting $C_W = 0$ and $m_{\tau} = m_{\ell}$. Here, we are assuming that new effects are present only in $\bdtau$ and the $\bdell$ channels for $\ell = \mu$ and $e$ are free from any NP effects. $H^s_{V, Y}(q^2)$ and $H_{V, Y}(q^2)$ are the helicity amplitudes for $\bar{B}\rightarrow D$ and $\bar{B}\rightarrow D^*$ transitions respectively (with $Y = \pm, ~0$ and $t$). These amplitudes can be expressed in terms of form factors in $B\rightarrow D^{(*)}$ transitions. The details of the form factors and their parametrisations can be looked up in 
	\cite{Jaiswal:2017rve} and the references in there.

	In the present work, we have followed the CLN \cite{Caprini:1997mu} parametrisation of the $B\to D^{(*)}$ form factors and have used both the fitted and predicted values of these parameters obtained in \cite{Jaiswal:2017rve}.
	In terms of the differential distributions, the ratios $\mathcal{R}_{D^{(*)}}$ %and $\mathcal{R}_{J/\psi}$
	are defined as
	\begin{align}\label{Rth}
	\mathcal{R}_{D^{(*)}} &= \left.\Big[\int^{q^2_{max}}_{m^2_{\tau}} \frac{d\Gamma\left(\overline{B} \rightarrow D^{(*)}
		\tau \overline{\nu}\right)}{d q^2} d q^2\right.\Big]\times  \nn \\ & \left.\Big[\int^{q^2_{max}}_{m^2_{\ell}} 
	\frac{d\Gamma\left(\overline{B} \rightarrow D^{(*)} \ell \overline{\nu}\right)}{d q^2} d q^2\right.\Big]^{-1},
	\end{align}
	with $q^2_{max}= (m_B - m_{D^{(*)}})^2$, and $\ell=e$ or $\mu$. 
	
	Along with these ratios, there are a number of other observables, that can be constructed in these channels, which are sensitive to NP. Most of them are not yet measured experimentally. These are:
	\begin{itemize}
		\item $\tau$-polarization is defined by studying further $\tau$ decays:
		\begin{align}\label{ptaudef1}
		P_{\tau}(D^{(*)}) = \frac{\Gamma^{(*)\lambda_{\tau}=1/2} - \Gamma^{(*)\lambda_{\tau}= -1/2}}{\Gamma^{(*)\lambda_{\tau}=1/2} + 
		\Gamma^{(*)\lambda_{\tau}= -1/2}}\,,
		\end{align}
		where $\Gamma^{(*)\lambda_{\tau}=\pm \frac{1}{2}} = \int_{m_{\tau}^2}^{q^2_{max}}\frac{d\Gamma^{\lambda_{\tau}=\pm 1/2}
		(\bar{B}\rightarrow D^{(*)}\tau \bar{\nu})}{dq^2}$, $\lambda_{\tau}$ is the $\tau$ helicity, and $q^2_{max}= (m_B - m_{D^{(*)}})^2$.
		\item $D^*$ longitudinal polarization can be extracted from the angular distribution in $D^* \to D \pi$ decays:
		\begin{align}\label{dstpol}
		F_L^{D^*} = \frac{\Gamma^{\lambda_{D^*}=0}}{\Gamma^{\lambda_{D^*}=0} + \Gamma^{\lambda_{D^*}=1} + \Gamma^{\lambda_{D^*}=-1}}\,,
		\end{align}
		where $\Gamma^{\lambda_{D^*}=0,\pm 1} = \int_{m_{\tau}^2}^{q^2_{max}}\frac{d\Gamma^{\lambda_{D^*}=0,\pm 1}(\bar{B}\rightarrow 
		D^{*}\tau \bar{\nu})}{dq^2}$. 
		\item If we write the double-differential decay distribution as
		\begin{align}
		{ d^2\Gamma\left(\overline{B} \rightarrow D^{(*)}
			\tau \overline{\nu}\right) \over dq^2 d\cos\theta } &= a^{(*)}_\theta(q^2) + b^{(*)}_\theta(q^2) \cos\theta \nn \\ & + c^{(*)}_\theta(q^2) \cos^2\theta \,,
		\end{align}
		where $\theta$ is the angle between the three-momenta of $\tau$ and $\bar B$ in the $\tau\bar{\nu}$ rest frame, then $b^{(*)}_\theta(q^2)$ determines 
		the lepton forward-backward asymmetry in the following way:
		\begin{align}
		\mathcal{A}^{(*)}_{\rm FB} &= { \int_0^1 {d\Gamma^{(*)} \over d\cos\theta}d\cos\theta-\int^0_{-1}{d\Gamma^{(*)} \over d\cos\theta}d\cos\theta 
		\over \int_{-1}^1 {d\Gamma^{(*)} \over d\cos\theta}d\cos\theta } \nn \\ &= { \int b^{(*)}_\theta(q^2) dq^2 \over \Gamma^{(*)} } \,,
		\end{align}
	\end{itemize}
	As mentioned earlier, in addition to these observables there are several other channels that will be affected by the same set of NP operators. We have used some of those most relevant observables in our analysis, either as fit inputs or as constraints and/or for prediction.

	\subsubsection{$B_c \to {J/\psi}~\ell~\nu_{\ell}$}\label{sec:Rjpsi}
	
	Ratios similar to those defined in eq. \ref{Rth} can be defined for the decay channel $\bar{B_c}\rightarrow J/\psi\ell \bar{\nu_\ell}$  by replacing the 
	respective mesons. For various form factors in $B_c \to J/\psi$ decays, see \cite{Watanabe:2017mip}.
		
	Given the unavailability of a precise calculation of $B_c\to J/\psi$ form factors till date, we have the option to choose from a collection of available parametrisations \cite{Wen-Fei:2013uea,Anisimov:1998uk,Ebert:2003cn,Hernandez:2006gt,Kiselev:2002vz,Ivanov:2006ni,Wang:2008xt}. 
	Choosing different parametrisations results in varying the central value of $\mathcal{R}_{J/\psi}$ within the range $0.25$ - $0.29$, which is considered theoretical range in recent experimental analyses. Taking the uncertainties from different parametrisations into consideration, we see that the allowed theoretical range of $\mathcal{R}_{J/\psi}$ is actually larger than that. We consider two parametrisations residing at two far ends of this range, namely perturbative QCD (PQCD \cite{Wen-Fei:2013uea}), and light-front covariant quark  model (LFCQ \cite{Wang:2008xt}) in this work. A preliminary result on the form factor $A_1(q^2_{max})$ (this is the only form factor contributing to the decay at zero recoil) is available from the HPQCD collaboration \cite{Lytle:2016ixw}. This result is consistent with the parametrisations used in this draft.

	\begin{table*}[!hbt]
	\caption{SM values of observables obtained and/or used in this paper, with correlations, wherever relevant. Due to considerable uncertainty and difference in central values, value of $\mathcal{R}_{J/\Psi}$ is quoted for both LFCQ and PQCD parametrizations. They are treated separately throughout the NP analysis as well.\\
			{\small $^*$ Based on \cite{Jaiswal:2017rve}}} 
			\label{tab:SMres}
	\begin{center}
		\scriptsize
%		\begin{ruledtabular}
			\begin{tabular}{cccccccccc}
				\hline
				Observable & \multicolumn{2}{c}{SM Prediction} & \multicolumn{7}{c}{Correlation}  \\
				\hline
				$\mathcal{R}_{D^*}$ & \text{0.260(6)} & \multirow{2}{*}{$\begin{rcases*} \\ \\ \end{rcases*}$\cite{Jaiswal:2017rve}} & 1. & 0.118 & 0.617 & 0.118 & 0.604 & 0.628 & -0.118 \\
				%\cline{1-2}
				$\mathcal{R}_{D}$ & \text{0.305(3)} & &  & 1. & -0.023 & 1. & 0.021 & 0.007 & -1. \\
				%\cline{1-2}
				$P_{\tau}(D^*)$ & \text{-0.491(25)} & \multirow{5}{*}{$\begin{rcases*} \\ \\ \\ \\ \\ \end{rcases*}$ New$^*$} &  &  & 1. & -0.023 & 0.803 & 0.895 & 0.023 \\
				%\cline{1-2}
				$P_{\tau}(D)$ & \text{0.3355(4)} &  &  &  &  & 1. & 0.021 & 0.007 & -1. \\
				%\cline{1-2}
				$F_L^{D^*}$ & \text{0.457(10)} &  &  &  &  &  & 1. & 0.921 & -0.021 \\
				%\cline{1-2}
				$\mathcal{A}_{FB}^*$ & \text{-0.058(14)} &  &  &  &  &  &  & 1. & -0.007 \\
				%\cline{1-2}
				$\mathcal{A}_{FB}$ & \text{0.3586(3)} &  &  &  &  &  &  &  & 1. \\
				\hline
				$\mathcal{R}_{J/\Psi}$ (LFCQ) & \text{0.249(42)} & \cite{Huang:2004vf} &  &  &  &  &  &  &  \\
				%\cline{1-2}
				$\mathcal{R}_{J/\Psi}$ (PQCD) & 0.289(28) & \cite{Wen-Fei:2013uea}    &  &  &  &  &  &  &  \\
				%\cline{1-2}
				$\mathcal{R}_{\Lambda}^\mu$ & \text{0.329(13)} & \cite{Detmold:2015aaa} &  &  &  &  &  &  &  \\
				%\cline{1-2}
				$\mathcal{R}_{\Lambda}^e$ & \text{0.328(13)} & \cite{Detmold:2015aaa} &  &  &  &  &  &  &  \\
				%\cline{1-2}
				$\mathcal{B}(B_c \to \tau \nu)$ & \text{0.0208(18)} & %\multirow{2}{*}{$\begin{rcases*} \\ \\ \\ \\ \end{rcases*}$
				(this work)
				%} 
				&  &  &  &  &  &  & \\
%				\cline{1-2}
%				$\mathcal{R}_{X_c}$ ({\it NLO} + $\frac{1}{m_b^2} + \frac{1}{m_b^3}$) &  &  &  &  &  &  &  & & \\
%				\cline{1-2}
%				$\overline{m_c}$(3 GeV) in $\overline{MS}$ scheme & 0.226(8) &  &  &  &  &  &  &  & \\
%				\cline{1-2}
%				$m_c $(1 GeV) in kinetic scheme & 0.214(4) &  &  &  &  &  &  &  & \\
				\hline	 
			\end{tabular}
			
%		\end{ruledtabular}
	\end{center}
\end{table*}

\begin{table*}[!htb]
\caption{\small SM Predictions for $\mathcal{R}_{X_c}$. Other relevant inputs are taken from table \ref{tab:corrmat} in the kinetic scheme, while those for the $\overline{MS}$ scheme are taken from table II of ref. \cite{Alberti:2014yda}.}
		\label{tab:SMres2}
	\centering
%	\begin{ruledtabular}
\scriptsize
		\begin{tabular}{ccc}
			\hline\noalign{\smallskip}
			\multicolumn{3}{c}{SM predictions for $\mathcal{R}_{X_c}$ }\\
			\noalign{\smallskip}\hline
			% & \multicolumn{4}{|c|}{$m_c$ in scheme:} \\
			%\hline
			Accuracy in $B\to X_c\tau\nu_{\tau}$  &  $\overline{m_c}$(3 {\it GeV}) = 0.987(13)    & $m_c^{kin}$ = 1.091 (20)  \\
			(at order $\alpha_s^n$ \& $\frac{1}{m_b^n}$)   & (in {\it GeV})  & (in {\it GeV})\\
			\hline
			 LO + NLO + NNLO + $\frac{1}{m_b^2}$& \text{0.238(5)}  & \text{0.232(4)}  \\
			\hline
			LO + NLO + NNLO + $\frac{1}{m_b^2} + \frac{1}{m_b^3}$ &   \text{0.214(4)}  & \text{0.209(4)}\\
			\noalign{\smallskip}\hline
		\end{tabular}
		
%	\end{ruledtabular}
\end{table*}

\begin{table*}[!htbp]
\caption{The correlations between various non-perturbative parameters and the masses. These were all obtained in the analysis of inclusive $B\to X_c\ell\nu_{\ell}$ decays in \cite{Alberti:2014yda}.} 
			\label{tab:corrmat}
	\begin{center}
%		\begin{ruledtabular}
			\begin{tabular}{ccccccccc}
				\hline
				Parameters & Value & \multicolumn{7}{c}{Correlation}  \\
				\hline
				$m_b^{Kin}$ & \text{4.561(21)} & 1. & 0.608 & -0.096 & 0.132 & 0.554 & -0.170 & -0.062 \\
				%\cline{1-2}
				$m_c$ & \text{1.092(20)} &  & 1. & -0.022 & 0.003 & -0.032 & 0.011 & 0.023 \\
				%\cline{1-2}
				$\mu_\pi^2$ & \text{0.464(67)} &  &  & 1. & 0.717 & -0.045 & 0.060 & 0.158 \\
				%\cline{1-2}
				$\rho_D^3$ & \text{0.175(40)} & &  &  & 1. & -0.077 & -0.134 & 0.076 \\
				%\cline{1-2}
				$\mu_G^2$ & \text{0.333(61)} &  &  &  &  & 1. & -0.042 & -0.022 \\
				%\cline{1-2}
				$\rho_{LS}^3$ & \text{-0.146(96)} &  &  &  &  &  & 1. & -0.020 \\
				%\cline{1-2}
				$\mathcal{B}(B\to X_c \ell\nu_{\ell})$ & $10.66(16)\%$ &  &  &  &  &  &  & 1. \\
				\hline
			\end{tabular}
			
%		\end{ruledtabular}
	\end{center}
\end{table*}

	\begin{table*}[!hbt]
		\scriptsize
		\caption{Present experimental status of the observables used in this analysis. First uncertainty is statistical and the second one is 
				systematic.\\
				{\small $^*$ This correlation is between $\mathcal{R}(D^*)$ and $P_{\tau}(D^*)$. Stat. corr. = 0.29 and syst. corr. = 0.55.\\
				$^\dagger$ This uncertainty originates from the uncertainties on $\mathcal{B}(B^0\to D^{*-}\pi^+\pi^-\pi^+)$ and $\mathcal{B}(B^0\to D^{*-}\mu^+\nu_{\mu})$.}} 
				\label{tab:RDRDsPtau}
		\begin{center}
%			\begin{ruledtabular}
				\begin{tabular}{cccccc}
					\hline
					& $\mathcal{R}_D$  & $\mathcal{R}_{D^*}$  &	$\to$ Correlation	& $P_{\tau}(D^*)$ & $\mathcal{R}_{J/\Psi}$ \\
					\hline
					\Babar~\cite{Lees:2013uzd}   	& $0.440(58)(42)$ 	& $0.332(24)(18)$ & $-0.27$ & - & - \\
					%\hline
					Belle (2015) \cite{Huschle:2015rga} & $0.375(64)(26)$ 	& $0.293(38)(15)$ & $-0.49$ & - & - \\
					%\hline
					Belle (2016) \cite{Abdesselam:2016cgx} & -			 	& $0.302(30)(11)$  & - & - & - \\
					%\hline
					Belle (2016) \cite{Hirose:2016wfn} & - 			& $0.270(35)(^{+ 0.028}_{-0.025})$ & 0.33 $^*$ & $ -0.38(51)(~^{+0.21}_{-0.16})$ & 
					- \\
					%\hline
					LHCb (2015) \cite{Aaij:2015yra} & - 				& $0.336(27)(30)$ & - & - & - \\
					%\hline
					LHCb (2017) \cite{Aaij:2017deq} & - 				& $0.286(19)(25)(21)^\dagger$ & - & - 
					& - \\
					%\hline
					LHCb (2017) \cite{Aaij:2017tyk} & -					& -				 & - & - & $0.71(17)(18)$
					\\
					\hline
				\end{tabular}
				
%			\end{ruledtabular}
		\end{center}
	\end{table*}

\subsubsection{$\Lambda_b \to \Lambda_c \ell\nu_{\ell} $}
	
	The $q^2$ distribution for the decay process ($\Lambda_{b}\rightarrow \Lambda_{c} \tau^{-} \nu_{\tau}$) can be written as \cite{Datta:2017aue}
	\begin{align}\label{dq1}
	\nn &\frac{d\Gamma(\Lambda_{b}\rightarrow \Lambda_{c} \tau^{-} \nu_{\tau})}{dq^2} =
	\frac{{G_F}^2 {|V_{cb}}|^2 q^2 |  {\bf p}_{\Lambda_{c}}|}{192\pi^3 {M_1}^2} \left(1-\frac{{m_\tau}^2}{q^2}\right)^2 \times\\
	&\left.[A_{1}^{VA} + \frac{{m_\tau}^2}{2q^2} A_{2}^{VA} + \frac{3}{2} A_{3}^{SP} + 2 (1+\frac{2 m_{\tau}^2}{q^2}) A_4^T \right. \nn \\ & \left. + \frac{3{m_\tau}}{\sqrt{q^2}} A_{5}^{VA-SP}+ \frac{ 6 {m_\tau}}{\sqrt{q^2}} A_{6}^{VA-T}\right.]
	\end{align}
	where $A_1^{VA}$ and $A_2^{VA}$ represent the contributions from the vector and axial vector currents respectively. Their origin could be either the SM or any NP model. 
	$A_3^{SP}$ and $A_4^{T}$ represent the contributions from the scalar-pseudoscalar and tensor currents, which will appear only in the NP models. $A_5^{VA-SP}$ and $A_6^{VA-T}$ are the interference terms which will have contributions from various operators in the SM, as well as an NP model.
	These are functions of combinations of the helicity amplitudes $H_{\lambda_{\Lambda_c},\lambda_{w}}$, which in turn can be expressed in terms of form factors and NP couplings.  
	Several instances, where these form factors have been studied using sum rules and quark models, can be found in the literature \cite{Cardarelli:1997sx,Dosch:1997zx,Huang:1998rq,MarquesdeCarvalho:1999bqs,Huang:2004vf,Pervin:2005ve,Ke:2007tg,Wang:2009hra,Azizi:2009wn,Khodjamirian:2011jp,Gutsche:2014zna,Gutsche:2015mxa}. For our purpose, helicity form factors have been calculated using the formula from lattice QCD in the relativistic heavy quark limit \cite{Detmold:2015aaa}.
	
	Similar to the ratios defined earlier, two observables can be defined here, motivated by the lepton flavor university violation elsewhere:
	\begin{eqnarray}
	\mathcal{R}^\mu_{\Lambda}  =  \frac{\mathcal{B} \left(\Lambda_b \to \Lambda_c \tau \bar{\nu}_{\tau}\right)}{\mathcal{B}\left(\Lambda_b \to \Lambda_c \mu \bar{\nu}_{\mu}\right)}\\ 
	\mathcal{R}^e_{\Lambda}  =  \frac{\mathcal{B} \left(\Lambda_b \to \Lambda_c \tau \bar{\nu}_{\tau}\right)}{\mathcal{B}\left(\Lambda_b \to \Lambda_c e \bar{\nu}_{e}\right)}
	\end{eqnarray}
	
	Along with these ratios, we have also considered the forward-backward asymmetry in $\Lambda_{b}\rightarrow \Lambda_{c} \tau^{-} \nu_{\tau}$, defined as
	\begin{align}
		\nn \mathcal{A}^{\Lambda}_{\rm FB} &= {\int_0^1 \Gamma^{(1)}~ d\cos\theta_\tau-\int^0_{-1} \Gamma^{(1)}~d\cos\theta_\tau 
		\over \int_{-1}^1 \Gamma^{(1)}~d\cos\theta_\tau} \\
	\end{align}
	where $\Gamma^{(1)}={d\Gamma \over d\cos\theta_\tau}$ and $\theta_\tau$ is the angle between the momenta of the $\tau$ lepton and $\Lambda_c$ baryon in the dilepton rest frame.
	
	\subsubsection{$ B \to X_c \tau \bar \nu_\tau$}\label{sec:binclu}

	Similar to the ratios $\mathcal{R}_{D^{(*)}}$, we can define the following ratio for the inclusive decay $B\to X_c\tau\nu_{\tau}$:
	\begin{equation}
		\mathcal{R}_{X_c} = \frac{\mathcal{B} \left(B \to X_c \tau \bar{\nu}_{\tau}\right)}{\mathcal{B}\left(B \to X_c \ell \bar{\nu}_{\ell}\right)},
	\end{equation}
	 with $\ell = \mu, e$. The decay $B\to X_c\ell\nu_{\ell}$ is well studied in the literature; for a comprehensive update see \cite{Nandi:2017wxs} and references therein. In the present work, the detailed mathematical expression of the decay width of $B\to X_c \ell\nu_{\ell}$ and all other relevant inputs are taken from \cite{Alberti:2014yda}.
	The simplified expression for the decay width of the inclusive semitaunic decay of $B$ meson in SM are given in  \cite{Mannel:2017jfk}:
	\begin{align}\label{dwidthincluSM}
		\nn \Gamma^{SM}&( \bar B \to X_c \tau \bar \nu)  = \\
		&\Gamma_0  \left.[C_0^{(0)} + \frac{\alpha_s}{\pi} C_0^{(1)} + \left(\frac{\alpha_s}{\pi}\right)^2 C_0^{(2)} +  C_{\mu_\pi^2} \cfrac{\mu_\pi^2}{m_b^2} \right.\nn \\
		& \left. + C _{\mu_G^2} \cfrac{\mu_G^2}{m_b^2}+ C_{\rho_D^3}\cfrac{\rho_D^3}{m_b^3} + C_{\rho_{LS}^3}\cfrac{\rho_{LS}^3}{m_b^3} \right.] 
	\end{align}
	Here, the terms involving $C_0^{(0)}$, $C_{0}^{(1)}$, and $C_{0}^{(2)}$ represent the contributions from the leading order(LO), next-to-leading order (NLO) \cite{Czarnecki:1994bn}, and next-to-next-to-leading order (NNLO) \cite{Biswas:2009rb} corrections in $\alpha_s$ respectively, whereas  $C_{\mu_\pi^2}$, $C_{\mu_G^2}$, and $C_{\rho_D^3}$, $C_{\rho_{LS}^3}$ are the contributions at order $1/{m_b}^2$ \cite{Falk:1994gw} and $1/{m_b}^3$ \cite{Mannel:2017jfk}, respectively. These coefficients depend on the quark and lepton masses and $\Gamma_0$, defined as 
	\begin{equation}
		\Gamma_0 = \frac{G_F^2 |V_{cb}|^2 \, m_{b}^5 A_{ew} }{192\pi^3}\,.
	\end{equation}
	The parameters like $\mu_{\pi}^2, \mu_G^2, \rho_D^3, \rho_{LS}^3$ are the matrix elements of the operators of dimension five and six, respectively, which are non-perturbative in nature. We have also included the well known electroweak correction $A_{ew}( = 1.014)$. As mentioned earlier, the values of the various non-perturbative parameters and the masses (eq. \ref{dwidthincluSM}) along with their correlations are taken from tables II and III of ref. \cite {Alberti:2014yda}. In our analysis, the $b$ quark mass is defined in the kinetic scheme, while the $c$ quark mass has been defined in both the kinetic ($m^{Kin}_c = 1.091 (20)$ GeV \cite{Chetyrkin:2009fv}) and the $\overline{MS}$ scheme ($\overline{m_c}(3 \text{GeV}) = 0.9843 (56)$ GeV \cite {2018arXiv180204248B}). Relations of the pole masses (eq. \ref{dwidthincluSM}) with the kinetic and 
	$\overline{ MS}$ masses are taken from \cite{Benson:2003kp} and \cite{Melnikov:2000qh}, respectively. We have also considered $\alpha_s = 0.22\pm 0.018$. 

We have given the predictions for the ratio $\mathcal{R}_{X_c}$ instead of $Br(B\to X_c\tau\nu_{\tau})$. As can be seen from the above expressions, this ratio is relatively clean, since the errors due to $|V_{cb}|$ and the mass of the $b$-quark cancel in the ratio. Our predictions for $\mathcal{R}_{X_c}$ in the SM are given in table \ref{tab:SMres2}. These predictions differ from each other due to the difference in the mass of the charm quark in two different schemes. We note that the central values of the two predictions change by $\approx$ 2\% due to scheme dependence, albeit being consistent within $1\sigma$ uncertainties. Also, we have checked our prediction for the $1S$ scheme masses of the $b$ and $c$ quark, and we agree with that given in ref. \cite{Freytsis:2015qca} which is also different from the predictions given in table \ref{tab:SMres2} (NLO and $\frac{1}{m_b^2}$). These results are clearly scheme dependent.
	
	In the case of $m_c^{kin}$ (1 GeV), the correlation matrix for the non-perturbative parameters and the masses are given in table \ref{tab:corrmat}, which are obtained from the analysis of \cite {Alberti:2014yda}. This is the first analysis which includes all the known corrections in the prediction of $R_{X_c}$. In ref. \cite{Mannel:2017jfk}, the analysis has been done with similar set of inputs, without considering the NNLO corrections. We have checked that our result agrees with them, within the error bar, at the same level of accuracy. The inputs for the analysis with $\overline{m_c}(3 \text{GeV})$ are taken from table II of ref. \cite {Alberti:2014yda}. In this scheme, the predictions have larger uncertainties compared to those in the kinetic scheme. This is due to the difference in the correlation matrix of parameters given in table \ref{tab:SMres2}.
	
	To calculate the effects of new physics in the inclusive decay $ B \to X_c \tau \bar \nu_\tau$,  we decompose the decay width as 
	\begin{equation}\label{dwidthinclutot}
		\Gamma( \bar B \to X_c \tau \bar \nu) = \Gamma^{SM} + \Gamma^{NP}_{(1)} + \Gamma^{NP}_{(2)}\,.
	\end{equation}
	Here, the first piece is arising solely from SM, while the second and third terms are the contributions from NP  with different powers of the new couplings. The expressions of $\Gamma^{NP}_{(1)}$ and $\Gamma^{NP}_{(2)}$ are taken from \cite{Goldberger:1999yh}. Some other recent works, discussing NP effects in the inclusive mode, are ref.s \cite{Kamali:2018fhr,Colangelo:2018cnj}.
	
	\subsubsection{$\mathcal{B}(B_c \to \tau \nu_\tau)$}\label{sec:bctn}
	
	In terms of the general hamiltonian defined in eq. \ref{eq1}, the branching fraction of $B_c \to \tau \nu_\tau$ can be expressed as \cite{Gonzalez-Alonso:2016etj},
	\begin{align}\label{Brbctn}
	\nn &\mathcal{B}(B_c \to \tau \nu_\tau) = \\
	&\tau_{B_c} \frac{m_{B_c} m^2_{\tau} f^2_{B_c} G^2_F \left|V_{cb}\right|}{8 \pi} \left(1 - \frac{m^2_{\tau}}{m^2_{B_c}}\right)^2 \left.|1 + \left(C_{V_1} - C_{V_2}\right) \right.\nn \\ & \left.+ \frac{m^2_{B_c}}{m_\tau (m_b + m_c)} \left(C_{S_1} - C_{S_2}\right)\right.|^2,
	\end{align}
	where $f_{B_c} = 0.434(15)$GeV and $\tau_{B_c} = 0.507(9)$ps are the $B_c$ decay constant and lifetime, respectively. Note that $C_T$ does not enter in 
	this decay.
	
	The SM predictions of all these observables are listed in table. \ref{tab:SMres}. The predictions of the $\mathcal{R}_{D^{(*)}}$ are based on the results 
	of the analysis in \cite{Jaiswal:2017rve}, here we have considered only the CLN parametrisations of the form factors. 
	The predictions of $P_{\tau}(D^{(*)})$, $A_{FB}^{(*)}$, and $P_{D^*}$ are new which are presented along with the correlations amongst all the observables. 
	All these predictions are based on the results of the analysis in \cite{Jaiswal:2017rve}.
	
%%%%%%%%%%%%%%%%%%%%%%%%%%%%%%%%%%%%%%%%%%%%%%%%%%%%
	\subsection{Experimental status}\label{sec:experiment}
%%%%%%%%%%%%%%%%%%%%%%%%%%%%%%%%%%%%%%%%%%%%%%%%%%%%
	
	All the experimental results used in the analysis of NP is tabulated in table \ref{tab:RDRDsPtau}. There have been quite a few measurements of the ratios 
	$\mathcal{R}_{D^{(*)}}$ in recent years. Apart from the most recent ones measuring $\mathcal{R}_{D^*}$, they are consistent with a sizeable deviation 
	from the SM. The experimental result most deviated from the SM predictions is still the first one reported by~\Babar. Though it is apparent from the 
	recent measurements that $\mathcal{R}_{D^*}$ values are coming down towards the SM, it is still too early to consider it as a trend for two reasons:
	\begin{enumerate}[label={(\alph*)}]
		\item The experimental uncertainties are still quite large.
		\item The actual deviation depends heavily on the correlation between $\mathcal{R}_{D}~ \&~ \mathcal{R}_{D^{*}}$, and any analysis bears the risk of being inconclusive without the simultaneous measurement of both of them. As an example, one can check the  Belle (2015) result \cite{Huschle:2015rga}, where the $\mathcal{R}_{D^*}$ is consistent with the SM result within $1\sigma$, but the combined result is at tension with the SM due to $\mathcal{R}_{D}$ and its correlation with $\mathcal{R}_{D^*}$.
	\end{enumerate} 
	The first, although quite imprecise, measurement of $\tau$ polarization asymmetry is done by Belle in 2015 \cite{Huschle:2015rga}. Though essentially it 
	is an upper limit, we have included this measurement as a data point in our analysis. Table \ref{tab:RDRDsPtau} also contains the recent measurement of 
	$\mathcal{R}_{J/\psi}$ by LHCb \cite{Aaij:2017tyk}. Not only is this result in tension with the theoretical predictions, the central measured-value is almost double of that predicted by SM. As the experimental uncertainty is large, they are still consistent with 90\% C.L. range. LHCb has used 
	a $z$-expansion parametrization \cite{Bourrely:2008za} for the shared form factors for the signal and normalization modes and has determined them directly 
	from the data.	As is evident from the theoretical results for $\mathcal{R}_{J/\psi}$ (table \ref{tab:SMres}), the PQCD result is a little closer to the 
	LHCb result. As has been pointed out in \cite{Alok:2017qsi}, and later also corroborated in \cite{Biswas:2018jun}, if uncertainty decreases but the central 
	value remains approximately the same in future experiments, NP effects which explain the increase in $\mathcal{R}_{D^{(*)}}$, will be unable to explain the 
	measured value of $\mathcal{R}_{J/\psi}$. This should, in essence, result in a worse fit while this value is considered.
	
	The decay $B_c \to \tau \nu$, despite being out of the experimental reach for now \cite{Gouz:2002kk}, can be used as an effective constraint on any NP 
	effects that could potentially explain the $\mathcal{R}_{D^{(*)}}$ and $\mathcal{R}_{J/\psi}$ excesses. A conservative upper limit quoted for 
	$\mathcal{B}(B_c \to \tau \nu)$, even after adding NP effects, is $\lesssim 30\%$ \cite{Alonso:2016oyd}. A stronger upper bound of $\lesssim 10\%$ is 
	obtained from LEP data taken at $Z$-peak \cite{Akeroyd:2017mhr} with a prospect of an even tighter bound from the full L3 data \cite{Acciarri:1996bv}.
	In our analysis, we have used these two constraints.

	%\begin{landscape}% Landscape page
\begin{table*}[!hbt]

	\centering
	\caption{\small Scenarios selected after passing the normality check and the criterion $\Delta$AIC$_c \le 4$, for all data available (with or without $\mathcal{R}_{J/\Psi}$). First and second columns of each dataset represent the reduced $\chi^2$ and corresponding $p$-value. Third, fourth and last columns represent the independent fit parameters, Akaike weights, and whether or not the fit results satisfy the constraint $\mathcal{B}(B_c\to \tau\nu_\tau) \le 30\%$ respectively. `\checkmark {\bf !}' means that only some of the multiple minima satisfy this limit for the scenario in question. The scenario where $C_{V_1}$ is complex (i.e. both  $\mathcal{I}m(C_{V_1})$ and $\mathcal{R}e(C_{V_1})$ present), we consider this as a single parameter case for model selection   }
		\label{tab:alldat1}
%	\begin{ruledtabular}
 \resizebox{\textwidth}{!}{
		\begin{tabular}{c||ccccc||ccccc||ccccc}
			\hline
			& \multicolumn{5}{c||}{Data Without $\mathcal{R}_{J/\Psi}$} & \multicolumn{5}{c||}{All Data ($\mathcal{R}_{J/\Psi}$ with LFCQ)} & \multicolumn{5}{c}{All Data ($\mathcal{R}_{J/\Psi}$ with PQCD)}\\
			\cline{2-16}
			& $\chi^2_{min}$ & $p$-val & Param.s & & $B_c\to$ & $\chi^2_{min}$ & $p$-val & Param.s & & $B_c\to$ & $\chi^2_{min}$ & $p$-val & Param.s & & $B_c\to$ \\
			Index & $/ $ DoF & (\%) & & $w^{\text{AIC}_c}$ & $\tau\nu$ & $/ $ DoF & (\%) & & $w^{\text{AIC}_c}$ & $\tau\nu$ & $/ $ DoF & (\%) & & $w^{\text{AIC}_c}$ & $\tau\nu$ \\
			\hline
			1 & \text{4.05/8} & 85.3 & $\mathcal{R}e(C_T)$ & 29.80 & \checkmark & \text{7.24/9} & 61.22 & $\mathcal{R}e(C_{V_1})$ & 20.44 & \checkmark & \text{6.46/9} & 69.34 & $\mathcal{R}e(C_{S_2})$ & 26.57 & $\pmb{\times}$ \\
			2 & \text{4.58/8} & 71.09 & $\mathcal{I}m(C_{V_1})$,$\mathcal{R}e(C_{V_1})$ & 17.44 & \checkmark & \text{7.24/9} & 51.1 & $\mathcal{I}m(C_{V_1})$,$\mathcal{R}e(C_{V_1})$ & 20.44 & \checkmark & \text{6.68/9} & 67.01 & $\mathcal{R}e(C_{V_1})$ & 21.21 &  \checkmark \\
			3 & \text{4.58/8} & 80.13 & $\mathcal{R}e(C_{V_1})$ & 17.44 & \checkmark & \text{7.28/9} & 60.78 & $\mathcal{R}e(C_{S_2})$ & 19.60 & $\pmb{\times}$ & \text{6.68/9} & 57.12  & $\mathcal{I}m(C_{V_1})$,$\mathcal{R}e(C_{V_1})$  & 21.21  & \checkmark\\
			4 & \text{4.64/8} & 79.54 & $\mathcal{R}e(C_{S_2})$ & 16.47 & $\pmb{\times}$ & \text{7.49/9} & 58.59 & $\mathcal{R}e(C_T)$ & 15.87 & \checkmark & \text{8.21/9} & 51.29 & $\mathcal{R}e(C_T)$ & 4.59 & \checkmark \\
			5 & \text{3.54/7} & 83.07 & $\mathcal{I}m(C_{S_2})$,$\mathcal{R}e(C_{S_2})$ & 1.59 & $\pmb{\times}$ & \text{6.18/8} & 62.68 & $\mathcal{R}e(C_T)$,$\mathcal{R}e(C_{V_2})$ & 2.37 & \checkmark & \text{5.63/8} & 68.82 & $\mathcal{R}e(C_{S_2})$,$\mathcal{R}e(C_{V_1})$ & 2.43 & \checkmark {\bf !}\\
			6 & \text{3.54/7} & 83.07 & $\mathcal{R}e(C_{S_1})$,$\mathcal{R}e(C_{S_2})$ & 1.59 & $\pmb{\times}$ & \text{6.38/8} & 60.43 & $\mathcal{R}e(C_{S_1})$,$\mathcal{R}e(C_T)$ & 1.93 & \checkmark {\bf !} & \text{5.65/8} & 68.6 & $\mathcal{R}e(C_{S_1})$,$\mathcal{R}e(C_{S_2})$ & 2.39 & $\pmb{\times}$ \\
			7 & \text{3.56/7} & 82.9 & $\mathcal{R}e(C_{S_2})$,$\mathcal{R}e(C_{V_1})$ & 1.57 & \checkmark {\bf !} & \text{6.4/8} & 60.22 & $\mathcal{R}e(C_{S_1})$,$\mathcal{R}e(C_{S_2})$ & 1.90 & $\pmb{\times}$ & \text{5.65/8} & 68.59 & $\mathcal{R}e(C_{S_2})$,$\mathcal{R}e(C_{V_2})$ & 2.38 & \checkmark {\bf !} \\
			8 & \text{3.56/7} & 82.9 & $\mathcal{R}e(C_{S_2})$,$\mathcal{R}e(C_T)$ & 1.57 & \checkmark {\bf !} & \text{6.4/8} & 60.21 & $\mathcal{I}m(C_{S_2})$,$\mathcal{R}e(C_{S_2})$ & 1.90 & $\pmb{\times}$ & \text{5.65/8} & 68.59 & $\mathcal{I}m(C_{S_2})$,$\mathcal{R}e(C_{S_2})$ & 2.38 & $\pmb{\times}$ \\
			9 & \text{3.56/7} & 82.88 & $\mathcal{R}e(C_{S_2})$,$\mathcal{R}e(C_{V_2})$ & 1.57 & \checkmark {\bf !} & \text{6.42/8} & 60.02 & $\mathcal{R}e(C_{S_2})$,$\mathcal{R}e(C_T)$ & 1.86 & \checkmark {\bf !} & \text{5.66/8} & 68.55 & $\mathcal{R}e(C_T)$,$\mathcal{R}e(C_{V_2})$ & 2.38 & \checkmark \\
			10 & \text{3.62/7} & 82.23 & $\mathcal{R}e(C_T)$,$\mathcal{R}e(C_{V_2})$ & 1.48 & \checkmark & \text{6.46/8} & 59.58 & $\mathcal{R}e(C_{S_2})$,$\mathcal{R}e(C_{V_1})$ & 1.79 & \checkmark {\bf !} & \text{5.68/8} & 68.31 & $\mathcal{R}e(C_{S_2})$,$\mathcal{R}e(C_T)$ & 2.32 & \checkmark {\bf !} \\
			11 & \text{3.69/7} & 81.45 & $\mathcal{R}e(C_{S_1})$,$\mathcal{R}e(C_T)$ & 1.37 & \checkmark {\bf !} & \text{6.46/8} & 59.54 & $\mathcal{R}e(C_{S_1})$,$\mathcal{R}e(C_{V_2})$ & 1.78 & \checkmark {\bf !} & \text{5.79/8} & 67.03 & $\mathcal{R}e(C_{S_1})$,$\mathcal{R}e(C_T)$ & 2.07 & \checkmark {\bf !} \\
			12 & \text{3.7/7} & 81.31 & $\mathcal{R}e(C_{S_1})$,$\mathcal{R}e(C_{V_2})$ & 1.36 & \checkmark {\bf !} & \text{6.47/8} & 59.45 & $\mathcal{R}e(C_{S_2})$,$\mathcal{R}e(C_{V_2})$ & 1.77 & \checkmark {\bf !} & \text{5.85/8} & 66.42 & $\mathcal{R}e(C_{S_1})$,$\mathcal{R}e(C_{V_2})$ & 1.96 & \checkmark {\bf !} \\
			13 & \text{3.76/7} & 80.71 & $\mathcal{R}e(C_{S_1})$,$\mathcal{R}e(C_{V_1})$ & 1.29 & \checkmark {\bf !} & \text{6.52/8} & 58.91 & $\mathcal{R}e(C_{S_1})$,$\mathcal{R}e(C_{V_1})$ & 1.69 & \checkmark {\bf !} & \text{5.96/8} & 65.22 & $\mathcal{R}e(C_{S_1})$,$\mathcal{R}e(C_{V_1})$ & 1.76 & \checkmark {\bf !} \\
			14 & \text{3.79/7} & 80.37 & $\mathcal{R}e(C_{V_1})$,$\mathcal{R}e(C_{V_2})$ & 1.25 & \checkmark & \text{6.55/8} & 58.58 & $\mathcal{I}m(C_{V_2})$,$\mathcal{R}e(C_{V_2})$ & 1.64 & \checkmark & \text{6.01/8} & 64.62 & $\mathcal{R}e(C_{V_1})$,$\mathcal{R}e(C_{V_2})$ & 1.67 & \checkmark \\
			15 & \text{3.79/7} & 80.37 & $\mathcal{I}m(C_{V_2})$,$\mathcal{R}e(C_{V_2})$ & 1.25 & \checkmark & \text{6.55/8} & 58.58 & $\mathcal{R}e(C_{V_1})$,$\mathcal{R}e(C_{V_2})$ & 1.64 & \checkmark & \text{6.01/8} & 64.62 & $\mathcal{I}m(C_{V_2})$,$\mathcal{R}e(C_{V_2})$ & 1.67 & \checkmark \\
			16 & \text{3.82/7} & 80.08 & $\mathcal{R}e(C_T)$,$\mathcal{R}e(C_{V_1})$ & 1.22 & \checkmark & \text{6.63/8} & 57.68 & $\mathcal{R}e(C_T)$,$\mathcal{R}e(C_{V_1})$ & 1.51 & \checkmark & \text{6.1/8} & 63.63 & $\mathcal{R}e(C_T)$,$\mathcal{R}e(C_{V_1})$ & 1.53 & \checkmark \\
			17 & \text{3.87/7} & 79.49 & $\mathcal{I}m(C_T)$,$\mathcal{R}e(C_T)$ & 1.16 & \checkmark & \text{7.13/8} & 52.25 & $\mathcal{I}m(C_T)$,$\mathcal{R}e(C_T)$ & 0.92 & \checkmark & $-$  & $-$ & $-$ & $-$ & $-$ \\
% 			17 & \text{4.58/7} & 71.09 & $\mathcal{I}m(C_{V_1})$,$\mathcal{R}e(C_{V_1})$ & 0.68 & \checkmark & \text{7.24/8} & 51.1 & $\mathcal{I}m(C_{V_1})$,$\mathcal{R}e(C_{V_1})$ & 1.02 & \checkmark & $-$ & $-$ & $-$ & $-$ & $-$ \\
			\hline
		\end{tabular}
		}
		
%	\end{ruledtabular}
\end{table*}

\begin{table*}[!htbp]
	\centering
	\caption{\small Results similar to table \ref{tab:alldat1}, but with $P_{\tau}(D^*)$ dropped (with or without $\mathcal{R}_{J/\Psi}$).}
		\label{tab:noptau1}
%	\begin{ruledtabular}
\resizebox{\textwidth}{!}{
		\begin{tabular}{c||ccccc||ccccc||ccccc}
			\hline
			& \multicolumn{5}{c||}{Data without $P_{\tau}(D^*)$ and $\mathcal{R}_{J/\Psi}$} & \multicolumn{5}{c||}{Data without $P_{\tau}(D^*)$ ($\mathcal{R}_{J/\Psi}$ with LFCQ)} & \multicolumn{5}{c}{Data without $P_{\tau}(D^*)$ ($\mathcal{R}_{J/\Psi}$ with PQCD)}\\
			\cline{2-16}
			& $\chi^2_{min}$ & $p$-val & Param.s & & $B_c\to$ & $\chi^2_{min}$ & $p$-val & Param.s & & $B_c\to$ & $\chi^2_{min}$ & $p$-val & Param.s & & $B_c\to$ \\
			Index & $/ $ DoF & (\%) & & $w^{\text{AIC}_c}$ &  $\tau\nu$ & $/ $ DoF & (\%) & & $w^{\text{AIC}_c}$ &  $\tau\nu$ & $/ $ DoF & (\%) & & $w^{\text{AIC}_c}$ &  $\tau\nu$ \\
			\hline
			1 & \text{3.55/7} & 82.95 & $\mathcal{R}e(C_T)$ & 37.91 & \checkmark & \text{6.92/8} & 54.53 & $\mathcal{R}e(C_{V_1})$ & 22.57 & \checkmark & 6.36/8  & 49.81 &  $\mathcal{I}m(C_{V_1})$,$\mathcal{R}e(C_{V_1})$ & 25.14 & \checkmark \\
			2 & \text{4.27/7} & 64.01 & $\mathcal{I}m(C_{V_1})$,$\mathcal{R}e(C_{V_1})$ & 18.51 & \checkmark & \text{6.92/8} & 43.72 & $\mathcal{I}m(C_{V_1})$,$\mathcal{R}e(C_{V_1})$ & 22.57 & \checkmark & \text{6.36/8} & 60.67 & $\mathcal{R}e(C_{V_1})$ & 25.14 & \checkmark\\
			3 & \text{4.27/7} & 74.81 & $\mathcal{R}e(C_{V_1})$ & 18.51 & \checkmark & \text{7.01/8} & 53.57 & $\mathcal{R}e(C_T)$ & 20.65 & \checkmark & \text{6.45/8} & 59.65 & $\mathcal{R}e(C_{S_2})$ & 22.94 & $\pmb{\times}$ \\
			4 & \text{4.64/7} & 70.44 & $\mathcal{R}e(C_{S_2})$ & 12.86 & $\pmb{\times}$ & \text{7.28/8} & 50.69 & $\mathcal{R}e(C_{S_2})$ & 15.77 & $\pmb{\times}$ & \text{7.73/8} & 46.05 & $\mathcal{R}e(C_T)$ & 6.41 & \checkmark \\
			5 & \text{3.54/6} & 73.88 & $\mathcal{R}e(C_{V_1})$,$\mathcal{R}e(C_{V_2})$ & 0.92 & \checkmark & \text{6.09/7} & 52.87 & $\mathcal{R}e(C_T)$,$\mathcal{R}e(C_{V_2})$ & 1.67 & \checkmark & \text{5.57/7} & 59.08 & $\mathcal{R}e(C_T)$,$\mathcal{R}e(C_{V_2})$ & 1.80 & \checkmark \\
			6 & \text{3.54/6} & 73.88 & $\mathcal{R}e(C_{S_1})$,$\mathcal{R}e(C_{V_1})$ & 0.92 & \checkmark {\bf !} & \text{6.11/7} & 52.66 & $\mathcal{R}e(C_{S_1})$,$\mathcal{R}e(C_T)$ & 1.64 & \checkmark {\bf !} & \text{5.62/7} & 58.47 & $\mathcal{R}e(C_{S_2})$,$\mathcal{R}e(C_{V_1})$ & 1.71 & \checkmark {\bf !} \\
			7 & \text{3.54/6} & 73.88 & $\mathcal{R}e(C_{S_2})$,$\mathcal{R}e(C_{V_1})$ & 0.92 & \checkmark {\bf !} & \text{6.18/7} & 51.95 & $\mathcal{R}e(C_{S_2})$,$\mathcal{R}e(C_T)$ & 1.54 & \checkmark {\bf !} & \text{5.63/7} & 58.36 & $\mathcal{R}e(C_{S_2})$,$\mathcal{R}e(C_T)$ & 1.70 &  \checkmark {\bf !} \\
			8 & \text{3.54/6} & 73.88 & $\mathcal{R}e(C_T)$,$\mathcal{R}e(C_{V_1})$ & 0.92 & \checkmark & \text{6.18/7} & 51.83 & $\mathcal{R}e(C_{S_1})$,$\mathcal{R}e(C_{V_2})$ & 1.53 & \checkmark {\bf !} & \text{5.63/7} & 58.35 & $\mathcal{R}e(C_{S_1})$,$\mathcal{R}e(C_T)$ & 1.69 & \checkmark {\bf !} \\
			9 & \text{3.54/6} & 73.88 & $\mathcal{R}e(C_{S_1})$,$\mathcal{R}e(C_{V_2})$ & 0.92 & \checkmark {\bf !} & \text{6.22/7} & 51.47 & $\mathcal{R}e(C_{S_1})$,$\mathcal{R}e(C_{V_1})$ & 1.48 & \checkmark {\bf !} & \text{5.64/7} & 58.23 & $\mathcal{R}e(C_{S_2})$,$\mathcal{R}e(C_{V_2})$ & 1.68 & \checkmark {\bf !} \\
			10 & \text{3.54/6} & 73.88 & $\mathcal{R}e(C_{S_2})$,$\mathcal{R}e(C_{V_2})$ & 0.92 & \checkmark {\bf !} & \text{6.24/7} & 51.25 & $\mathcal{R}e(C_{S_2})$,$\mathcal{R}e(C_{V_2})$ & 1.45 & \checkmark {\bf !} & \text{5.65/7} & 58.12 & $\mathcal{R}e(C_{S_1})$,$\mathcal{R}e(C_{S_2})$ & 1.66 & $\pmb{\times}$ \\
			11 & \text{3.54/6} & 73.88 & $\mathcal{R}e(C_T)$,$\mathcal{R}e(C_{V_2})$ & 0.92 & \checkmark & \text{6.27/7} & 50.81 & $\mathcal{R}e(C_{S_2})$,$\mathcal{R}e(C_{V_1})$ & 1.39 & \checkmark {\bf !} & \text{5.65/7} & 58.1 & $\mathcal{I}m(C_{S_2})$,$\mathcal{R}e(C_{S_2})$ & 1.66 & $\pmb{\times}$ \\
			12 & \text{3.54/6} & 73.88 & $\mathcal{R}e(C_{S_1})$,$\mathcal{R}e(C_{S_2})$ & 0.92 & $\pmb{\times}$ & \text{6.29/7} & 50.66 & $\mathcal{R}e(C_{V_1})$,$\mathcal{R}e(C_{V_2})$ & 1.38 & \checkmark & \text{5.67/7} & 57.87 & $\mathcal{R}e(C_{S_1})$,$\mathcal{R}e(C_{V_2})$ & 1.63 & \checkmark {\bf !} \\
			13 & \text{3.54/6} & 73.88 & $\mathcal{R}e(C_{S_1})$,$\mathcal{R}e(C_T)$ & 0.92 & \checkmark {\bf !} & \text{6.29/7} & 50.66 & $\mathcal{I}m(C_{V_2})$,$\mathcal{R}e(C_{V_2})$ & 1.38 & \checkmark & \text{5.72/7} & 57.26 & $\mathcal{R}e(C_{S_1})$,$\mathcal{R}e(C_{V_1})$ & 1.55 & \checkmark {\bf !} \\
			14 & \text{3.54/6} & 73.88 & $\mathcal{R}e(C_{S_2})$,$\mathcal{R}e(C_T)$ & 0.92 & \checkmark {\bf !} & \text{6.34/7} & 50.05 & $\mathcal{R}e(C_T)$,$\mathcal{R}e(C_{V_1})$ & 1.31 & \checkmark & \text{5.74/7} & 56.98 & $\mathcal{I}m(C_{V_2})$,$\mathcal{R}e(C_{V_2})$ & 1.51 & \checkmark \\
			15 & \text{3.54/6} & 73.88 & $\mathcal{I}m(C_{V_2})$,$\mathcal{R}e(C_{V_2})$ & 0.92 & \checkmark & \text{6.4/7} & 49.4 & $\mathcal{R}e(C_{S_1})$,$\mathcal{R}e(C_{S_2})$ & 1.23 & $\pmb{\times}$ & \text{5.74/7} & 56.98 & $\mathcal{R}e(C_{V_1})$,$\mathcal{R}e(C_{V_2})$ & 1.51 & \checkmark \\
			16 & \text{3.54/6} & 73.88 & $\mathcal{I}m(C_{S_2})$,$\mathcal{R}e(C_{S_2})$ & 0.92 & $\pmb{\times}$ & \text{6.4/7} & 49.39 & $\mathcal{I}m(C_{S_2})$,$\mathcal{R}e(C_{S_2})$ & 1.23 & $\pmb{\times}$ & \text{5.81/7} & 56.26 & $\mathcal{R}e(C_T)$,$\mathcal{R}e(C_{V_1})$ & 1.42 & \checkmark \\
			17 & \text{3.54/6} & 73.88 & $\mathcal{I}m(C_T)$,$\mathcal{R}e(C_T)$ & 0.92 & \checkmark & \text{6.9/7} & 43.92 & $\mathcal{I}m(C_T)$,$\mathcal{R}e(C_T)$ & 0.74 & \checkmark & $-$ & $-$ & $-$ & $-$ & $-$ \\
% 			18 & $-$ & $-$ & $-$ & $-$ & $-$ & \text{6.92/7} & 43.72 & $\mathcal{I}m(C_{V_1})$,$\mathcal{R}e(C_{V_1})$ & 0.94 & \checkmark & $-$ & $-$ & $-$ & $-$ & $-$ \\
			\hline
		\end{tabular}
		}
%	\end{ruledtabular}
\end{table*}

\begin{table*}[t]
	\centering
	\caption{\small Results similar to table \ref{tab:alldat1}, but with data from \Babar~ dropped (i.e., only with Belle and LHCb data; with or without $\mathcal{R}_{J/\Psi}$).}
		\label{tab:beL1}
%	\begin{ruledtabular}
\resizebox{\textwidth}{!}{
		\begin{tabular}{c||ccccc||ccccc||ccccc}
			\hline
			& \multicolumn{5}{c||}{Belle + LHCb (Except $\mathcal{R}_{J/\Psi}$) } & \multicolumn{5}{c||}{Belle + LHCb ($\mathcal{R}_{J/\Psi}$ with LFCQ)} & \multicolumn{5}{c}{Belle + LHCb ($\mathcal{R}_{J/\Psi}$ with PQCD)}\\
			\cline{2-16}
			& $\chi^2_{min}$ & $p$-val & Param.s & & $B_c\to$ & $\chi^2_{min}$ & $p$-val & Param.s & & $B_c\to$ & $\chi^2_{min}$ & $p$-val & Param.s & & $B_c\to$ \\
			Index & $/ $ DoF & (\%) & & $w^{\text{AIC}_c}$ & $\tau\nu$ & $/ $ DoF & (\%) & & $w^{\text{AIC}_c}$ & $\tau\nu$ & $/ $ DoF & (\%) & & $w^{\text{AIC}_c}$ & $\tau\nu$ \\
			\hline
			1 & \text{1.74/6} & 88.4 &  $\mathcal{I}m(C_{V_1})$,$\mathcal{R}e(C_{V_1})$ & 31.36 & \checkmark & \text{4.56/7} & 71.35 & $\mathcal{R}e(C_{V_1})$ & 32.07 & \checkmark & \text{4.01/7} & 67.55 & $\mathcal{I}m(C_{V_1})$,$\mathcal{R}e(C_{V_1})$ & 30.64 & \checkmark\\
			2 & \text{1.74/6} & 94.2 & $\mathcal{R}e(C_{V_1})$ & 31.36 & \checkmark & \text{4.56/7} & 60.14 & $\mathcal{I}m(C_{V_1})$,$\mathcal{R}e(C_{V_1})$  & 32.07 & \checkmark & \text{4.01/7} & 77.87 & $\mathcal{R}e(C_{V_1})$ & 30.64 & \checkmark \\
			3 & \text{2.41/6} & 87.8 & $\mathcal{R}e(C_{T})$ & 15.99 & \checkmark & \text{5.47/7} & 60.25 & $\mathcal{R}e(C_{S_2})$ & 12.86 & $\pmb{\times}$ & \text{4.66/7} & 70.08 & $\mathcal{R}e(C_{S_2})$ & 15.90 & $\pmb{\times}$ \\
			4 & \text{2.78/6} & 83.57 & $\mathcal{R}e(C_{S_2})$ & 11.06 & $\pmb{\times}$ & \text{5.77/7} & 56.65 & $\mathcal{R}e(C_{T})$ & 9.53 & \checkmark & \text{5.21/7} & 63.45 & $\mathcal{R}e(C_{T})$ & 9.23 & \checkmark \\
			5 & \text{4.74/6} & 57.81 & $\mathcal{R}e(C_{S_1})$ & 1.57 & \checkmark & \text{7.92/7} & 33.97 & $\mathcal{R}e(C_{V_2})$ & 1.11 & \checkmark & \text{7.35/7} & 39.34 & $\mathcal{R}e(C_{V_2})$ & 1.08 & \checkmark \\
			6 & \text{5.03/6} & 54. & $\mathcal{R}e(C_{V_2})$ & 1.17 & \checkmark & \text{4.23/6} & 64.62 & $\mathcal{R}e(C_{T})$,$\mathcal{R}e(C_{V_2})$ & 1.07 & \checkmark & \text{3.65/6} & 72.33 & $\mathcal{R}e(C_{S_2})$,$\mathcal{R}e(C_{V_1})$ & 1.04 & \checkmark {\bf !} \\
			7 & \text{1.45/5} & 91.9 & $\mathcal{I}m(C_{S_2})$,$\mathcal{R}e(C_{S_2})$ & 0.63 & $\pmb{\times}$ & \text{8.04/7} & 32.94 & $\mathcal{R}e(C_{S_1})$ & 0.99 & \checkmark & \text{3.67/6} & 72.11 & $\mathcal{R}e(C_{S_1})$,$\mathcal{R}e(C_{S_2})$ & 1.03 & $\pmb{\times}$ \\
			8 & \text{1.45/5} & 91.9 & $\mathcal{R}e(C_{S_1})$,$\mathcal{R}e(C_{S_2})$ & 0.63 & $\pmb{\times}$ & \text{4.36/6} & 62.76 & $\mathcal{R}e(C_{S_1})$,$\mathcal{R}e(C_{T})$ & 0.93 & \checkmark {\bf !} & \text{3.67/6} & 72.09 & $\mathcal{I}m(C_{S_2})$,$\mathcal{R}e(C_{S_2})$ & 1.03 & $\pmb{\times}$ \\
			9 & \text{1.48/5} & 91.58 & $\mathcal{R}e(C_{S_2})$,$\mathcal{R}e(C_{T})$ & 0.61 & \checkmark {\bf !} & \text{4.38/6} & 62.49 & $\mathcal{R}e(C_{S_2})$,$\mathcal{R}e(C_{T})$ & 0.91 & \checkmark {\bf !} & \text{3.68/6} & 71.99 & $\mathcal{R}e(C_{S_2})$,$\mathcal{R}e(C_{V_2})$ & 1.02 & \checkmark {\bf !} \\
			10 & \text{1.48/5} & 91.58 & $\mathcal{R}e(C_{S_2})$,$\mathcal{R}e(C_{V_1})$ & 0.61 & \checkmark {\bf !} & \text{4.39/6} & 62.42 & $\mathcal{R}e(C_{S_1})$,$\mathcal{R}e(C_{S_2})$ & 0.91 & $\pmb{\times}$ & \text{3.69/6} & 71.85 & $\mathcal{R}e(C_{T})$,$\mathcal{R}e(C_{V_2})$ & 1.01 & \checkmark \\
			11 & \text{1.48/5} & 91.55 & $\mathcal{R}e(C_{S_2})$,$\mathcal{R}e(C_{V_2})$ & 0.61 & \checkmark {\bf !} & \text{4.39/6} & 62.4 & $\mathcal{I}m(C_{S_2})$,$\mathcal{R}e(C_{S_2})$ & 0.91 & $\pmb{\times}$ & \text{3.71/6} & 71.57 & $\mathcal{R}e(C_{S_2})$,$\mathcal{R}e(C_{T})$ & 0.99 & \checkmark {\bf !} \\
			12 & \text{1.53/5} & 91. & $\mathcal{R}e(C_{T})$,$\mathcal{R}e(C_{V_2})$ & 0.58 & \checkmark & \text{4.43/6} & 61.83 & $\mathcal{R}e(C_{S_1})$,$\mathcal{R}e(C_{V_2})$ & 0.87 & \checkmark {\bf !} & \text{7.49/7} & 38. & $\mathcal{R}e(C_{S_1})$ & 0.95 & \checkmark \\
			13 & \text{1.53/5} & 90.96 & $\mathcal{I}m(C_{T})$,$\mathcal{R}e(C_{T})$ & 0.58 & \checkmark & \text{4.47/6} & 61.38 & $\mathcal{R}e(C_{S_2})$,$\mathcal{R}e(C_{V_1})$ & 0.84 & \checkmark {\bf !} & \text{3.79/6} & 70.57 & $\mathcal{R}e(C_{S_1})$,$\mathcal{R}e(C_{T})$ & 0.92 & \checkmark {\bf !} \\
			14 & $-$ & $-$ & $-$ & $-$ & $-$ & \text{4.48/6} & 61.21 & $\mathcal{R}e(C_{S_1})$,$\mathcal{R}e(C_{V_1})$ & 0.83 & \checkmark {\bf !} & \text{3.83/6} & 69.91 & $\mathcal{R}e(C_{S_1})$,$\mathcal{R}e(C_{V_2})$ & 0.87 & \checkmark {\bf !} \\
			15 & $-$ & $-$ & $-$ & $-$ & $-$ & \text{4.48/6} & 61.2 & $\mathcal{R}e(C_{S_2})$,$\mathcal{R}e(C_{V_2})$ & 0.83 & \checkmark {\bf !} & \text{3.92/6} & 68.68 & $\mathcal{R}e(C_{S_1})$,$\mathcal{R}e(C_{V_1})$ & 0.80 & \checkmark {\bf !} \\
			16 & $-$ & $-$ & $-$ & $-$ & $-$ & \text{4.49/6} & 61.06 & $\mathcal{R}e(C_{V_1})$,$\mathcal{R}e(C_{V_2})$ & 0.82 & \checkmark & \text{3.95/6} & 68.39 & $\mathcal{I}m(C_{V_2})$,$\mathcal{R}e(C_{V_2})$ & 0.78 & \checkmark \\
			17 & $-$ & $-$ & $-$ & $-$ & $-$ & \text{4.49/6} & 61.06 & $\mathcal{I}m(C_{V_2})$,$\mathcal{R}e(C_{V_2})$ & 0.82 & \checkmark & \text{3.95/6} & 68.39 & $\mathcal{R}e(C_{V_1})$,$\mathcal{R}e(C_{V_2})$ & 0.78 & \checkmark \\
			18 & $-$ & $-$ & $-$ & $-$ & $-$ & \text{4.52/6} & 60.66 & $\mathcal{R}e(C_{T})$,$\mathcal{R}e(C_{V_1})$ & 0.80 & \checkmark & \text{3.98/6} & 67.99 & $\mathcal{R}e(C_{T})$,$\mathcal{R}e(C_{V_1})$ & 0.76 & \checkmark \\
			19 & $-$ & $-$ & $-$ & $-$ & $-$ & \text{4.67/6} & 58.65 & $\mathcal{I}m(C_{T})$,$\mathcal{R}e(C_{T})$ &  0.69 & \checkmark  & $-$ & $-$ & $-$ & $-$ & $-$  \\
% 			19 & $-$ & $-$ & $-$ & $-$ & $-$ &  &  &  & &  & $-$ & $-$ & $-$ & $-$ & $-$ \\
			\hline
		\end{tabular}
		}
%	\end{ruledtabular}
\end{table*}%\end{turnpage}

\begin{table*}[htbp]
	\centering
	\caption{\small Results similar to table \ref{tab:alldat1}, but only with all $\mathcal{R}_{D^*}$ data (with or without $\mathcal{R}_{J/\Psi}$).}
		\label{tab:allrdst1}
%	\begin{ruledtabular}
\resizebox{\textwidth}{!}{
		\begin{tabular}{c||ccccc||ccccc||ccccc}
			\hline
			& \multicolumn{5}{c||}{All $\mathcal{R}_{D^*}$} & \multicolumn{5}{c||}{All $\mathcal{R}_{D^*}$ + $\mathcal{R}_{J/\Psi}$ (LFCQ)} & \multicolumn{5}{c|}{All $\mathcal{R}_{D^*}$ + $\mathcal{R}_{J/\Psi}$ (PQCD)}\\
			\cline{2-16}
			& $\chi^2_{min}$ & $p$-val & Param.s & & $B_c\to$ & $\chi^2_{min}$ & $p$-val & Param.s & & $B_c\to$ & $\chi^2_{min}$ & $p$-val & Param.s & & $B_c\to$ \\
			Index & $/ $ DoF & (\%) & & $w^{\text{AIC}_c}$ & $\tau\nu$ & $/ $ DoF & (\%) & & $w^{\text{AIC}_c}$ & $\tau\nu$ & $/ $ DoF & (\%) & & $w^{\text{AIC}_c}$ & $\tau\nu$ \\
			\hline
			1 & \text{2.43/5} & 78.76 & $\mathcal{R}e(C_{V_1})$ & 16.41 & \checkmark & \text{5.08/6} & 53.35 & $\mathcal{R}e(C_{T})$ & 17.80 & \checkmark & \text{4.51/6} & 60.75 & $\mathcal{R}e(C_{T})$ & 16.82 & \checkmark \\
			2 & \text{2.43/5} & 78.76 & $\mathcal{R}e(C_{V_2})$ & 16.41 & \checkmark & \text{5.15/6} & 52.53 & $\mathcal{R}e(C_{V_2})$ & 16.68 & \checkmark & \text{4.53/6} & 60.49 & $\mathcal{R}e(C_{S_1})$ & 16.49 & $\pmb{\times}$ \\
			3 & \text{2.43/5} & 78.76 & $\mathcal{R}e(C_{S_1})$ & 16.41 & $\pmb{\times}$ & \text{5.16/6} & 39.61 & $\mathcal{I}m(C_{V_1})$,$\mathcal{R}e(C_{V_1})$ & 16.36 & \checkmark & \text{4.53/6} & 60.49 & $\mathcal{R}e(C_{S_2})$ & 16.49 & $\pmb{\times}$ \\
			4 & \text{2.43/5} & 78.76 & $\mathcal{R}e(C_{S_2})$ & 16.41 & $\pmb{\times}$ & \text{5.16/6} & 52.29 & $\mathcal{R}e(C_{V_1})$ & 16.36 & \checkmark & \text{4.56/6} & 60.13 & $\mathcal{R}e(C_{V_2})$ & 16.05 & \checkmark \\
			5 & \text{2.43/5} & 78.76 & $\mathcal{R}e(C_{T})$ & 16.41 & \checkmark & \text{5.28/6} & 50.80 & $\mathcal{R}e(C_{S_1})$ & 14.52 & $\pmb{\times}$ & \text{4.61/6} & 46.53 & $\mathcal{I}m(C_{V_1})$,$\mathcal{R}e(C_{V_1})$ &15.27 & \checkmark \\
			6 & 2.43/5 & 65.80 & $\mathcal{I}m(C_{V_1})$,$\mathcal{R}e(C_{V_1})$ & 16.41 &  \checkmark & \text{5.28/6} & 50.80 & $\mathcal{R}e(C_{S_2})$ & 14.52 & $\pmb{\times}$ & 4.61/6 & 59.47 & $\mathcal{R}e(C_{V_1})$ & 15.27 &\checkmark \\
			7 & $-$ & $-$ & $-$ & $-$ & $-$ & \text{4.8/5} & 44.03&$\mathcal{R}e(C_{T})$,$\mathcal{R}e(C_{V_2})$& 0.35 &\checkmark & $-$ & $-$ & $-$ & $-$ & $-$ \\
			\hline
		\end{tabular}
		}
%	\end{ruledtabular}
\end{table*}
%%%%%%%%%%%%%%%%%%%%%%%%%%%%%%%%%%%%%%%%%%%%%%%%%%%%%%%%%%%%%%%%	
\section{Methodology}
%\subsection{Methodology}
%%%%%%%%%%%%%%%%%%%%%%%%%%%%%%%%%%%%%%%%%%%%%%%%%%%%%%%%%%%%%%%%

\begin{figure*}[t]
	\centering
	\subfloat[Scenario $1$]{\includegraphics[height=3.5cm]{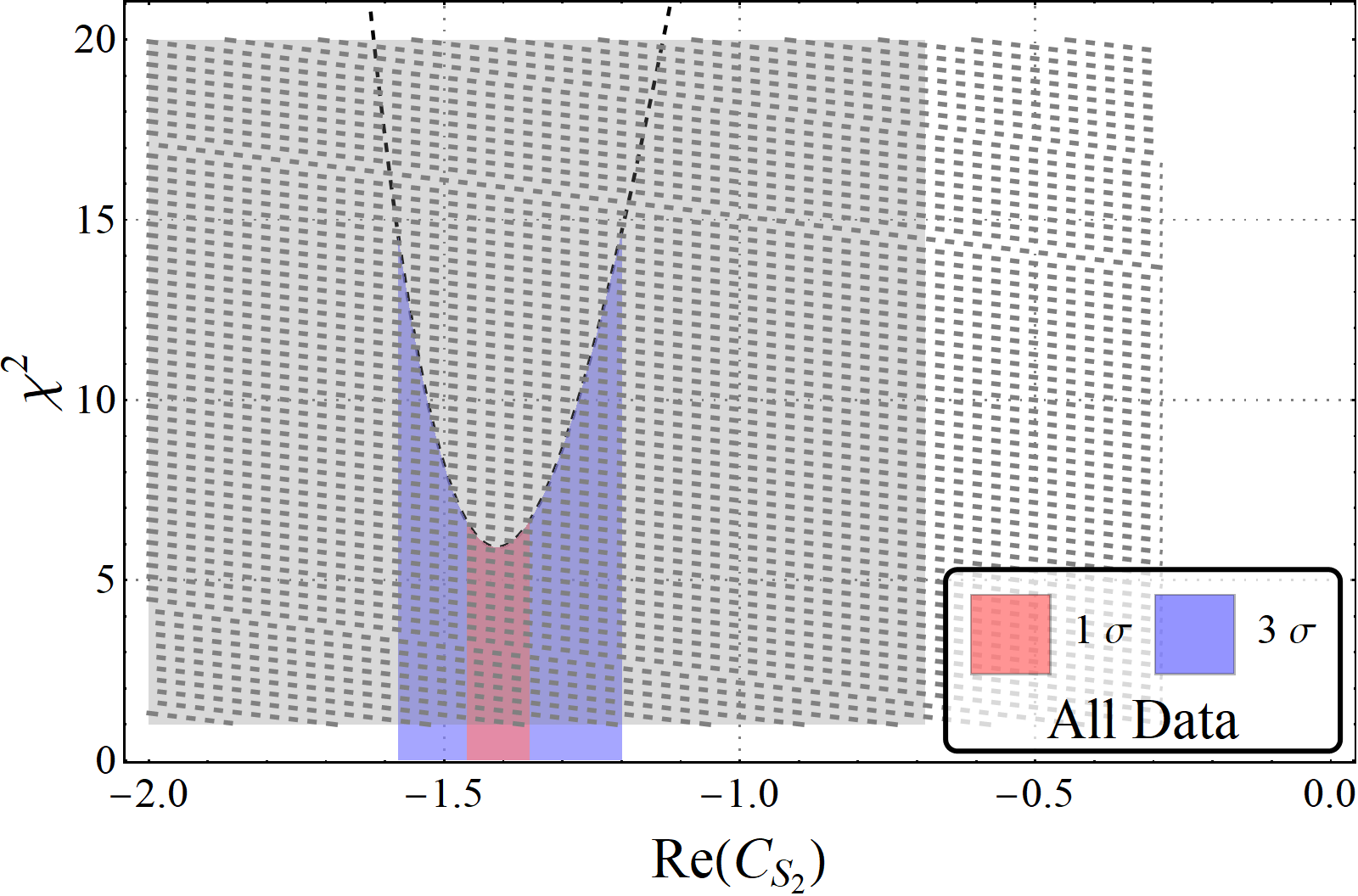}\label{fig:alldat1JPQ}}~
	\subfloat[Scenario $2$]{\includegraphics[height=3.5cm]{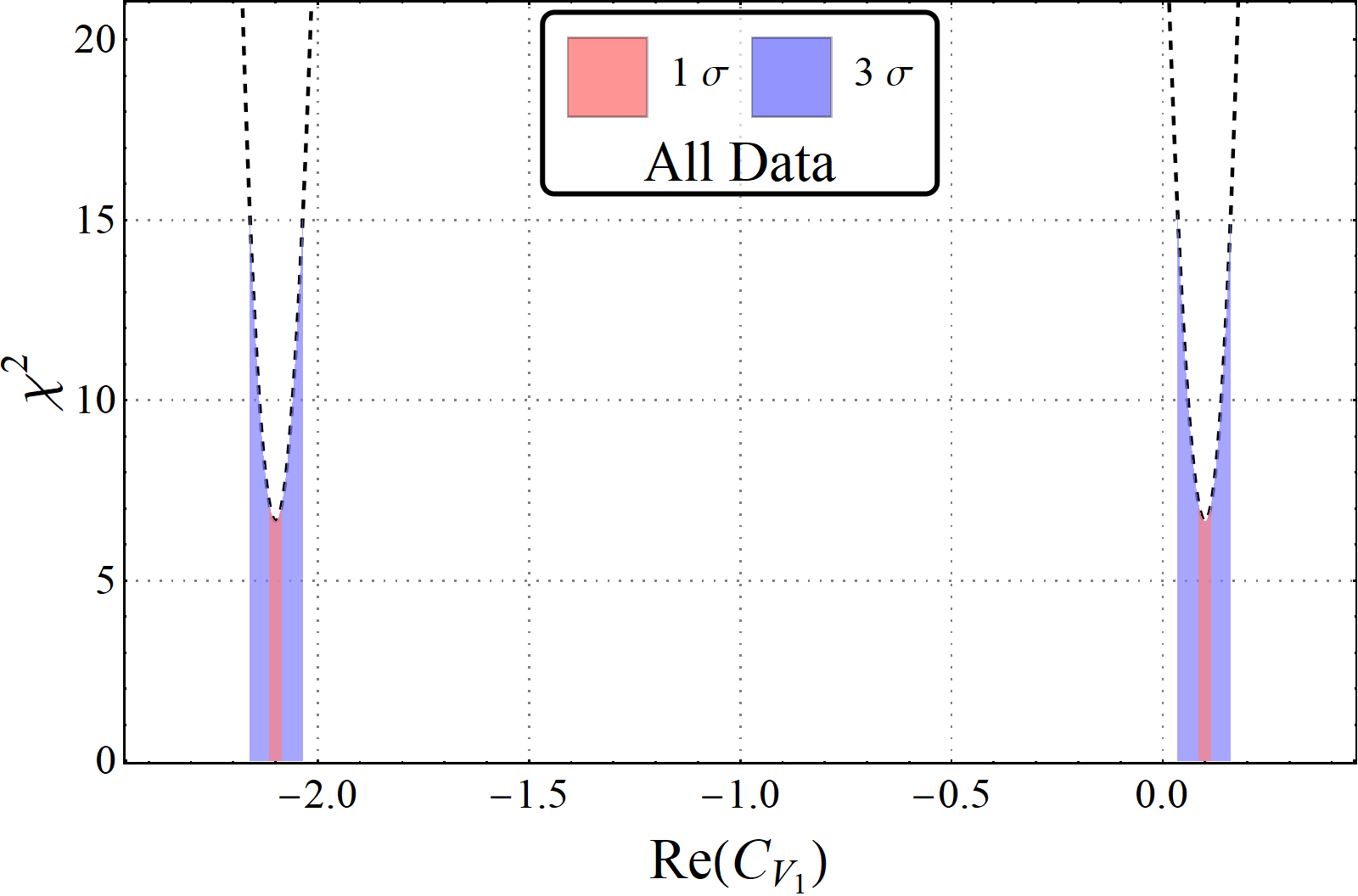}\label{fig:alldat2JPQ}}~
	\subfloat[Scenario $3$]{\includegraphics[height=3.5cm]{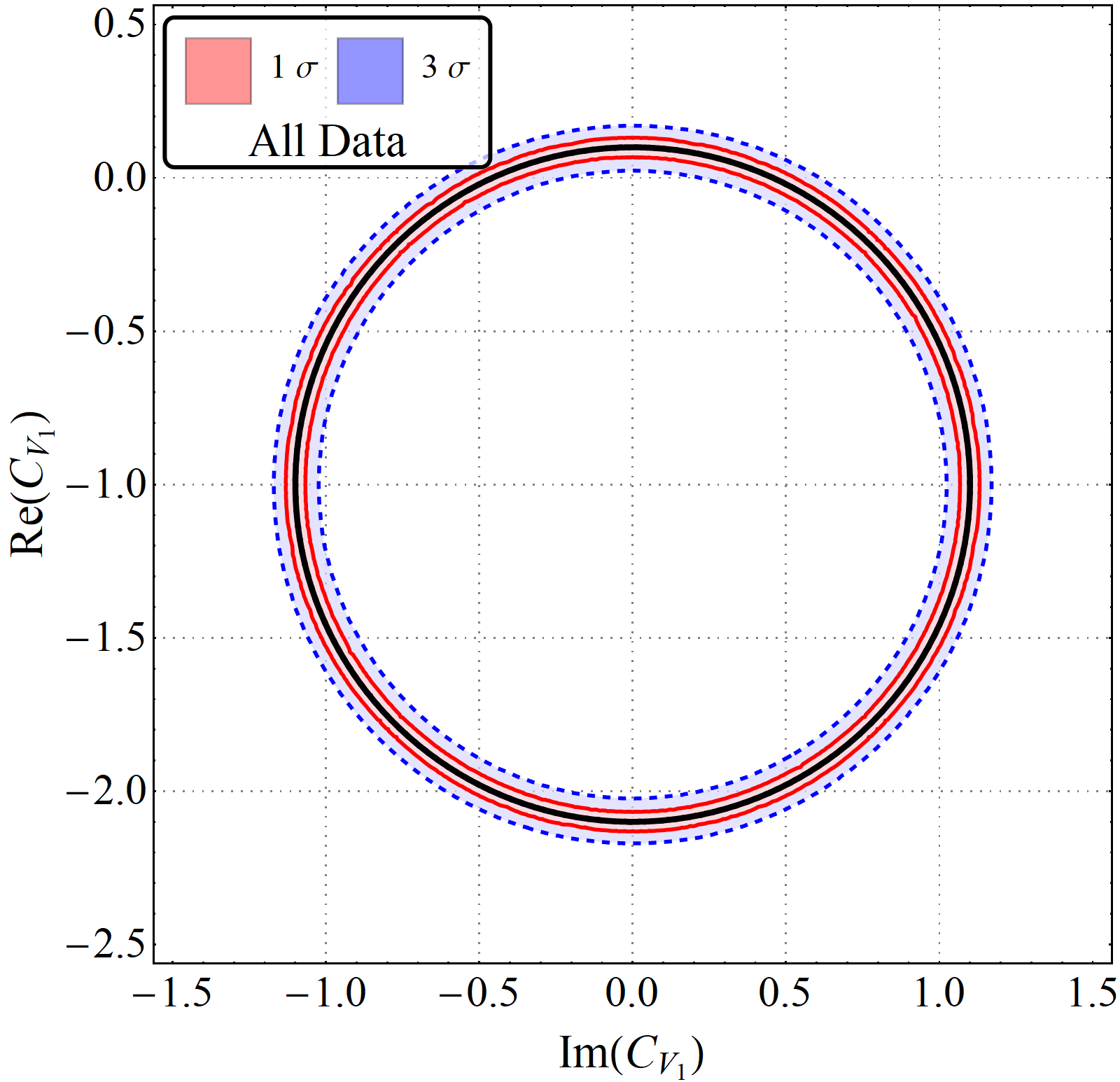}\label{fig:alldat16JPQ}}\\
	\subfloat[Scenario $4$]{\includegraphics[height=4cm]{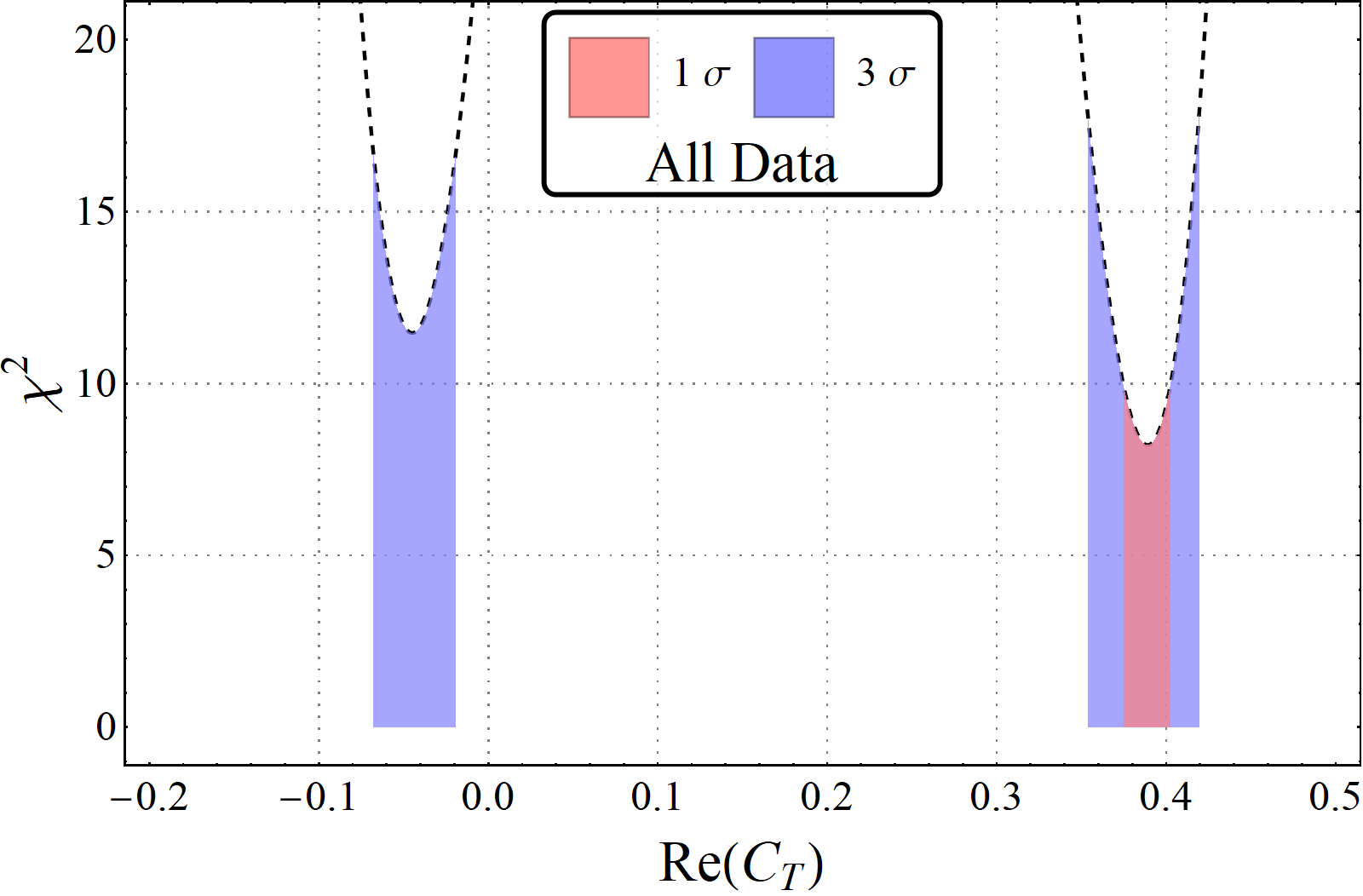}\label{fig:alldat3JPQ}}~
	\subfloat[Scenario $5$]{\includegraphics[height=5cm]{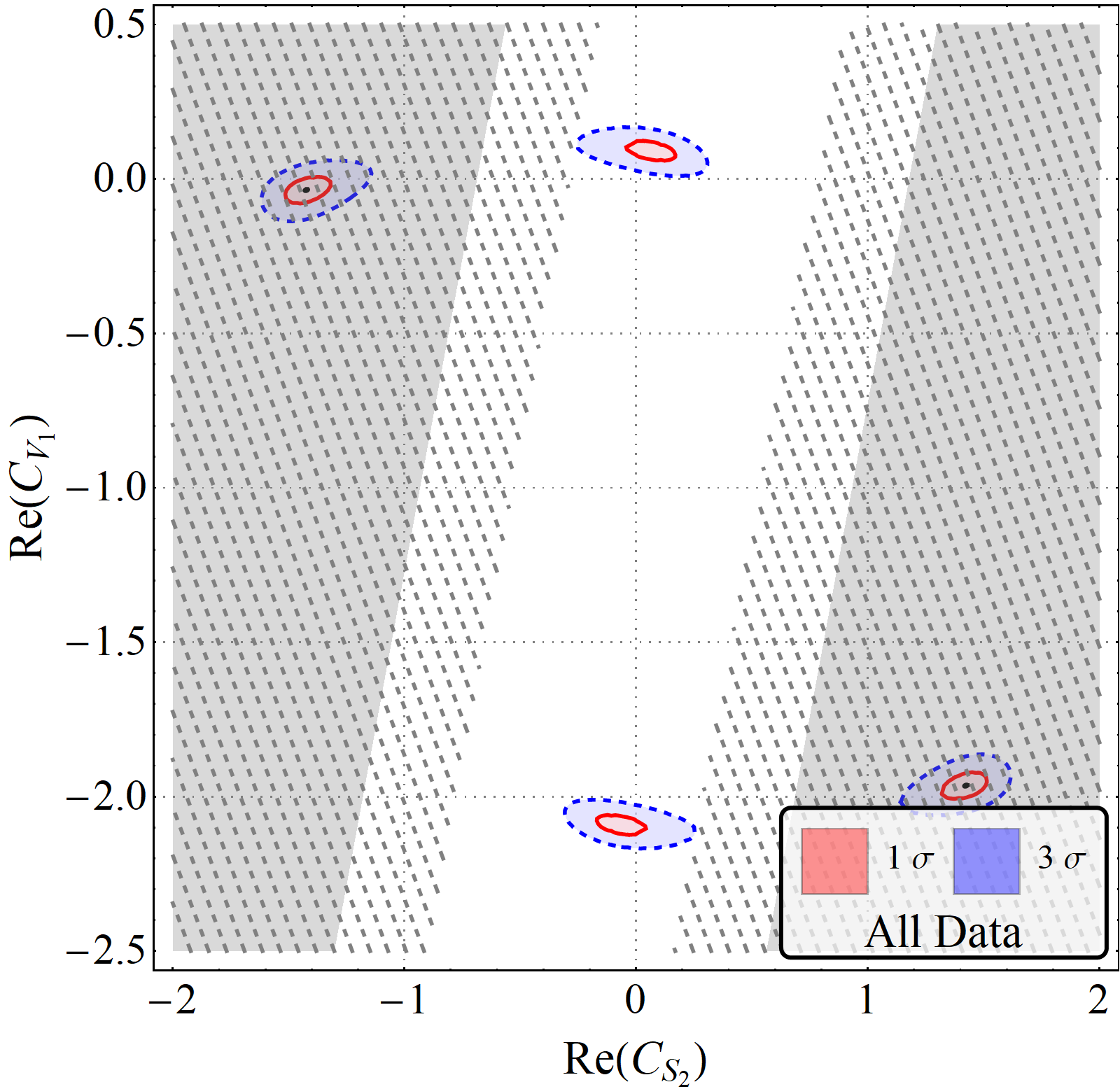}\label{fig:alldat4JPQ}}~
	\subfloat[Scenario $6$]{\includegraphics[height=5cm]{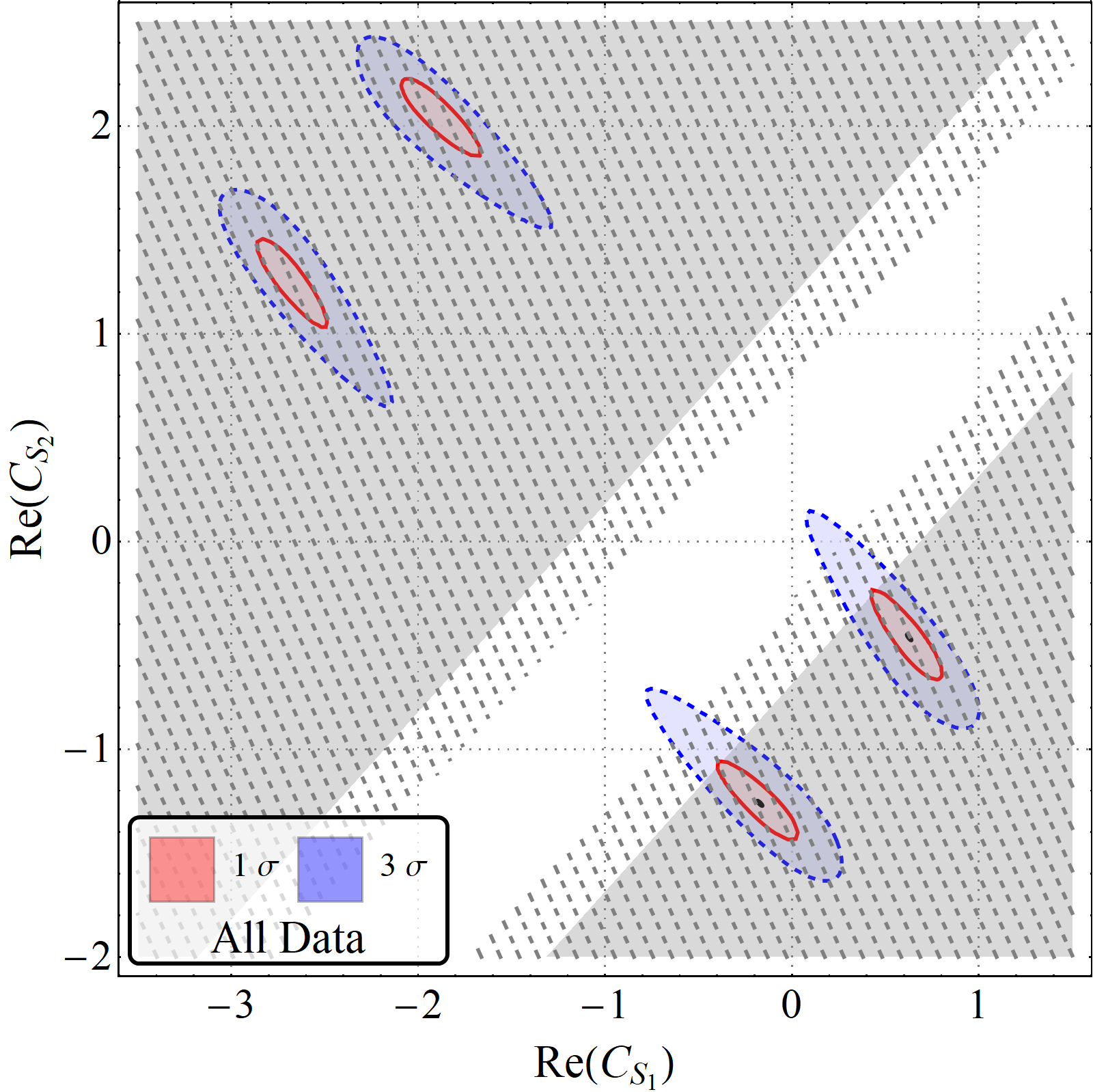}\label{fig:alldat5JPQ}}\\
	\subfloat[Scenario $7$]{\includegraphics[height=5cm]{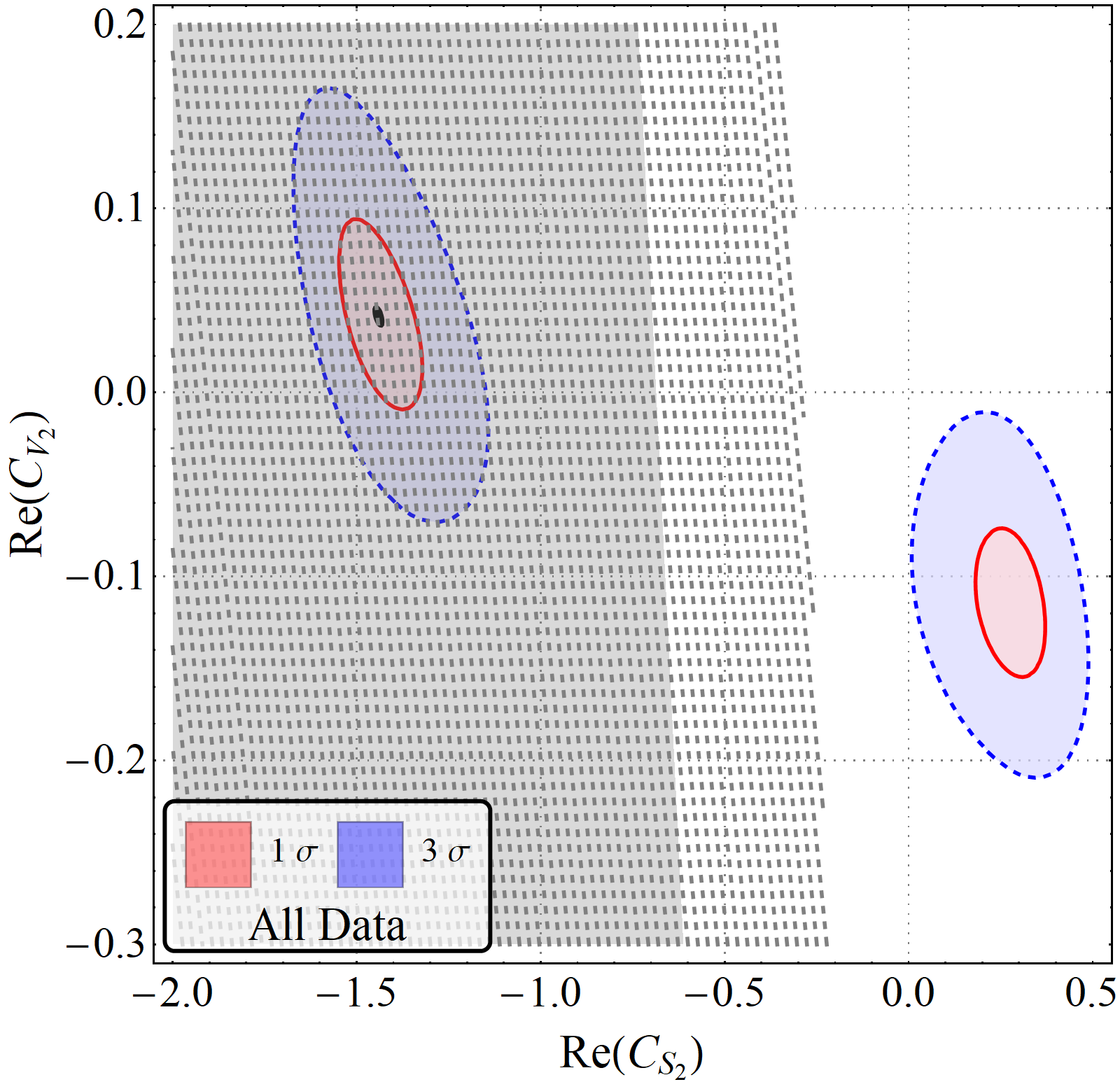}\label{fig:alldat6JPQ}}~
	\subfloat[Scenario $8$]{\includegraphics[height=5cm]{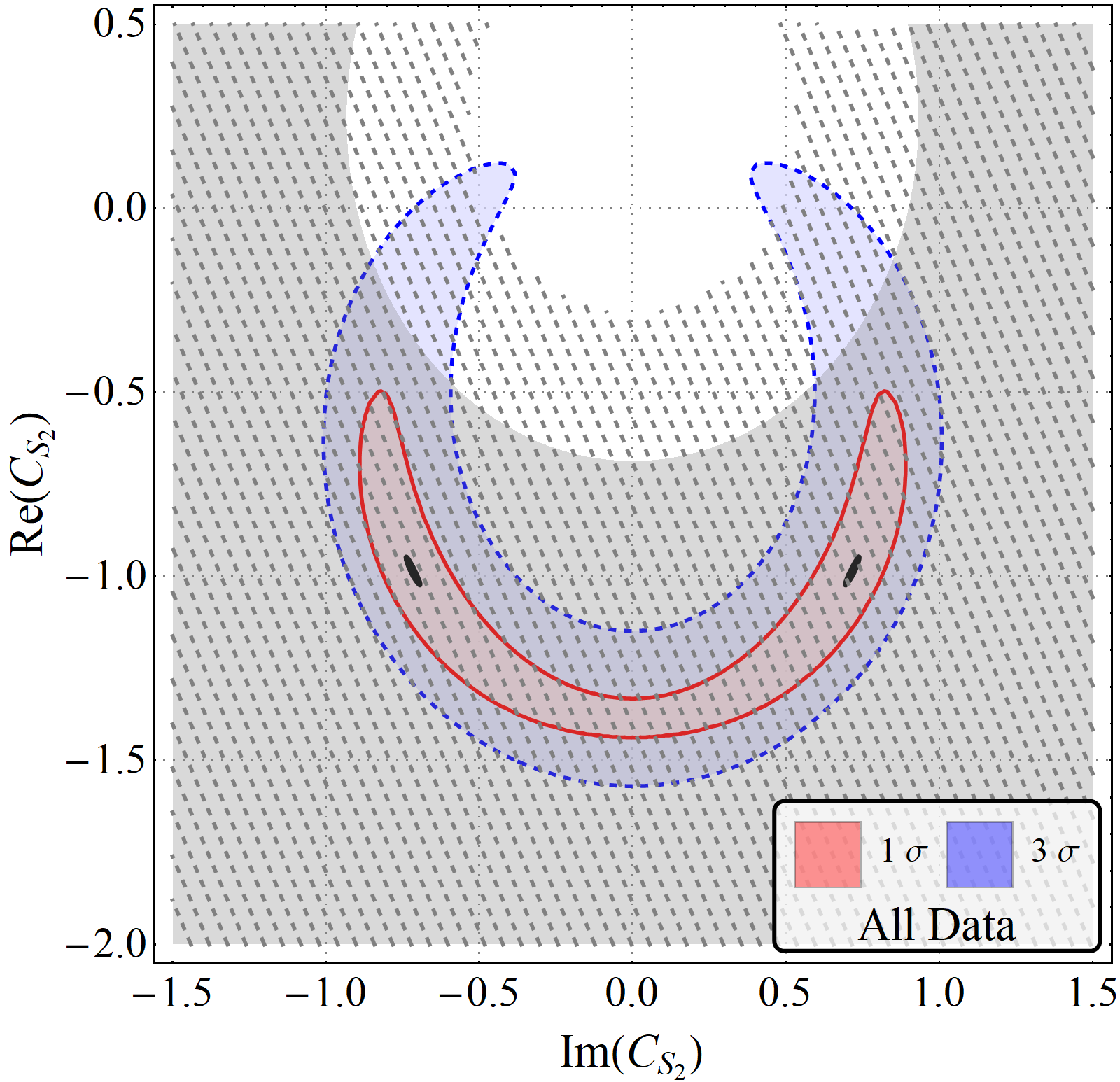}\label{fig:alldat7JPQ}}~
	\subfloat[Scenario $9$]{\includegraphics[height=5cm]{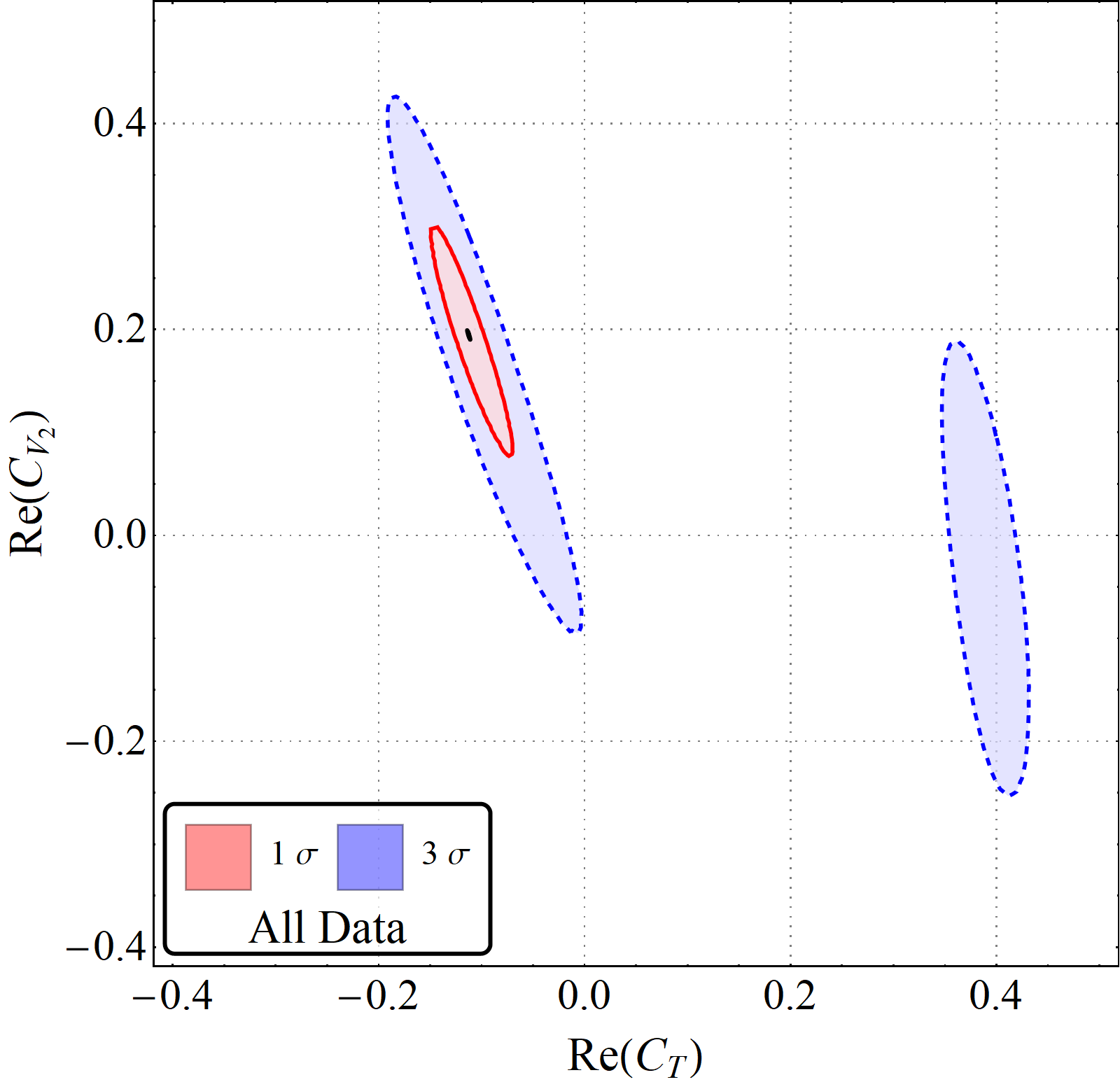}\label{fig:alldat8JPQ}}
	
	\caption{\small The allowed parameter space of NP Wilson coefficients and their correlations considered in different scenarios for the dataset with all data, where $\mathcal{R}_{J/\Psi}$ is calculated in PQCD (last dataset of tables \ref{tab:alldat1} and \ref{tab:alldat2}). Red (solid) and blue (dashed) contours enclose respectively $1\sigma$ and $3\sigma$ confidence levels(C.L.), as defined in section \ref{sec:res2}. Shaded and diagonally hatched overlay regions represents parameter space disallowed by constraints $\mathcal{B}(B_c\to\tau\nu_\tau) \le 30\%$ and $10\%$ respectively. These plots are continued to the next figure \ref{fig:alldat2}.}
	\label{fig:alldat1}
\end{figure*}

\begin{figure*}[htbp]
	\centering
	\subfloat[Scenario $10$]{\includegraphics[height=5cm]{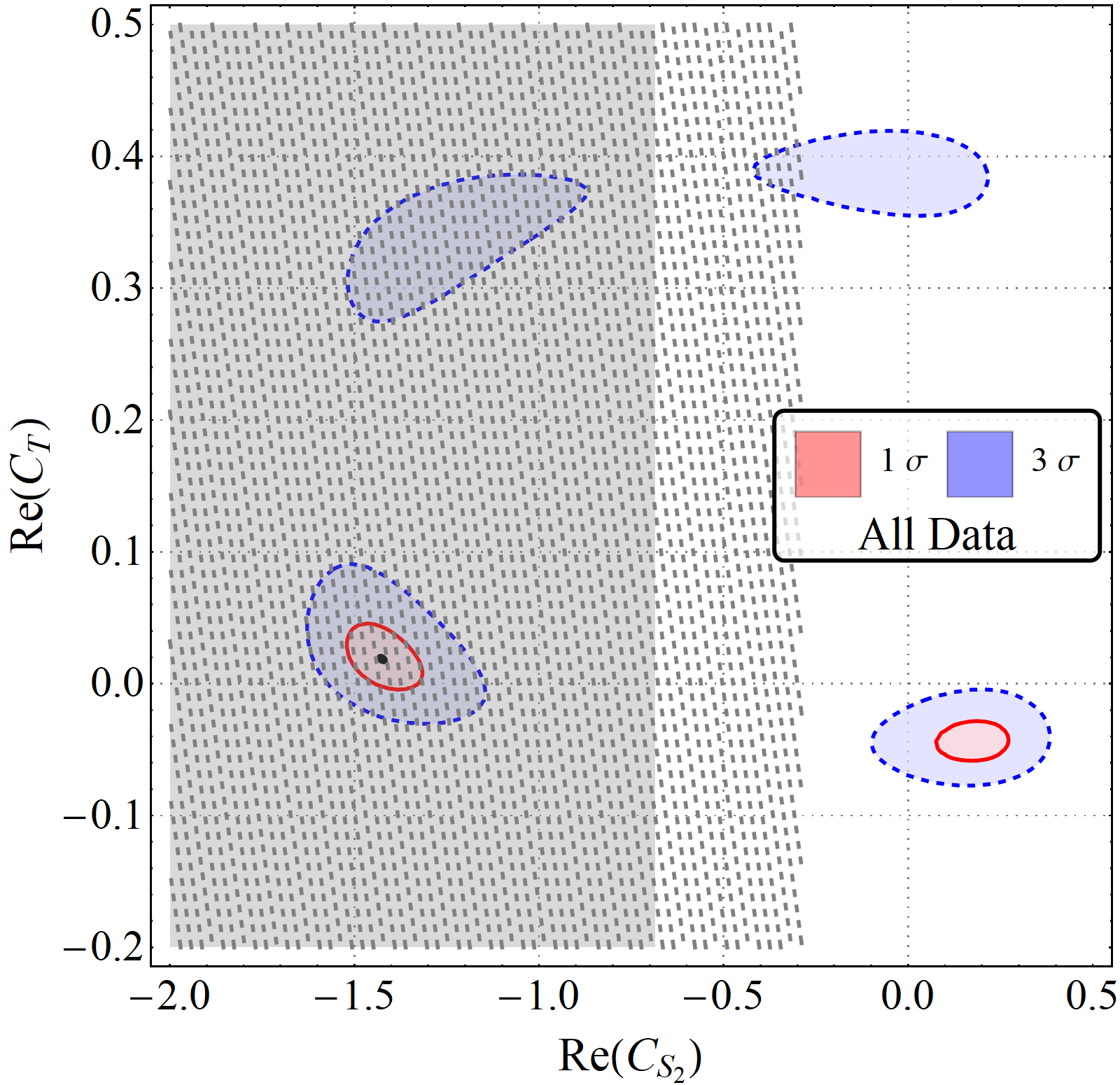}\label{fig:alldat9JPQ}}~
	\subfloat[Scenario $11$]{\includegraphics[height=5cm]{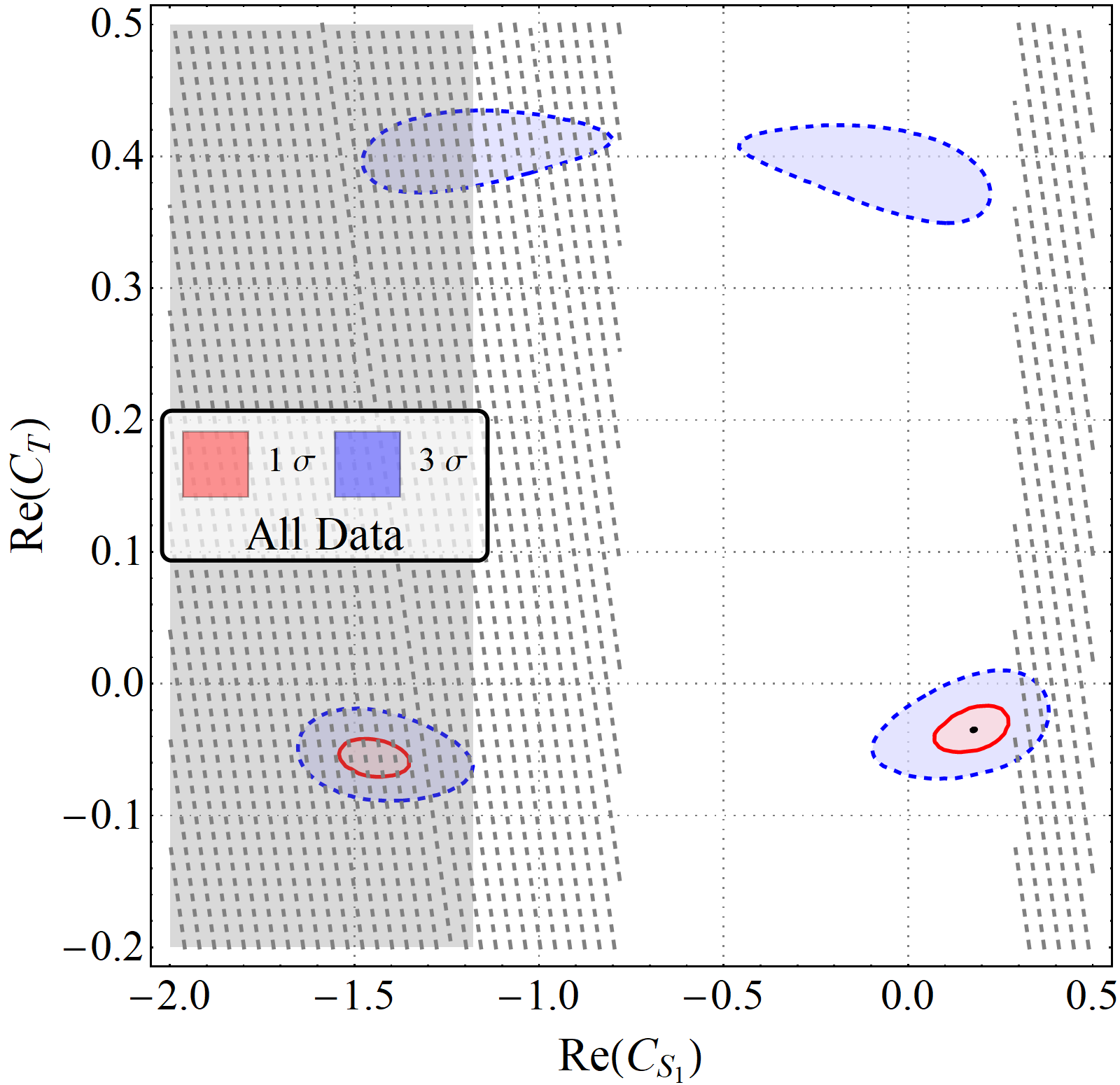}\label{fig:alldat10JPQ}}~
	\subfloat[Scenario $12$]{\includegraphics[height=5cm]{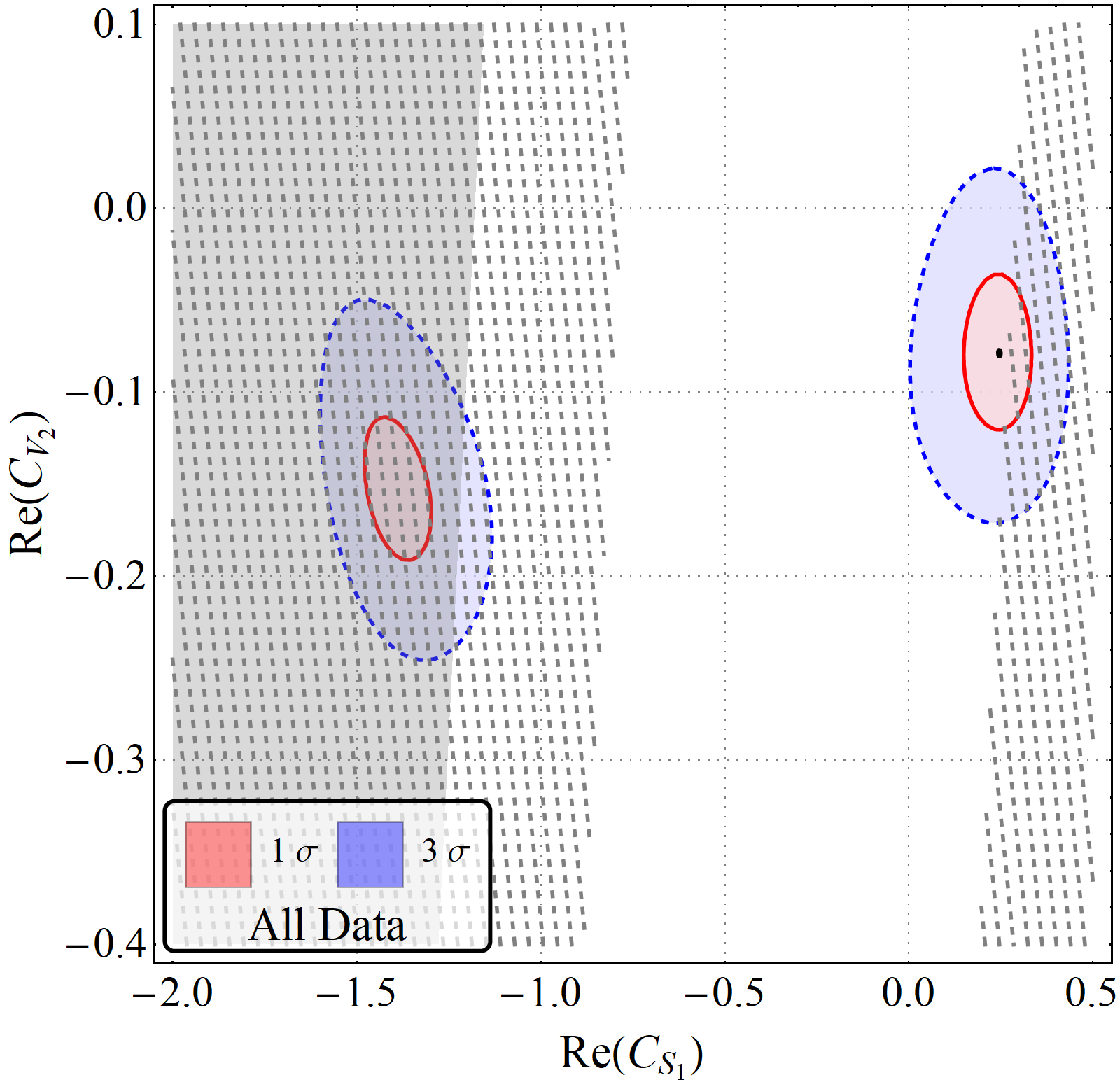}\label{fig:alldat11JPQ}}\\
	\subfloat[Scenario $13$]{\includegraphics[height=5cm]{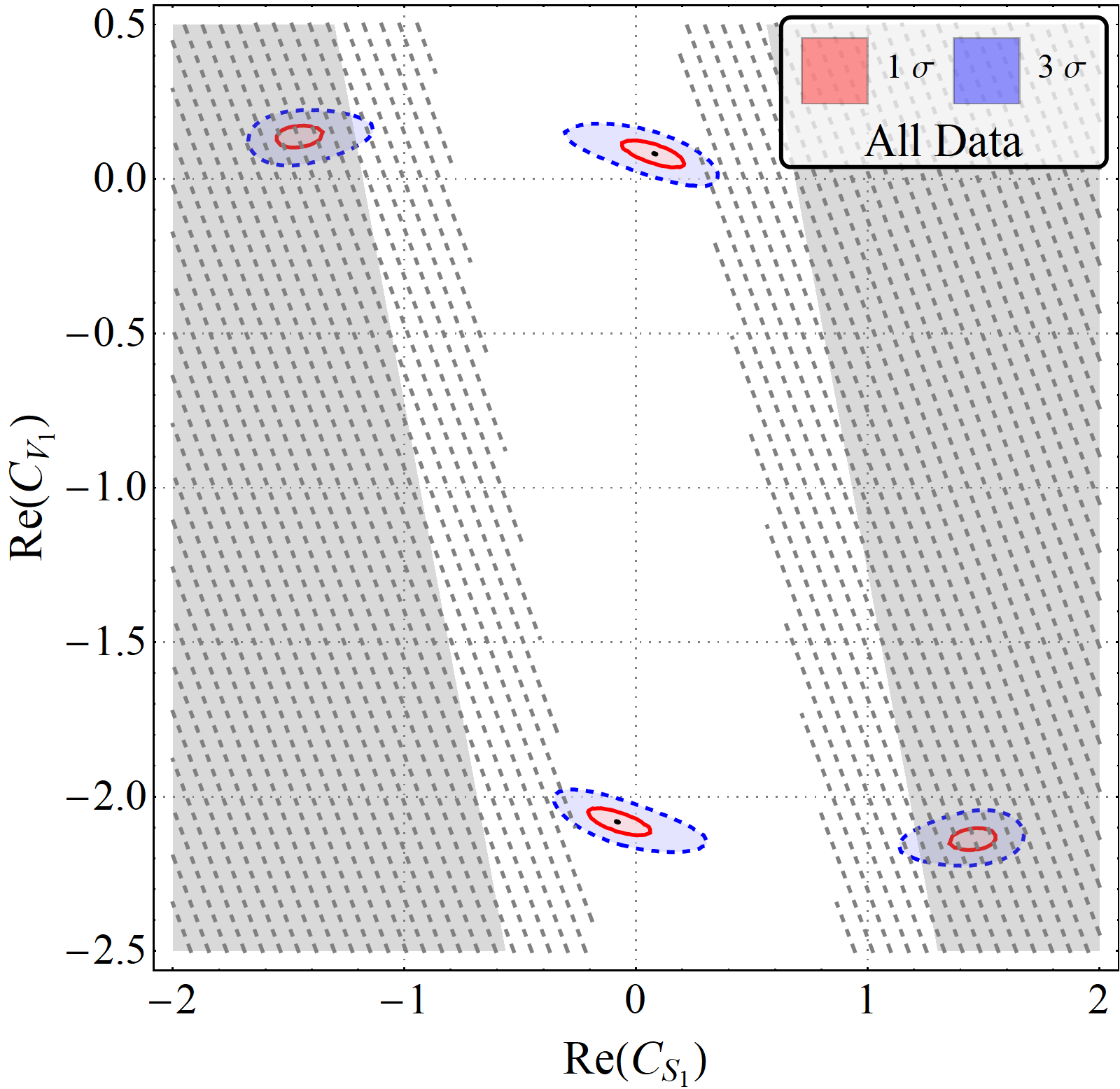}\label{fig:alldat12JPQ}}~
	\subfloat[Scenario $14$]{\includegraphics[height=5cm]{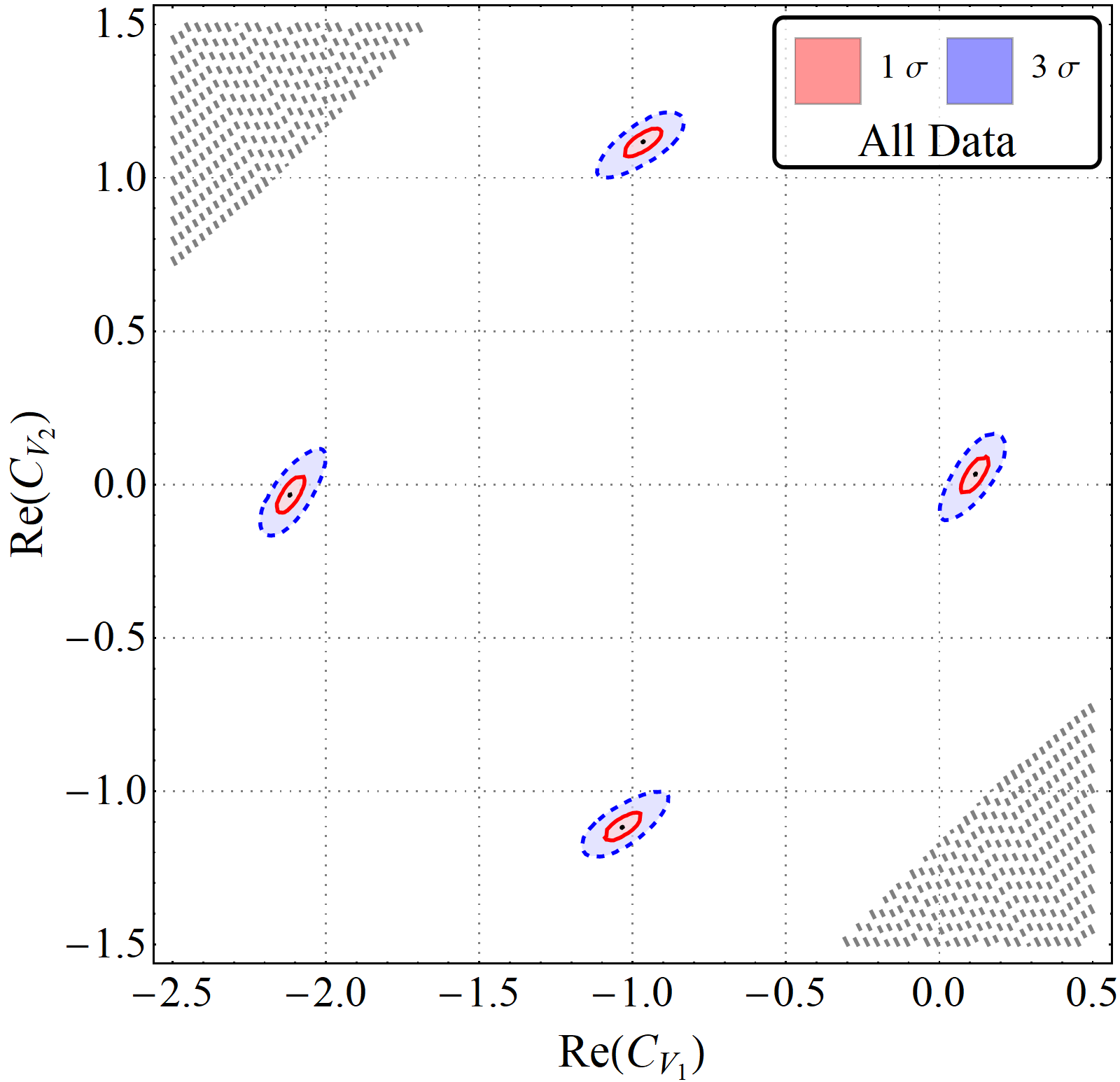}\label{fig:alldat13JPQ}}~
	\subfloat[Scenario $15$]{\includegraphics[height=5cm]{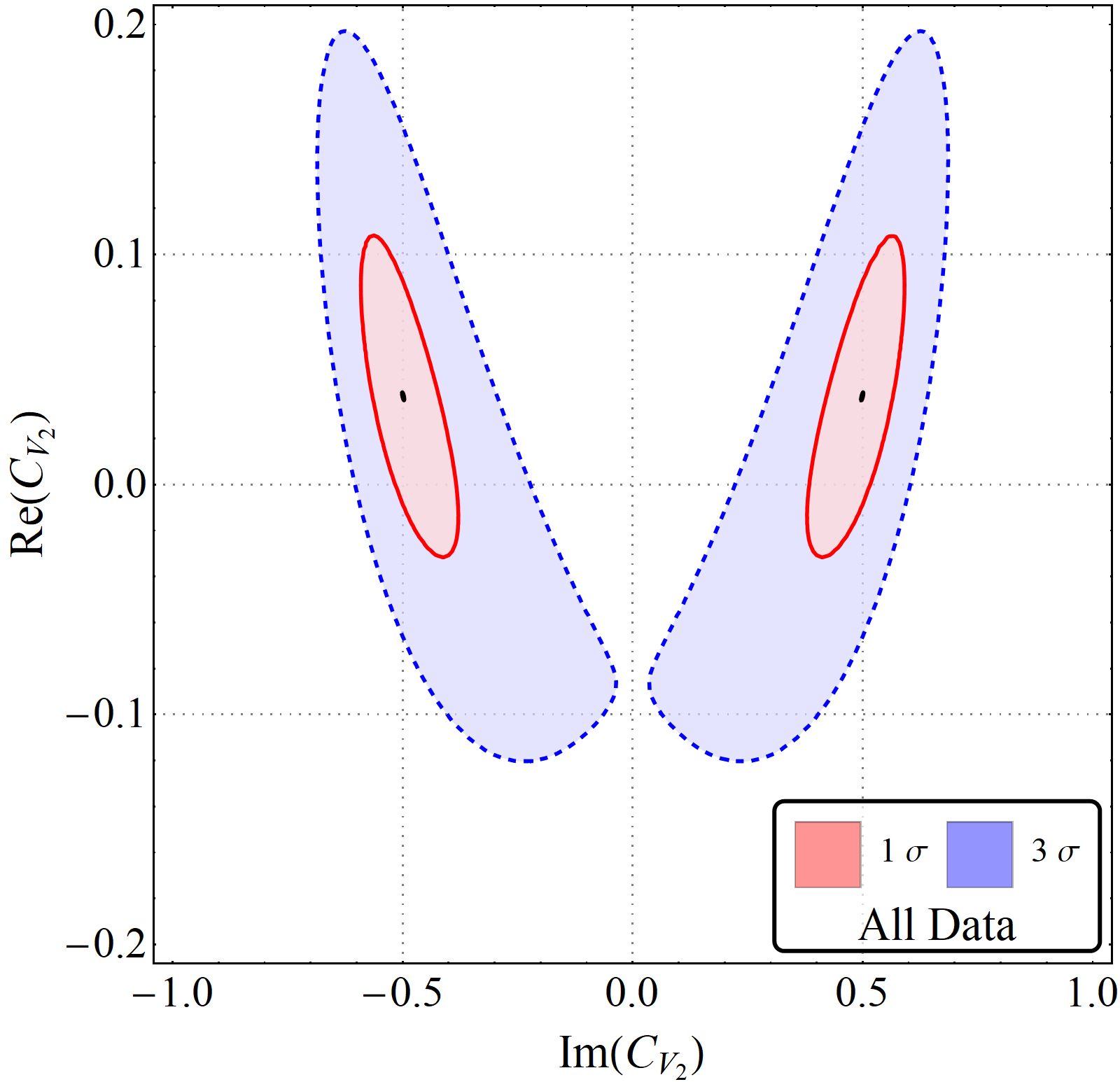}\label{fig:alldat14JPQ}}\\
	\subfloat[Scenario $16$]{\includegraphics[height=5cm]{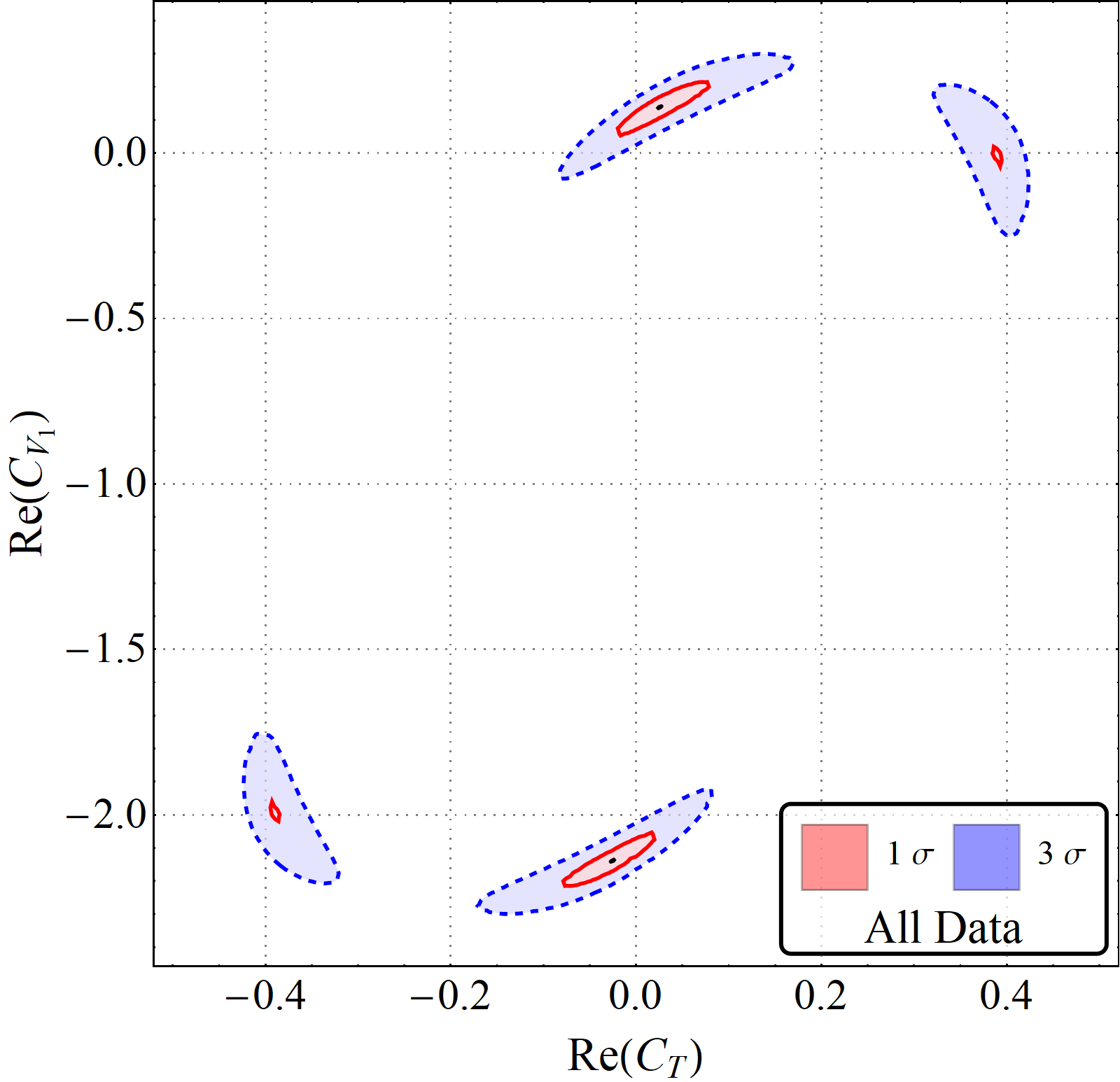}\label{fig:alldat15JPQ}}~
	\caption{\small Plots for the remaining scenarios, continued from figure \ref{fig:alldat1}.}
	\label{fig:alldat2}
\end{figure*}

\subsection{Numerical Optimization}\label{sec:chidef}
	
		\begin{table}[t]
		\begin{center}
		\caption{Nuisance inputs to create $\chi^2_{nuis.}$ defined in eq. \ref{chidefnui}. These are obtained from the analysis in ref. \cite{Jaiswal:2017rve}.} 
			\label{tab:nuisance}
			%			\begin{ruledtabular}
			\begin{tabular}{ccccccc}
				\hline\noalign{\smallskip}
				Parameters 	& Value		&\multicolumn{5}{c}{Correlation}\\
				\noalign{\smallskip}\hline\noalign{\smallskip}
				$\rho_D^2$ 	& 1.138(23) 	& 	1. 	& 0.15 	& -0.01 & -0.07 & 0 \\
				%\cline{1-2}
				$\rho_{D^*}^2$ & 1.251(113)	&  		& 1. 	& 0.08 	& -0.80 & 0 \\
				%\cline{1-2}
				$R_1(1)$ 	& 1.370(36) 	&  		& 		& 1. 	& -0.08 & 0 \\
				%\cline{1-2}
				$R_2(1)$ 	& 0.888(65) 	&  		& 		& 		& 1. 	& 0 \\
				%\cline{1-2}
				$R_0(1)$ 	& 1.196(102) 	&  		&  		&  		&  		& 1 \\
				%			\hline
				%			$m_B$		& 5.27962(15) GeV	& & & & & \\
				%			$m_{D^*}$	& 2.01026(5) GeV	& & & & & \\
				%			$m_W$ 		& 80.385(15) GeV	& & & & & \\
				%			$m_W$ 		& 80.385(15) GeV	& & & & & \\
				%			$m_c$ 		& 1.28(3) GeV	& & & & & \\
				%			$m_b$ 		& $4.18^{+0.04}_{-0.03}$ GeV	& & & & & \\
				%			$m_{\tau}$	& 1.77682(16) GeV	& & & & & \\
				\hline
			\end{tabular}
			
			%			\end{ruledtabular}
		\end{center}
	\end{table}

As mentioned earlier, the goal of this paper is to perform a model independent multi-scenario analysis with the experimentally available results on the charged current anomalies, 
	in conjunction with other relevant results, to obtain a data-based selection of a `best' scenario and ranking and weighting of the remaining scenarios from a 
	predefined set. If we consider the NP Wilson coefficients occurring in eq. \ref{eq1} to be complex, all possible combinations of the real and imaginary parts 
	of the coefficients (10 parameters in total) should constitute such a predefined set, from which we can choose different scenarios. Scenarios containing only 
	imaginary Wilson coefficients are neglected. 
	
	For each such scenario $k$, we define a $\chi^2$ statistic, which is a function of the real and/or imaginary parts of the Wilson coefficients ($C^k_W$) 
	associated with the scenario in question, and is defined as:
	\begin{align}\label{chidef}
		\nn \chi^2_k (C^k_W) = &\sum^{{\rm data}}_{i,j = 1} \left(O^{{exp}}_i - O^{th}_i(C^k_W)\right) \left(V^{stat} + V^{syst} \right)^{-1}_{i j} \nn \\ &\left(O^{{exp}}_j - O^{th}_j(C^k_W)\right) + \chi^2_{Nuis.}\,.
	\end{align}
	Here, $O^{th}_p(C^k_W)$ are given by eqns.~\ref{Rth}, \ref{ptaudef1} and sec. \ref{sec:Rjpsi} as applicable and $O^{{exp}}_p$ is the central value of 
	the $p^\text{th}$ experimental result. Statistical (systematic) covariance matrices $V^{stat (syst)}$, are constructed by taking separate correlations, 
	wherever available. The nuisance parameters (Table~\ref{tab:nuisance}) occurring in the theoretical expressions are tuned in to the fit using a term
	\begin{equation}\label{chidefnui}
		\nn \chi^2_{Nuis.} = \sum^{{\rm theory}}_{i,j = 1} \left({\rm I}_i^p - {\rm v}_i^p\right)~\left(V^{Nuis}\right)^{-1}_{i j}
		\left({\rm I}_j^p - {\rm v}_j^p\right)\,,
	\end{equation}
	where $I_k^p$ and $v_k^p$ are the $k^{th}$ input parameter and its respective value. 
	For each scenario, we perform two sets of fits. First, we use different combinations of the experimental results of $\mathcal{R}_{D^{(*)}}$ (and $P_{\tau}(D^*)$).
	For the second set, we redo the fits including $\mathcal{R}_{J/\psi}$. As the form factor parametrization, as well as the single experimental result for 
	$\mathcal{R}_{J/\psi}$ are quite imprecise, instead of defining a $\chi^2_{nuis.}(\mathcal{R}_{J/\psi})$, we add the SM uncertainty of $\mathcal{R}_{J/\psi}$ 
	in quadrature to the experimental one. Following the discussion in sec. \ref{sec:Rjpsi}, we do two sets of fits in this stage, with two different sets of form 
	factor parametrization for $B_c\to J/\Psi$, namely LFCQ and PQCD.
	
	After each fit, we determine the quality of it in terms of the $p$-value obtained corresponding to the $\chi^2_{min}$ values and the degrees of freedom (DoF) 
	for that fit. We also double check the quality of the fit and existence of outliers in the fitted dataset by constructing a `Pull' 
	($= \left(O^{{exp}}_i - O^{th}_i(C^k_W)\right)/ (\Delta O^{{exp}}_i)$) for each data-point and checking the normality (i.e. the probability that 
	it is consistent with a Gaussian of $\mu=0$ and $\sigma=1$) of their distribution. Due to the small number of data-points in this analysis, no readily 
	available normality test can perform with certainty and it is necessary to scrutinize each individual pull distribution. Still, we perform a variant of the
	``Shapiro-Wik" normality test as an extra criteria for elimination of scenarios. In other words, we drop the fits which have a pull distribution with probability 
	to be a normal distribution $\leq 5\%$.  Finally, we add the constraints according to sec. \ref{sec:bctn} and \ref{sec:experiment} to our analysis and obtain 
	the allowed parameter space. Next, we perform a model-selection procedure on the remaining set of viable scenarios for each data-set. In the following 
	sub-section, we elaborate the method used to do the multi-model selection procedure.

\subsection{Model Selection Criteria}
	
	One measure of the degree of structure in a model, inferred from the data, is its dimension, i.e. the number of parameters in it. In general, bias 
	decreases and variance increases as the dimension of the model increases. Requirement of the optimum dimension of the model is called the `concept of 
	parsimony' \cite{boxjenkins}, which is essentially a bias versus variance trade-off in statistical terms. All model selection methods, to some extent, 
	depend on the principle of parsimony \cite{breiman}.
	
	The most generally applicable and reliable method for model comparison is `cross-validation', which, in addition to testing the predictive power of the model, 
	minimizes the bias and variance together by minimizing the mean-squared-error (MSE). The problem of applying cross validation to the present analysis is that 
	its applicability to very small sample sizes (as is the case here) is questionable \cite{BELEITES,Varoquaux}.	Thus, to the goal of model selection, we have 
	used information-theoretic approaches, especially the second order variant of Akaike Information Criterion (AIC$_c$) \cite{sugiura78} in the present work. 
	This criterion and other competing criteria have previously been applied in one work of ours analyzing the charged current anomalies of $b$-decay 
	\cite{Bhattacharya:2016zcw}. In that analysis, we had worked with the binned data on the differential decay distribution of these channels. 
	
	Given the notation for full reality or truth is $f$ and an approximating model in terms of probability distribution is $g$, we can create a $\chi^2$ function 
	in terms of the parameters of $g$ and empirical results, following sec. \ref{sec:chidef}. For each model $g_i$ in a set with $R$ competing scenarios, we can 
	define an AIC$_c$ in terms of the $\chi^2_{min}$ in the parameter space which is equivalent to the maximum point of the empirical log-likelihood function:
	
	\begin{align}
	{\rm AIC}_c = \chi^2_{min} + 2 K + \frac{2 K (K+1)}{n - K -1}\,
	\label{aicc}
	\end{align}
	where $n$ is the number of data points and $K$ is the number of estimable parameters
	\footnote{There is a subtle point here about the number of `estimable parameters'. As an example, there are two competing scenarios in the present analysis: (a) with non-zero $\mathcal{R}e(C_{V_1})$ and (b) with non-zero $\mathcal{R}e(C_{V_1})$ and $\mathcal{I}m(C_{V_1})$. Now, the two parameters of case (b) always appear together in an identical manner in the expressions of our observables, essentially making the number of `estimable parameters' = 1. Thus, in the first column of table \ref{tab:alldat1}, cases 1-4 has the same DOF = 8. We follow this throughout the analysis.
		
	Another way of finding the `number of estimable parameters' is to calculate the $p$-value of the fit from toy Monte-Carlo (MC) method. This value, in conjunction to the approximation that the fit-statistic follows a $\chi^2$ distribution, can give us the number of degrees of freedom, and thus the number of estimable parameters (as we have also checked). As we need $\Delta$AIC$_c$ instead of the absolute value of the AIC$_c$ in our analysis, the naive way of estimation of the number of parameters (except the case $C_{V_1}$, as explained above and which we treat as a special case) works just fine.}
	in $g_i$. As a rule of thumb, use of AIC$_c$ is preferred in literature
	when $n/K < 40$. In application, the model with the smallest value of AIC$_c$ is estimated to be the `closest' to the unknown reality generating the data, 
	among the considered models. Whereas all AIC$^i_c$ are on a relative scale and are strongly dependent on sample size, simple differences of them 
	($\Delta^{AIC}_i = {\rm AIC}^i_c - {\rm AIC}^{min}_c$) estimate the relative expected information loss between $f$ and $g_i$ allowing comparison and 
	ranking of candidate models in increasing order of $\Delta^{AIC}_i$. It is also possible to quantify the weight of evidence in favor of model $i$ by 
	defining a set  of positive ``Akaike weights", $w_i = (e^{(-\Delta^{AIC}_i / 2)})/(\sum_{r = 1}^R e^{(-\Delta^{AIC}_r / 2)})$, adding up to $1$ 
	\cite{Burnham}. As these depend on the entire set, adding or dropping a model during a post hoc analysis requires re-computation for all models in 
	the new set.
\begin{table*}[t]
	\centering
	\caption{\small Best fit results and correlations of the scenarios listed in table \ref{tab:alldat1}. We omit the scenarios disallowed by the constraint $\mathcal{B}(B_c\to \tau\nu_\tau) \le 30\%$. For the cases where only some of the minima are allowed by the constraint, we quote only those of them allowed by the constraint and closest to the SM point at the same time. Elaboration is in section \ref{sec:res3}. For scenarios where the best fit, instead of being an isolated point, is actually a contour in the parameter-space, we ask the reader  to check the corresponding plot. Figure \ref{fig:alldat16JPQ} is the plot for scenario 3 in the last dataset of this table.}
	  \label{tab:alldat2}
%	\begin{ruledtabular}
		 \resizebox{\textwidth}{!}{
		\begin{tabular}{cccc||cccc||cccc}
			\hline
		      \multicolumn{4}{c||}{Data Without $\mathcal{R}_{J/\Psi}$} & \multicolumn{4}{c||}{All Data ($\mathcal{R}_{J/\Psi}$ with LFCQ)} & \multicolumn{4}{c}{All Data ($\mathcal{R}_{J/\Psi}$ with PQCD)}\\
		      \hline
		      Index & Param.s & Best-fit & Correlation & Index & Param.s & Best-fit & Correlation & Index & Param.s & Best-fit & Correlation \\
		      \hline
		      1 & $\mathcal{R}e(C_T)$ & \text{0.387(11)} & -- & 1 & $\mathcal{R}e(C_{V_1})$ & \text{0.100(22)} & -- & 2 & $\mathcal{R}e(C_{V_1})$ & \text{0.100(21)} & -- \\
		      %\hline
		      3 & $\mathcal{R}e(C_{V_1})$ & \text{0.098(22)} & -- & 4 & $\mathcal{R}e(C_T)$ & \text{0.389(10)} & -- & 4 & $\mathcal{R}e(C_T)$ & \text{0.389(10)} & -- \\
		      %\hline
		      7 & $\mathcal{R}e(C_{S_2})$ & \text{0.073(79)} & -0.409 & 5 & $\mathcal{R}e(C_T)$ & \text{-0.113(26)} & -0.93 & 5 & $\mathcal{R}e(C_{S_2})$ & \text{0.072(79)} & -0.411 \\
			& $\mathcal{R}e(C_{V_1})$ & \text{0.089(24)} &   &   & $\mathcal{R}e(C_{V_2})$ & \text{0.195(74)} &   &   & $\mathcal{R}e(C_{V_1})$ & \text{0.090(24)} &   \\
		     % \hline
		      8 & $\mathcal{R}e(C_{S_2})$ & \text{0.181(67)} & 0.075 & 6 & $\mathcal{R}e(C_{S_1})$ & \text{0.179(66)} & 0.35 & 7 & $\mathcal{R}e(C_{S_2})$ & \text{0.280(68)} & -0.3 \\
			& $\mathcal{R}e(C_T)$ & \text{-0.043(11)} &   &   & $\mathcal{R}e(C_T)$ & \text{-0.034(12)} &   &   & $\mathcal{R}e(C_{V_2})$ & \text{-0.112(29)} &   \\
		     % \hline
		      9 & $\mathcal{R}e(C_{S_2})$ & \text{0.279(68)} & -0.302 & 9 & $\mathcal{R}e(C_{S_2})$ & \text{0.181(67)} & 0.074 & 9 & $\mathcal{R}e(C_T)$ & \text{-0.113(26)} & -0.93 \\
			& $\mathcal{R}e(C_{V_2})$ & \text{-0.111(29)} &   &   & $\mathcal{R}e(C_T)$ & \text{-0.043(11)} &   &   & $\mathcal{R}e(C_{V_2})$ & \text{0.195(74)} &   \\
		      %\hline
		      10 & $\mathcal{R}e(C_T)$ & \text{-0.112(26)} & -0.93 & 10 & $\mathcal{R}e(C_{S_2})$ & \text{0.072(79)} & -0.411 & 10 & $\mathcal{R}e(C_{S_2})$ & \text{0.181(67)} & 0.078 \\
			& $\mathcal{R}e(C_{V_2})$ & \text{0.196(74)} &   &   & $\mathcal{R}e(C_{V_1})$ & \text{0.090(24)} &   &   & $\mathcal{R}e(C_T)$ & \text{-0.043(11)} &   \\
		      %\hline
		      11 & $\mathcal{R}e(C_{S_1})$ & \text{0.179(66)} & 0.351 & 11 & $\mathcal{R}e(C_{S_1})$ & \text{0.246(61)} & -0.009 & 11 & $\mathcal{R}e(C_{S_1})$ & \text{0.179(66)} & 0.35 \\
			& $\mathcal{R}e(C_T)$ & \text{-0.033(12)} &   &   & $\mathcal{R}e(C_{V_2})$ & \text{-0.078(28)} &   &   & $\mathcal{R}e(C_T)$ & \text{-0.034(12)} &   \\
		      %\hline
		      12 & $\mathcal{R}e(C_{S_1})$ & \text{0.245(60)} & -0.01 & 12 & $\mathcal{R}e(C_{S_2})$ & \text{0.280(68)} & -0.3 & 12 & $\mathcal{R}e(C_{S_1})$ & \text{0.247(61)} & -0.008 \\
			& $\mathcal{R}e(C_{V_2})$ & \text{-0.075(28)} &   &   & $\mathcal{R}e(C_{V_2})$ & \text{-0.112(29)} &   &   & $\mathcal{R}e(C_{V_2})$ & \text{-0.079(28)} &   \\
		      %\hline
		      13 & $\mathcal{R}e(C_{S_1})$ & \text{0.086(90)} & -0.684 & 13 & $\mathcal{R}e(C_{S_1})$ & \text{0.085(90)} & -0.684 & 13 & $\mathcal{R}e(C_{S_1})$ & \text{0.085(90)} & -0.684 \\
			& $\mathcal{R}e(C_{V_1})$ & \text{0.078(30)} &   &   & $\mathcal{R}e(C_{V_1})$ & \text{0.079(30)} &   &   & $\mathcal{R}e(C_{V_1})$ & \text{0.079(30)} &   \\
		      %\hline
		      14 & $\mathcal{R}e(C_{V_1})$ & \text{0.117(31)} & 0.709 & 14 & $\mathcal{I}m(C_{V_2})$ & \text{0.499(68)} & 0.718 & 14 & $\mathcal{R}e(C_{V_1})$ & \text{0.118(31)} & 0.712 \\
			& $\mathcal{R}e(C_{V_2})$ & \text{0.037(41)} &   &   & $\mathcal{R}e(C_{V_2})$ & \text{0.039(46)} &   &   & $\mathcal{R}e(C_{V_2})$ & \text{0.034(41)} &   \\
		      %\hline
		      15 & $\mathcal{I}m(C_{V_2})$ & \text{0.497(68)} & 0.716 & 15 & $\mathcal{R}e(C_{V_1})$ & \text{0.117(31)} & 0.709 & 15 & $\mathcal{I}m(C_{V_2})$ & \text{0.499(68)} & 0.718 \\
			& $\mathcal{R}e(C_{V_2})$ & \text{0.042(46)} &   &   & $\mathcal{R}e(C_{V_2})$ & \text{0.037(41)} &   &   & $\mathcal{R}e(C_{V_2})$ & \text{0.038(46)} &   \\
		      %\hline
		      16 & $\mathcal{R}e(C_T)$ & \text{0.030(34)} & 0.917 & 16 & $\mathcal{R}e(C_T)$ & \text{0.026(34)} & 0.918 & 16 & $\mathcal{R}e(C_T)$ & \text{0.026(34)} & 0.918 \\
			& $\mathcal{R}e(C_{V_1})$ & \text{0.142(54)} &   &   & $\mathcal{R}e(C_{V_1})$ & \text{0.140(54)} &   &   & $\mathcal{R}e(C_{V_1})$ & \text{0.139(54)} &   \\
		      %\hline
		      17 & $\mathcal{I}m(C_{T})$ & \text{0.16(15)} & -0.995 & 17 & $\mathcal{I}m(C_{T})$ & \text{-0.18(11)} & 0.993 & 3 & \multicolumn{3}{c}{See Fig. \ref{fig:alldat16JPQ}} \\
			& $\mathcal{R}e(C_T)$ & \text{0.32(15)} &   &   & $\mathcal{R}e(C_T)$ & \text{0.30(15)} &   & & \multicolumn{3}{c}{} \\
			%\hline
% 		      18 & \multicolumn{3}{c||}{See Plot} & 18 & \multicolumn{3}{c||}{See Plot} & - & - & - & - \\
		      \hline
	  \end{tabular}
	    }
%      \end{ruledtabular}
\end{table*}

\begin{table*}[htbp]
	\centering
	\caption{Results for the scenarios listed in table \ref{tab:noptau1}, following the convention of table \ref{tab:alldat2}.}
		\label{tab:noptau2}
%	\begin{ruledtabular}
 \resizebox{\textwidth}{!}{
		\begin{tabular}{cccc||cccc||cccc}
			\hline
		 \multicolumn{4}{c||}{Data without $P_{\tau}(D^*)$ and $\mathcal{R}_{J/\Psi}$} & \multicolumn{4}{c||}{Data without $P_{\tau}(D^*)$ ($\mathcal{R}_{J/\Psi}$ with LFCQ)} & \multicolumn{4}{c}{Data without $P_{\tau}(D^*)$ ($\mathcal{R}_{J/\Psi}$ with PQCD)}\\
			\hline
			Index & Param.s & Best-fit & Correlation & Index & Param.s & Best-fit & Correlation & Index & Param.s & Best-fit & Correlation \\
			\hline
			1 & $\mathcal{R}e(C_T)$ & \text{0.387(11)} & -- & 1 & $\mathcal{R}e(C_{V_1})$ & \text{0.101(22)} & -- & 2 & $\mathcal{R}e(C_{V_1})$ & \text{0.101(22)} & -- \\
			%\hline
			3 & $\mathcal{R}e(C_{V_1})$ & \text{0.099(22)} & -- & 3 & $\mathcal{R}e(C_T)$ & \text{0.389(11)} & -- & 4 & $\mathcal{R}e(C_T)$ & \text{0.389(11)} & -- \\
			%\hline
			5 & $\mathcal{R}e(C_{V_1})$ & \text{0.118(31)} & 0.707 & 5 & $\mathcal{R}e(C_T)$ & \text{-0.112(26)} & -0.93 & 5 & $\mathcal{R}e(C_T)$ & \text{-0.112(26)} & -0.93 \\
			  & $\mathcal{R}e(C_{V_2})$ & \text{0.036(41)} &   &   & $\mathcal{R}e(C_{V_2})$ & \text{0.193(74)} &   &   & $\mathcal{R}e(C_{V_2})$ & \text{0.193(74)} &   \\
			%\hline
			6 & $\mathcal{R}e(C_{S_1})$ & \text{0.082(90)} & -0.687 & 6 & $\mathcal{R}e(C_{S_1})$ & \text{0.178(67)} & 0.353 & 6 & $\mathcal{R}e(C_{S_2})$ & \text{0.070(79)} & -0.412 \\
			  & $\mathcal{R}e(C_{V_1})$ & \text{0.080(30)} &   &   & $\mathcal{R}e(C_T)$ & \text{-0.034(12)} &   &   & $\mathcal{R}e(C_{V_1})$ & \text{0.091(24)} &   \\
			%\hline
			7 & $\mathcal{R}e(C_{S_2})$ & \text{0.071(79)} & -0.409 & 7 & $\mathcal{R}e(C_{S_2})$ & \text{0.180(67)} & 0.075 & 7 & $\mathcal{R}e(C_{S_2})$ & \text{0.180(67)} & 0.074 \\
			  & $\mathcal{R}e(C_{V_1})$ & \text{0.090(24)} &   &   & $\mathcal{R}e(C_T)$ & \text{-0.044(11)} &   &   & $\mathcal{R}e(C_T)$ & \text{-0.044(11)} &   \\
			%\hline
			8 & $\mathcal{R}e(C_T)$ & \text{0.029(34)} & 0.916 & 8 & $\mathcal{R}e(C_{S_1})$ & \text{0.245(61)} & -0.01 & 8 & $\mathcal{R}e(C_{S_1})$ & \text{0.178(67)} & 0.352 \\
			  & $\mathcal{R}e(C_{V_1})$ & \text{0.143(54)} &   &   & $\mathcal{R}e(C_{V_2})$ & \text{-0.077(28)} &   &   & $\mathcal{R}e(C_T)$ & \text{-0.034(12)} &   \\
			%\hline
			9 & $\mathcal{R}e(C_{S_1})$ & \text{0.245(61)} & -0.009 & 9 & $\mathcal{R}e(C_{S_1})$ & \text{0.081(90)} & -0.687 & 9 & $\mathcal{R}e(C_{S_2})$ & \text{0.282(68)} & -0.302 \\
			  & $\mathcal{R}e(C_{V_2})$ & \text{-0.077(28)} &   &   & $\mathcal{R}e(C_{V_1})$ & \text{0.081(30)} &   &   & $\mathcal{R}e(C_{V_2})$ & \text{-0.114(30)} &   \\
			%\hline
			10 & $\mathcal{R}e(C_{S_2})$ & \text{0.280(68)} & -0.304 & 10 & $\mathcal{R}e(C_{S_2})$ & \text{0.282(68)} & -0.302 & 12 & $\mathcal{R}e(C_{S_1})$ & \text{0.247(61)} & -0.007 \\
			  & $\mathcal{R}e(C_{V_2})$ & \text{-0.113(30)} &   &   & $\mathcal{R}e(C_{V_2})$ & \text{-0.116(29)} &   &   & $\mathcal{R}e(C_{V_2})$ & \text{-0.080(28)} &   \\
			%\hline
			11 & $\mathcal{R}e(C_T)$ & \text{-0.111(26)} & -0.93 & 11 & $\mathcal{R}e(C_{S_2})$ & \text{0.067(79)} & -0.407 & 13 & $\mathcal{R}e(C_{S_1})$ & \text{0.081(90)} & -0.686 \\
			  & $\mathcal{R}e(C_{V_2})$ & \text{0.194(75)} &   &   & $\mathcal{R}e(C_{V_1})$ & \text{0.093(24)} &   &   & $\mathcal{R}e(C_{V_1})$ & \text{0.081(30)} &   \\
			%\hline
			13 & $\mathcal{R}e(C_{S_1})$ & \text{0.178(67)} & 0.353 & 12 & $\mathcal{R}e(C_{V_1})$ & \text{0.118(31)} & 0.707 & 14 & $\mathcal{I}m(C_{V_2})$ & \text{-0.500(68)} & -0.716 \\
			  & $\mathcal{R}e(C_T)$ & \text{-0.034(12)} &   &   & $\mathcal{R}e(C_{V_2})$ & \text{0.036(41)} &   &   & $\mathcal{R}e(C_{V_2})$ & \text{0.037(46)} &   \\
			%\hline
			14 & $\mathcal{R}e(C_{S_2})$ & \text{-0.011(89)} & -0.129 & 13 & $\mathcal{I}m(C_{V_2})$ & \text{0.500(68)} & 0.716 & 15 & $\mathcal{R}e(C_{V_1})$ & \text{0.118(31)} & 0.707 \\
			  & $\mathcal{R}e(C_T)$ & \text{0.387(11)} &   &   & $\mathcal{R}e(C_{V_2})$ & \text{0.037(46)} &   &   & $\mathcal{R}e(C_{V_2})$ & \text{0.036(41)} &   \\
			%\hline
			15 & $\mathcal{I}m(C_{V_2})$ & \text{0.499(68)} & 0.714 & 14 & $\mathcal{R}e(C_T)$ & \text{0.025(34)} & 0.917 & 16 & $\mathcal{R}e(C_T)$ & \text{0.025(34)} & 0.917 \\
			  & $\mathcal{R}e(C_{V_2})$ & \text{0.040(46)} &   &   & $\mathcal{R}e(C_{V_1})$ & \text{0.140(54)} &   &   & $\mathcal{R}e(C_{V_1})$ & \text{0.139(54)} &   \\
			%\hline
			17 & $\mathcal{I}m(C_{T})$ & \text{0.094(359)} & -0.998 & 17 & $\mathcal{I}m(C_{T})$ & \text{0.15(19)} & -0.997 & $-$ & \multicolumn{3}{c}{$-$} \\
			  & $\mathcal{R}e(C_T)$ & \text{0.37(17)} &   &   & $\mathcal{R}e(C_T)$ & \text{0.33(17)} & & & \multicolumn{3}{c}{} \\
			  \hline
			%- & - & - & - & 18 & \multicolumn{3}{c||}{See Plot} & - & - & - & - \\
% 			- & - & - & - &   & $\mathcal{R}e(C_{V_1})$ & \text{-1.9(9957)} &   & - & - & - & - \\
        
	      \end{tabular}
	      }
%	\end{ruledtabular}
\end{table*}

\begin{table*}[htbp]
	\centering
	\caption{ Results for the scenarios listed in table \ref{tab:beL1}, following the convention of table \ref{tab:alldat2}.}
		\label{tab:beL2}
%	\begin{ruledtabular}
     \resizebox{\textwidth}{!}{
\begin{tabular}{cccc||cccc||cccc}
	\hline
 \multicolumn{4}{c||}{Belle + LHCb (Except $\mathcal{R}_{J/\Psi}$) } & \multicolumn{4}{c||}{Belle + LHCb ($\mathcal{R}_{J/\Psi}$ with LFCQ)} & \multicolumn{4}{c}{Belle + LHCb ($\mathcal{R}_{J/\Psi}$ with PQCD)}\\
			\hline
			Index & Param.s & Best-fit & Correlation & Index & Param.s & Best-fit & Correlation & Index & Param.s & Best-fit & Correlation \\
			\hline
 2 & $\mathcal{R}e(C_{V_1})$ & \text{0.075(26)} & -- & 1 & $\mathcal{R}e(C_{V_1})$ & \text{0.078(26)} & -- & 2 & $\mathcal{R}e(C_{V_1})$ & \text{0.078(26)} & -- \\
 %\hline
 3 & $\mathcal{R}e(C_T)$ & \text{0.379(13)} & -- & 4 & $\mathcal{R}e(C_T)$ & \text{-0.037(13)} & -- & 4 & $\mathcal{R}e(C_T)$ & \text{-0.037(13)} & -- \\
 %\hline
 5 & $\mathcal{R}e(C_{S_1})$ & \text{0.179(81)} & -- & 5 & $\mathcal{R}e(C_{V_2})$ & \text{-0.074(34)} & -- & 5 & $\mathcal{R}e(C_{V_2})$ & \text{-0.074(33)} & -- \\
 %\hline
 6 & $\mathcal{R}e(C_{V_2})$ & \text{-0.069(34)} & -- & 6 & $\mathcal{R}e(C_T)$ & \text{-0.086(38)} & -0.951 & 6 & $\mathcal{R}e(C_{S_2})$ & \text{0.033(108)} & -0.309 \\
   &   &   &   &   & $\mathcal{R}e(C_{V_2})$ & \text{0.14(10)} &   &   & $\mathcal{R}e(C_{V_1})$ & \text{0.073(27)} &   \\
 %\hline
 9 & $\mathcal{R}e(C_{S_2})$ & \text{0.126(96)} & 0.015 & 7 & $\mathcal{R}e(C_{S_1})$ & \text{0.180(81)} & -- & 9 & $\mathcal{R}e(C_{S_2})$ & \text{0.209(98)} & -0.295 \\
   & $\mathcal{R}e(C_T)$ & \text{-0.035(13)} &   &   &   &   &   &   & $\mathcal{R}e(C_{V_2})$ & \text{-0.092(34)} &   \\
 %\hline
 10 & $\mathcal{R}e(C_{S_2})$ & \text{0.034(108)} & -0.306 & 8 & $\mathcal{R}e(C_{S_1})$ & \text{0.125(94)} & 0.361 & 10 & $\mathcal{R}e(C_T)$ & \text{-0.086(38)} & -0.952 \\
   & $\mathcal{R}e(C_{V_1})$ & \text{0.072(27)} &   &   & $\mathcal{R}e(C_T)$ & \text{-0.029(14)} &   &   & $\mathcal{R}e(C_{V_2})$ & \text{0.14(10)} &   \\
 %\hline
 11 & $\mathcal{R}e(C_{S_2})$ & \text{0.207(98)} & -0.298 & 9 & $\mathcal{R}e(C_{S_2})$ & \text{0.126(96)} & 0.015 & 11 & $\mathcal{R}e(C_{S_2})$ & \text{0.126(96)} & 0.019 \\
   & $\mathcal{R}e(C_{V_2})$ & \text{-0.091(34)} &   &   & $\mathcal{R}e(C_T)$ & \text{-0.035(13)} &   &   & $\mathcal{R}e(C_T)$ & \text{-0.036(13)} &   \\
 %\hline
 12 & $\mathcal{R}e(C_T)$ & \text{-0.085(38)} & -0.951 & 12 & $\mathcal{R}e(C_{S_1})$ & \text{0.183(86)} & 0.062 & 12 & $\mathcal{R}e(C_{S_1})$ & \text{0.181(81)} & -- \\
   & $\mathcal{R}e(C_{V_2})$ & \text{0.14(10)} &   &   & $\mathcal{R}e(C_{V_2})$ & \text{-0.065(33)} &   &   &   &   &   \\
 %\hline
 13 & $\mathcal{I}m(C_{T})$ & \text{0.206(29)} & -0.898 & 13 & $\mathcal{R}e(C_{S_2})$ & \text{0.033(108)} & -0.309 & 13 & $\mathcal{R}e(C_{S_1})$ & \text{0.125(94)} & 0.361 \\
   & $\mathcal{R}e(C_T)$ & \text{0.20(19)} &   &   & $\mathcal{R}e(C_{V_1})$ & \text{0.073(27)} &   &   & $\mathcal{R}e(C_T)$ & \text{-0.029(14)} &   \\
 %\hline
 - & - & - & - & 14 & $\mathcal{R}e(C_{S_1})$ & \text{0.043(122)} & -0.656 & 14 & $\mathcal{R}e(C_{S_1})$ & \text{0.183(86)} & 0.062 \\
 - & - & - & - &   & $\mathcal{R}e(C_{V_1})$ & \text{0.067(34)} &   &   & $\mathcal{R}e(C_{V_2})$ & \text{-0.065(33)} &   \\
 %\hline
 - & - & - & - & 15 & $\mathcal{R}e(C_{S_2})$ & \text{0.208(98)} & -0.295 & 15 & $\mathcal{R}e(C_{S_1})$ & \text{0.043(122)} & -0.656 \\
 - & - & - & - &   & $\mathcal{R}e(C_{V_2})$ & \text{-0.092(34)} &   &   & $\mathcal{R}e(C_{V_1})$ & \text{0.067(34)} &   \\
 %\hline
 - & - & - & - & 16 & $\mathcal{R}e(C_{V_1})$ & \text{0.086(42)} & 0.788 & 16 & $\mathcal{I}m(C_{V_2})$ & \text{0.42(11)} & 0.794 \\
 - & - & - & - &   & $\mathcal{R}e(C_{V_2})$ & \text{0.018(54)} &   &   & $\mathcal{R}e(C_{V_2})$ & \text{0.015(59)} &   \\
 %\hline
 - & - & - & - & 17 & $\mathcal{I}m(C_{V_2})$ & \text{-0.42(11)} & -0.794 & 17 & $\mathcal{R}e(C_{V_1})$ & \text{0.086(42)} & 0.788 \\
 - & - & - & - &   & $\mathcal{R}e(C_{V_2})$ & \text{0.016(59)} &   &   & $\mathcal{R}e(C_{V_2})$ & \text{0.018(54)} &   \\
 %\hline
 - & - & - & - & 18 & $\mathcal{R}e(C_T)$ & \text{0.0086(433)} & 0.941 & 18 & $\mathcal{R}e(C_T)$ & \text{0.0078(429)} & 0.941 \\
 - & - & - & - &   & $\mathcal{R}e(C_{V_1})$ & \text{0.092(77)} &   &   & $\mathcal{R}e(C_{V_1})$ & \text{0.091(76)} &   \\
%  \hline
%  - & - & - & - &  & \multicolumn{3}{c||}{} & $-$ & \multicolumn{3}{c|}{$-$} \\
%  - & - & - & - &   & \multicolumn{3}{c||}{} &   &  \multicolumn{3}{c|}{} \\
 %\hline
 - & - & - & - & 19 & $\mathcal{I}m(C_{T})$ & \text{0.210(15)} & 0.55 & - & - & - & - \\
 - & - & - & - &   & $\mathcal{R}e(C_T)$ & \text{0.16(19)} &   & - & - & - & - \\
 \hline
\end{tabular}
}
%	\end{ruledtabular}
\end{table*}

\begin{table*}[htbp]
	\centering
	\caption{Results for the scenarios listed in table \ref{tab:allrdst1}, following the convention of table \ref{tab:alldat2}.}
		\label{tab:allrdst2}
%	\begin{ruledtabular}
     \resizebox{\textwidth}{!}{
\begin{tabular}{cccc||cccc||cccc}
	\hline
 \multicolumn{4}{c||}{All $\mathcal{R}_{D^*}$} & \multicolumn{4}{c||}{All $\mathcal{R}_{D^*}$ + $\mathcal{R}_{J/\Psi}$ (LFCQ)} & \multicolumn{4}{c}{All $\mathcal{R}_{D^*}$ + $\mathcal{R}_{J/\Psi}$ (PQCD)}\\
			\hline
			Index & Param.s & Best-fit & Correlation & Index & Param.s & Best-fit & Correlation & Index & Param.s & Best-fit & Correlation \\
			\hline
 1 & $\mathcal{R}e(C_{V_1})$ & \text{0.086(26)} & -- & 1 & $\mathcal{R}e(C_T)$ & \text{-0.040(11)} & -- & 1 & $\mathcal{R}e(C_T)$ & \text{-0.040(11)} & -- \\
 %\hline
 2 & $\mathcal{R}e(C_{V_2})$ & \text{-0.095(29)} & -- & 2 & $\mathcal{R}e(C_{V_2})$ & \text{-0.099(28)} & -- & 4 & $\mathcal{R}e(C_{V_2})$ & \text{-0.099(28)} & -- \\
 %\hline
 5 & $\mathcal{R}e(C_T)$ & \text{-0.039(11)} & -- & 4 & $\mathcal{R}e(C_{V_1})$ & \text{0.089(26)} & -- & 6 & $\mathcal{R}e(C_{V_1})$ & \text{0.089(26)} & -- \\
 \hline
%  - & - & - & - & 6 & \multicolumn{3}{c||}{See Plot} & - & - & - & - \\
%  \hline
%  - & - & - & - &   & $\mathcal{R}e(C_{V_2})$ & \text{1.0(21)} &   & - & - & - & - \\
\end{tabular}
}
%	\end{ruledtabular}
\end{table*}

%%%%%%%%%%%%%%%%%%%%%%%%%%%%
\section{Results}
%%%%%%%%%%%%%%%%%%%%%%%%%%%%
	
	Following the methodology described in the previous subsection, we have taken several combinations of the available data and have performed the analysis in
	the following stages for each dataset.
	
%%%%%%%%%%%%%%%%%%%%%%%%%%%%%
\subsection{Model selection}\label{sec:res2}
%%%%%%%%%%%%%%%%%%%%%%%%%%%%%
	
	We have created the $\chi^2_k$ statistic for the $k^\text{th}$ scenario containing real and imaginary parts of Wilson coefficients $C^k_W$, and repeated 
	that for all $k$ (let us reiterate here that scenarios containing all imaginary $C_W$s are neglected). We have taken scenarios with as many as 4 individual 
	components of $\mathcal{R}e(\text{or }\mathcal{I}m)(C_W)$. Then we have minimized each of those over the corresponding parameter space (with the form factor
	parameters as nuisance parameters). After checking normality for each fit and dropping scenarios with $\le 5\%$ significance, we have arranged the remaining 
	scenarios in ascending order of AIC$_c$ and have kept only those with $\Delta$AIC$_c \le 4$. These are, essentially, the best scenarios to explain the data 
	in that specific dataset under the present experimental constraints. 
	
	Tables \ref{tab:alldat1}, \ref{tab:noptau1}, \ref{tab:beL1} and \ref{tab:allrdst1} contain the listed scenarios of the data sets which are obtained from table \ref{tab:RDRDsPtau}. Each table essentially contains three variations of similar datasets: the first one is data without $\mathcal{R}_{J/\Psi}$ and the rest two, with it. The reason for treating $\mathcal{R}_{J/\Psi}$ separately is the apparent tension of the measured central value with that of the SM one, as explained in section \ref{sec:experiment}. Moreover, the theoretical values of $\mathcal{R}_{J/\Psi}$ are heavily dependent on form factor parametrisation and differ considerably over different choices of it, as explained in section \ref{sec:Rjpsi}. Thus without showing bias to a particular type of parametrisation, we treat LFCQ and PQCD separately in second and third datasets of each table respectively. As PQCD predicts relatively higher values for $\mathcal{R}_{J/\Psi}$, fits are generally better for these sets than those corresponding to the LFCQ ones, as can be checked from the $p$-values listed in the second column of these datasets. The measured value of $P_{\tau}(D^*)$ has large error. Therefore,we have dropped $\tau$-polarization asymmetry from the list of inputs in one of the fits (table \ref{tab:noptau1}). The measured values	of $\mathcal{R}_{D^{(*)}}$ by \Babar~are relatively old, and are largely deviated from the respective SM predictions. Therefore, in order to check the impact of \Babar~data on our model selections, in one of the fit scenario we have dropped the \Babar~data (table \ref{tab:beL1}). We have also done the analysis with the measured $\mathcal{R}_{D^*}$ alone, which will help us to figure out the sensitivity of this observable towards a particular type of NP scenario.      
	
	We notice that for all datasets, the maximum number of independent fit parameters (for listed scenarios) is 2. As is explained in the previous sections, 
	this is natural, because AIC$_c$ penalises the increased variance associated with increase in number of independent parameters. The fourth column in these tables, for 
	each dataset, lists the $w^{\text{AIC}_c}$ for each scenario, which estimates the relative likelihood for that scenario (among the given set of scenarios, 
	the number of which is $\approx 90 \sim 95$ for our analysis) to explain the data. As can be seen, the first few scenarios take up a large chunk of the total 
	likelihood and it is evident that all unselected scenarios together constitute a very small fraction of it. This is another way of understanding why the listed 
	scenarios are the best ones suited to describe the given dataset.
		
	Once we have listed the best scenarios, we scrutinise the allowed parameter space for each of them. As all the finally selected models have a most of 
	two parameters (other than the nuisance parameters), we can plot the marginal confidence levels in the fit-parameter-space with the help of the defined 
	$\chi^2$ function. We have prepared plots for each of these scenarios either in terms of the goodness-of-fit contours (for two parameter scenarios), or 
	by directly plotting the $\chi^2$ with respect to the parameter (for single parameter scenarios). The contour plots are prepared with constant $\chi^2$ 
	contours equivalent to $1\sigma$ and $3\sigma$ that correspond to confidence levels of $68.27\%$ and $99.73\%$ respectively. For two parameters, the 
	constant $\chi^2$ values are $= \chi^2_{min} + \Delta\chi^2$, where $\chi^2_{min}$ is the minimum value of $\chi^2$ obtained after minimization over 
	the parameter space, and $\Delta\chi^2 = 2.296$ and $11.83$ for $1\sigma$ and $3\sigma$ respectively. Similarly, for single parameter case, $1\sigma$ and 
	$3\sigma$ intervals are shown with $\Delta\chi^2 = 1$ and $9$ respectively. As a representative case, we have shown the plots for all scenarios listed in 
	the last dataset of table \ref{tab:alldat1} (`All data', where the theoretical value of $\mathcal{R}_{J/\Psi}$ is calculated in PQCD) in figures 
	\ref{fig:alldat1} and \ref{fig:alldat2}. %\hlab{All other plots corresponding to other datasets are given in the ancillary files  downloadable from the arXiv	preprint page (detail given in appendix).} 
	
	We then use the limits on $\mathcal{B}(B_c\to\tau\nu)$ mentioned in section \ref{sec:experiment} as our conservative constraint on each scenario. 
	In the aforementioned tables (\ref{tab:alldat1}, \ref{tab:noptau1}, \ref{tab:beL1} and \ref{tab:allrdst1}), the last column of each dataset indicates 
	whether the corresponding scenario passes the constraint of $\mathcal{B}(B_c\to\tau\nu)\le 30\%$ (`\checkmark') or not (`$\pmb{\times}$'). For many scenarios,
	there are multiple best-fit points (or at least, multiple 68\% confidence regions). In some cases, some of these multiple minima are ruled out from the 
	$\mathcal{B}(B_c\to\tau\nu) \le 30\%$ constraints, while the rest are allowed. These scenarios are marked as `\checkmark {\bf !}' in the said column. 
	By observing the nature of the confidence levels in the parameter space, we can pick out the minima which are allowed. 
	
	In case of the plots depicting the parameter space, we show two limits: regions disallowed by $\mathcal{B}(B_c\to\tau\nu) \le 30\%$ (gray shaded region) 
	and $\mathcal{B}(B_c\to\tau\nu) \le 10\%$ (diagonally hatched region) respectively. The first one is the conservative limit from $B_c$ decay width and 
	the second aggressive one is motivated from the studies quoted in section \ref{sec:experiment} (though we are calling it aggressive, it is a perfectly 
	reasonable upper bound, given the $\sim 2\%$ SM prediction as shown in table \ref{tab:SMres}). We know that even if present, any NP effect is going to 
	be small. Thus when multiple minima are allowed by these constraints, we quote the results for those which are closest to the origin (corresponding to SM)
	in further analysis\footnote{As an example, for scenario 4 given in the last dataset of table \ref{tab:alldat1} (with $\mathcal{R}e(C_{S_2})$ and
	$\mathcal{R}e(C_{V_1})$), the symbol for the constraint column is `\checkmark {\bf !}'. The corresponding plot in figure \ref{fig:alldat4JPQ} shows that out of the 4 possible minima, only 2 are outside the shaded regions. Interestingly, the global minima of the $\chi^2$ in this case lies within the shaded region (thus disallowed), but there are two more allowed local $1\sigma$ C.L regions. In table \ref{tab:alldat2} (last dataset), the corresponding result is quoted for the one closest to zero.}. One more thing to note here is that scenarios with either real or imaginary part of $C_T$ will not have any constraint on the corresponding axis in the parameter space, as $\mathcal{B}(B_c\to\tau\nu)$ is unaffected by tensor interactions.
	
	Our results of the model selection with all the available data are shown in table \ref{tab:alldat1}. We note that the best possible scenarios are	the models with either of the operator ${\cal O}^{\ell}_T$ or ${\cal O}^{\ell}_{V_1}$ with real $C_W$. The explanations with the operator ${\cal O}^{\ell}_{S_1}$ are disfavoured by the bound from $\mathcal{B}(B_c \to \tau\nu_{\tau})$. These one-operator scenarios are allowed even if we choose $\Delta$AIC$_c \le 2$. The same criterion does not allow the two-operator scenarios. However, the constrain $\Delta$AIC$_c \le 4$ allows a few of the two-operator scenarios. In all these less preferred two-operator scenarios, the operators like ${\cal O}^{\ell}_{V_2}$, ${\cal O}^{\ell}_{S_1}$,  ${\cal O}^{\ell}_{S_2}$ appears in combination with either of ${\cal O}^{\ell}_T$ and ${\cal O}^{\ell}_{V_1}$. Also, we note that the operator ${\cal O}^{\ell}_{V_2}$ with complex $C_W$ is favoured by the data. 

	As can be seen in table \ref{tab:noptau1}, our conclusions on the selected models will not change if we drop the experimental input on $P_{\tau}(D^*)$ from our fit. Also, if we drop the \Babar~data on $\mathcal{R}_{D^{(*)}}$ from the list of the inputs for the fit, the best preferred scenarios are still the one with the operator ${\cal O}^{\ell}_T$ or ${\cal O}^{\ell}_{V_1}$ with real $C_W$s. However, there are other one-operator scenarios with ${\cal O}^{\ell}_{V_2}$ or ${\cal O}^{\ell}_{S_1}$ with real $C_W$s which are then allowed by the criterion $\Delta$AIC$_c \le 4$; see table \ref{tab:beL1} for detail. Here too, there are a few two-operator scenarios which successfully pass the above mentioned criterion. Finally, in order to understand the impact of the $\mathcal{R}_{D^{*}}$, we have also done an analysis considering only the data on $\mathcal{R}_{D^{*}}$ (table \ref{tab:allrdst1}). It shows that all of ${\cal O}^{\ell}_{V_1}$, ${\cal O}^{\ell}_T$, ${\cal O}^{\ell}_{V_2}$, ${\cal O}^{\ell}_{S_1}$ and ${\cal O}^{\ell}_{S_2}$ are allowed by the criterion $\Delta$AIC$_c \le 4$. However, the constraints from $\mathcal{B}(B_c\to \tau\nu_{\tau})$ disfavours the scenarios with scalar operators. As can be seen across all the tables, our conclusions will not change much if we incorporate $\mathcal{R}_{J/\psi}$ as input in our fit. This could be due to the large uncertainties present both in the predictions and measured values of $\mathcal{R}_{J/\psi}$.
	
	Our goal in this analysis is to find out what type new operators can best explain the existing data. Our operator basis consists of all the linearly independent operators that are relevant for the $b\to c \tau\nu_{\tau}$ decays. Therefore, we did not extend our analysis to look for new physics models whose effects in $b\to c\tau\nu_{\tau}$ can be parametrised by one or more of the operators of our operator basis. Our priority is the analysis of the data, and the data can guide us to build models.

\begin{table*}[t]
	\centering
	\caption{\small Predictions for observables listed in table \ref{tab:SMres}, for different scenarios listed in the first dataset of table \ref{tab:alldat2}. We have omitted the scenario where the best fit is a contour instead of a curve (scenario 2). Each result contains two uncertainties: the first one for uncertainties in form factors and the second one for uncertainties of the fitted NP Wilson coefficients. Though in most cases the NP error is predominant compared to the form factor ones, we quote both of them. These results continue in the next table \ref{tab:pred2}.}
		\label{tab:pred1}
%	\begin{ruledtabular}
		\scriptsize
		\begin{tabular}{ccccccc}
			\hline
			\multicolumn{7}{c}{Data without $\mathcal{R}_{J/\Psi}$}\\
			\hline
			\text{Scenario} & $\mathcal{R}_{D^*}$ & $\mathcal{R}_{D}$ & $P_{\tau}(D^*)$ & $P_{\tau}(D)$ & $F^{D^*}_L$ & $\mathcal{A}^*_{FB}$ \\
			\hline
			1 & \text{0.306(12)(14)} & \text{0.412(4)(3)} & \text{0.125(11)(11)} & \text{0.1799(4)(33)} & \text{0.134(15)(6)} & \text{0.027 (8)(12)} \\
			%\hline
			3 & \text{0.312(7)(12)} & \text{0.368(4)(14)} & \text{-0.493(25)(0)} & \text{0.3355(4)(0)} & \text{0.456(10)(0)} & \text{-0.059(14)(0)} \\
			%\hline
			7 & \text{0.305(7)(15)} & \text{0.406(4)(47)} & \text{-0.504(23)(13)} & \text{0.4080(5)(746)} & \text{0.452(10)(5)} & \text{-0.063(13)(5)} \\
			%\hline
			8 & \text{0.305(5)(15)} & \text{0.406(4)(47)} & \text{-0.485(16)(15)} & \text{0.5321 (5)(531)} & \text{0.431(7)(5)} & \text{-0.021(10)(13)} \\
			%\hline
			9 & \text{0.305(6)(15)} & \text{0.406(4)(47)} & \text{-0.534(19)(10)} & \text{0.6055 (6)(493)} & \text{0.450(9)(4)} & \text{-0.038(11)(10)} \\
			%\hline
			10 & \text{0.305(5)(15)} & \text{0.407(4)(46)} & \text{-0.381(12)(35)} & \text{0.3845 (4)(92)} & \text{0.396(7)(18)} & \text{0.006(8)(11)} \\
			%\hline
			11 & \text{0.305(7)(15)} & \text{0.407(4)(46)} & \text{-0.436(25)(12)} & \text{0.5263(5)(512)} & \text{0.457(10)(8)} & \text{-0.007(13)(12)} \\
			%\hline
			12 & \text{0.305(8)(15)} & \text{0.407(4)(46)} & \text{-0.443(32)(12)} & \text{0.5737(6)(447)} & \text{0.480(12)(5)} & \text{-0.017(15)(10)} \\
			%\hline
			13 & \text{0.305(7)(15)} & \text{0.407(4)(46)} & \text{-0.475(27)(17)} & \text{0.4212 (5)(840)} & \text{0.463(11)(6)} & \text{-0.052(14)(6)} \\
			%\hline
			14 & \text{0.305(7)(15)} & \text{0.407(4)(46)} & \text{-0.492(25)(1)} & \text{0.3355(4)(0)} & \text{0.454(10)(4)} & \text{-0.070(14)(14)} \\
			%\hline
			15 & \text{0.305(7)(15)} & \text{0.407(4)(46)} & \text{-0.492(25)(1)} & \text{0.3355(4)(0)} & \text{0.454(10)(4)} & \text{0.010(12)(13)} \\
			%\hline
			16 & \text{0.305(8)(15)} & \text{0.406(4)(47)} & \text{-0.505(28)(12)} & \text{0.3227(4)(1411)} & \text{0.463(12)(5)} & \text{-0.090(15)(37)} \\
			%\hline
			17 & \text{0.308(11)(14)} & \text{0.394(4)(42)} & \text{0.036(9)(211)} & \text{0.2001(4)(499)} & \text{0.181(14)(111)} & \text{0.023(8)(17)} \\
			\hline
%			17 & \text{0.312(7)(12)} & \text{0.368(4)(14)} & \text{-0.493(25)(0)} & \text{0.3355(4)(0)} & \text{0.456(10)(0)} & \text{-0.059(14)(0)} \\
		 \end{tabular}
		 %	\end{ruledtabular}
\end{table*}

\begin{table*}[htbp]
	\centering
	\caption{\small Prediction of observables, continued from table \ref{tab:pred1}.}
		\label{tab:pred2}
%	\begin{ruledtabular}
		\scriptsize
		\begin{tabular}{ccccccc}
			\hline
		 \multicolumn{7}{c}{Data without $\mathcal{R}_{J/\Psi}$}\\
			\hline
			\text{Scenario} & $\mathcal{A}_{FB}$ & $\mathcal{R}_{J/\Psi}$ & $\mathcal{R}^{\mu}_{\Lambda}$ & $\mathcal{R}^{e}_{\Lambda}$ & $\mathcal{A}^{\lambda}_{FB}$ & $\mathcal{B}(B_c\to\tau\nu) \times 10^2$ \\
			\hline
			1 & \text{0.4400(0)(11)} & \text{0.201(17)(10)} & \text{0.462(18)(18)} & \text{0.460(18)(18)} & \text{0.1674(0)(48)} & \text{2.08(18)(0)} \\
			%\hline
			3 & \text{0.3586(3)(0)} & \text{0.348(34)(14)} & \text{0.397(16)(16)} & \text{0.396(16)(16)} & \text{0.0231(0)(0)} & \text{2.51(21)(10)} \\
			%\hline
			7 & \text{0.3489(3)(116)} & \text{0.339(33)(16)} & \text{0.401(16)(17)} & \text{0.400(16)(17)} & \text{0.0273(0)(45)} & \text{1.31(11)(110)} \\
			%\hline
			8 & \text{0.3124(4)(128)} & \text{0.348(35)(18)} & \text{0.406(17)(18)} & \text{0.405(17)(18)} & \text{0.0434(0)(30)} & \text{0.15(1)(30)} \\
			%\hline
			9 & \text{0.3053 (4)(144)} & \text{0.343(34)(17)} & \text{0.400(16)(16)} & \text{0.399(16)(16)} & \text{0.0318(0)(20)} & \text{0}  \\
			%\hline
			10 & \text{0.3233(3)(70)} & \text{0.353(36)(18)} & \text{0.418(17)(21)} & \text{0.417(17)(21)} & \text{0.0614(0)(70)} & \text{1.35(11)(25)} \\
			%\hline
			11 & \text{0.3164(3)(114)} & \text{0.348(34)(18)} & \text{0.404(16)(17)} & \text{0.402(16)(17)} & \text{0.0608(0)(78)} & \text{6.21(53)(194)} \\
			%\hline
			12 & \text{0.3142 (4)(120)} & \text{0.344(33)(17)} & \text{0.398(16)(16)} & \text{0.397(16)(16)} & \text{0.0573(0)(76)} & \text{8.90(76)(213)} \\
			%\hline
			13 & \text{0.3468(3)(137)} & \text{0.340(33)(16)} & \text{0.400(16)(16)} & \text{0.399(16)(16)} & \text{0.0370(0)(141)} & \text{4.24(36)(205)} \\
			%\hline
			14 & \text{0.3586(3)(0)} & \text{0.338(33)(17)} & \text{0.402(16)(17)} & \text{0.400(16)(17)} & \text{0.0251(0)(26)} & \text{2.43(20)(13)} \\
			%\hline
			15 & \text{0.3586(3)(0)} & \text{0.338(33)(17)} & \text{0.402(16)(17)} & \text{0.400(16)(17)} & \text{0.0889(0)(172)} & \text{2.43(20)(13)} \\
			%\hline
			16 & \text{0.3672(2)(93)} & \text{0.332(32)(23)} & \text{0.398(16)(16)} & \text{0.397(16)(16)} & \text{0.0151(0)(77)} & \text{2.72(23)(26)} \\
			%\hline
			17 & \text{0.4290(0)(272)} & \text{0.225(19)(58)} & \text{0.450(17)(33)} & \text{0.449(17)(33)} & \text{0.1512(0)(400)} & \text{2.08(18)(0)} \\
			\hline
%			17 & \text{0.3586(3)(0)} & \text{0.348(34)(14)} & \text{0.397(16)(16)} & \text{0.396(16)(16)} & \text{0.0231(0)(0)} & \text{2.51(21)(10)} \\
		  \end{tabular}
		
%	\end{ruledtabular}
	
\end{table*}

\begin{figure*}[!htbp]
	\centering
	\subfloat[]{\includegraphics[height=4.5cm]{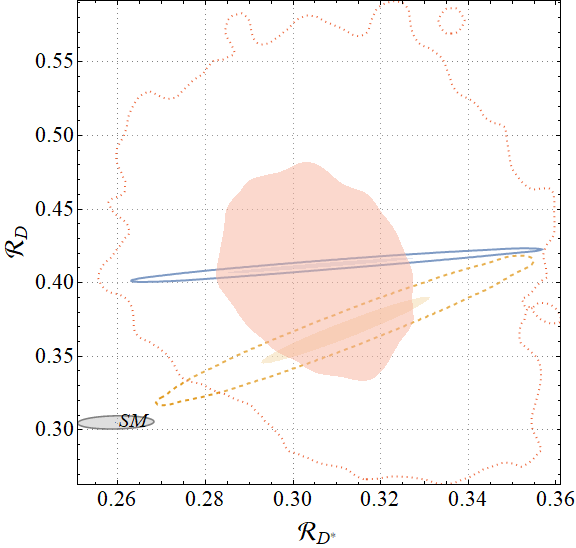}\label{fig:alldatcorrpltRDstRD1}}~
	\subfloat[]{\includegraphics[height=4.5cm]{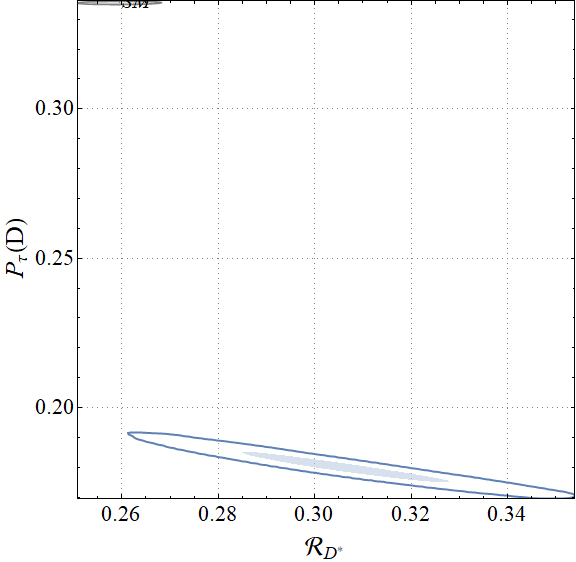}\label{fig:alldatcorrpltRDstPtauD1}}~
	\subfloat[]{\includegraphics[height=4.5cm]{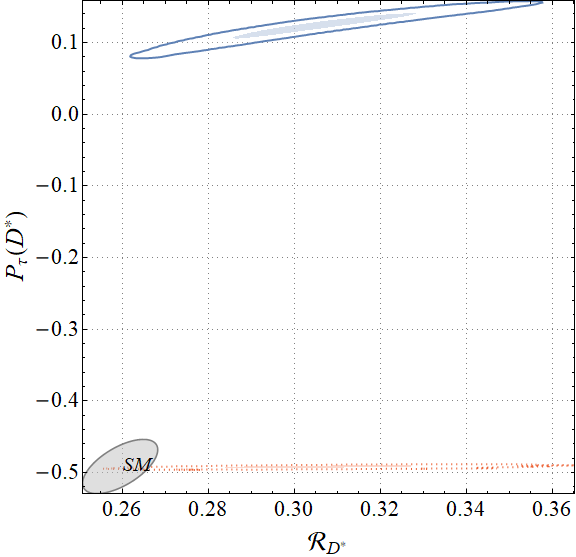}\label{fig:alldatcorrpltRDstPtauDst1}}\\
	\subfloat[]{\includegraphics[height=4.5cm]{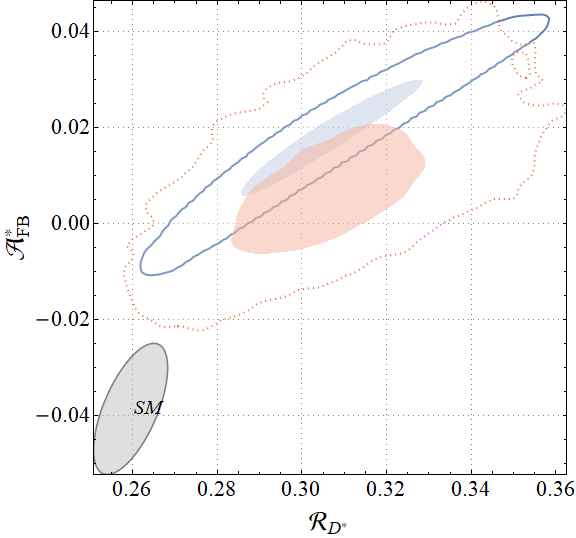}\label{fig:alldatcorrpltRDstAFBDst1}}~
	\subfloat[]{\includegraphics[height=4.5cm]{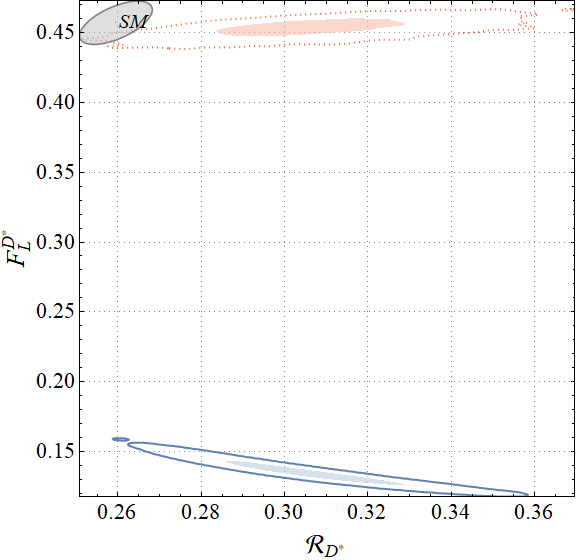}\label{fig:alldatcorrpltRDstFLDst1}}~
	\subfloat[]{\includegraphics[height=4.5cm]{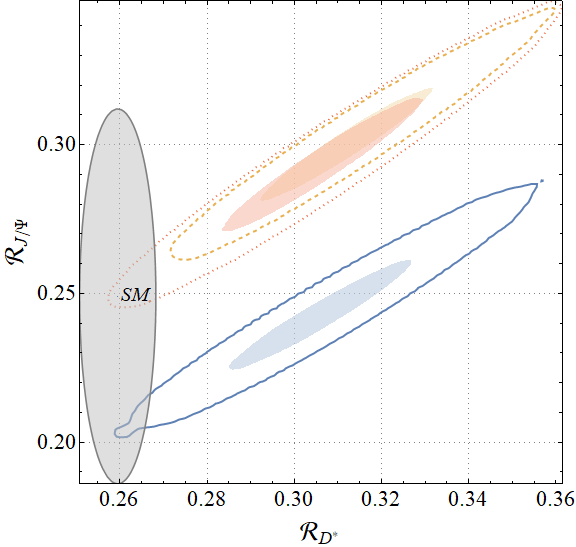}\label{fig:alldatcorrpltRDstRJ1}}\\
	\subfloat[]{\includegraphics[height=4.5cm]{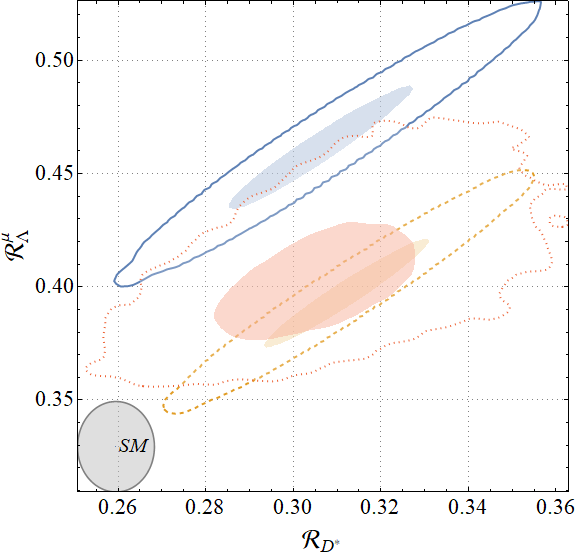}\label{fig:alldatcorrpltRDstRlamMu1}}~~
	\subfloat[]{\includegraphics[height=4.5cm]{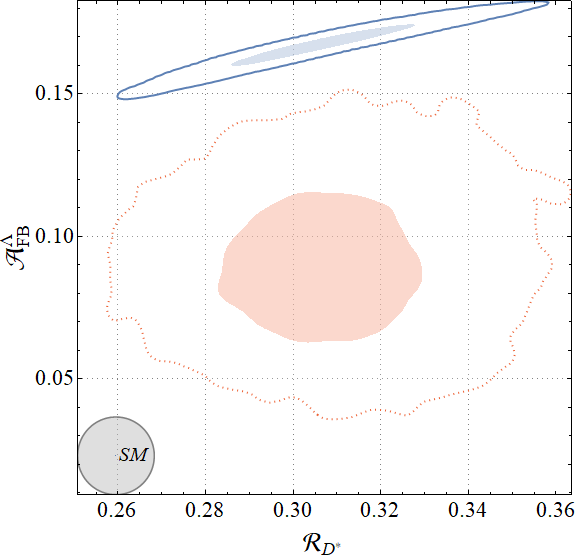}\label{fig:alldatcorrpltRDstAFBlam1}}~
	\subfloat{\includegraphics[height=5cm]{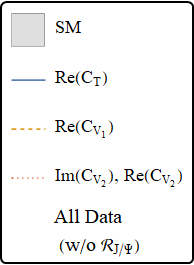}\label{fig:alldatcorrpltLegend1}}
	\caption{\small Correlation plots among different observables for one-operator scenarios listed in the first column of Table \ref{tab:alldat2}. Blue (solid), Orange(dashed), and Red(dotted) contours correspond to the scenarios with 
		$\mathcal{R}e(C_{T})$, $\mathcal{R}e(C_{V_1})$, and complex $C_{V_2}$ respectively. For each of these scenarios, $1\sigma$ (filled region) and $3\sigma$ contours are given. }
	\label{fig:alldatcorrplt1}
\end{figure*}

\begin{figure*}[!htbp]
	\centering
	\subfloat[]{\includegraphics[height=4.5cm]{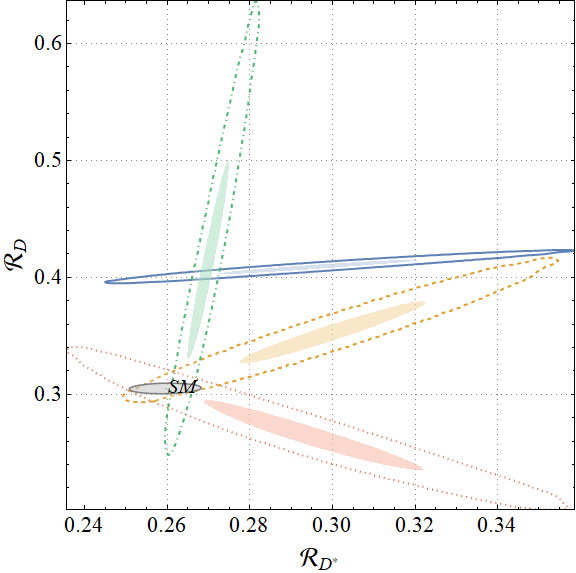}\label{fig:BELLHCBcorrpltRDstRD1}}~
	\subfloat[]{\includegraphics[height=4.5cm]{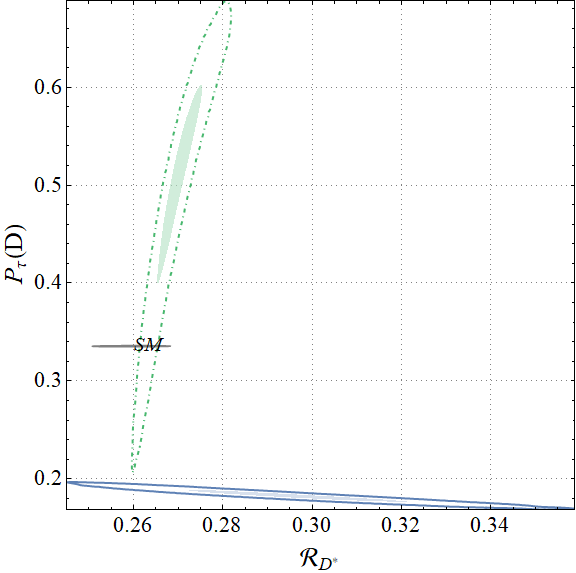}\label{fig:BELLHCBcorrpltRDstPtauD1}}~
	\subfloat[]{\includegraphics[height=4.5cm]{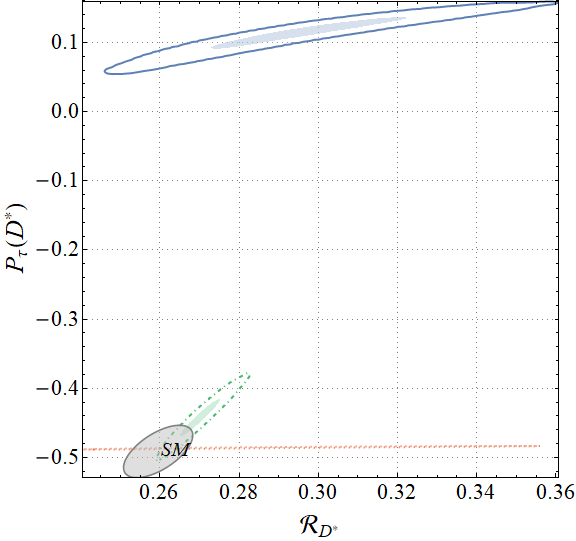}\label{fig:BELLHCBcorrpltRDstPtauDst1}}\\
	\subfloat[]{\includegraphics[height=4.5cm]{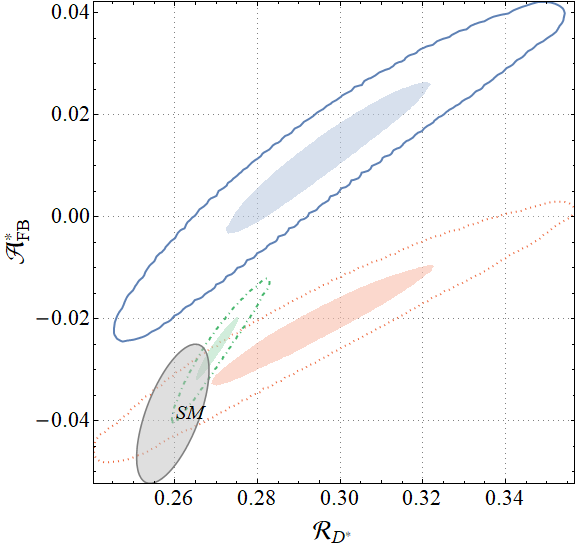}\label{fig:BELLHCBcorrpltRDstAFBDst1}}~
	\subfloat[]{\includegraphics[height=4.5cm]{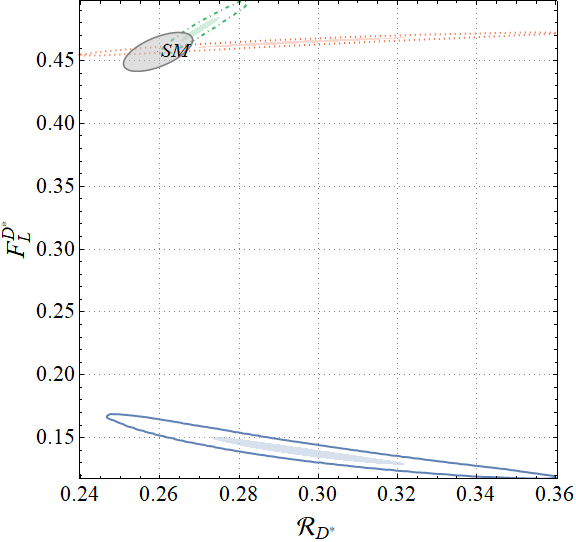}\label{fig:BELLHCBcorrpltRDstFLDst1}}~
	\subfloat[]{\includegraphics[height=4.5cm]{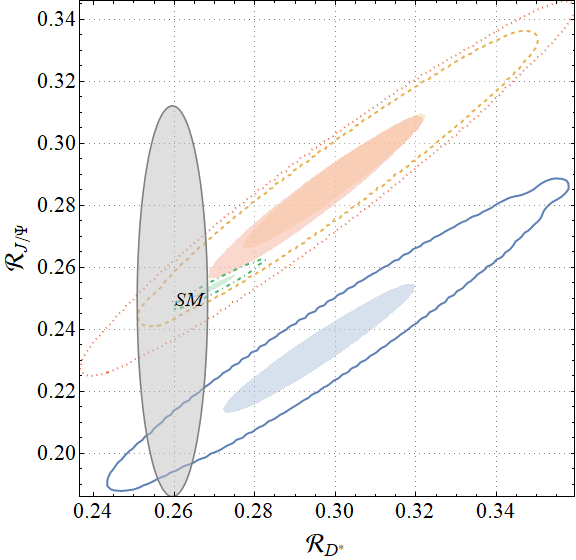}\label{fig:BELLHCBcorrpltRDstRJ1}}\\
	\subfloat[]{\includegraphics[height=4.5cm]{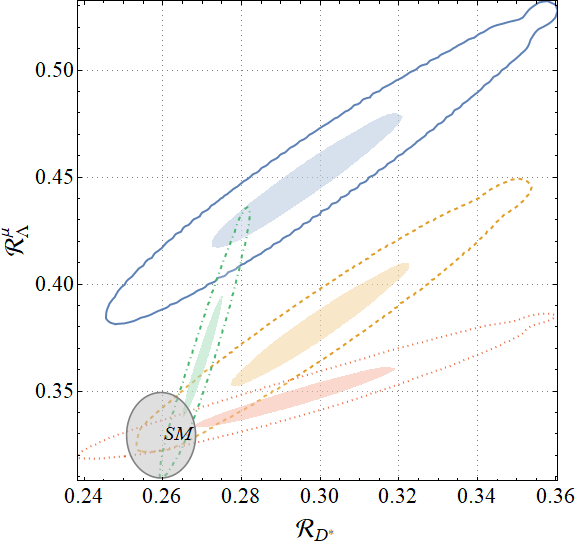}\label{fig:BELLHCBcorrpltRDstRlamMu1}}~
	\subfloat[]{\includegraphics[height=4.5cm]{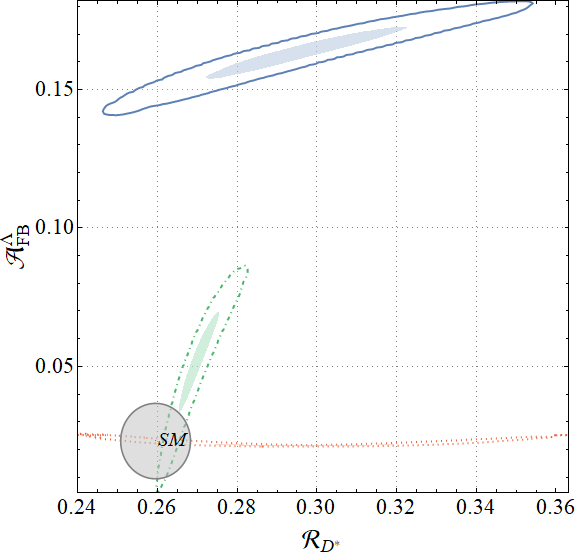}\label{fig:BELLHCBcorrpltRDstAFBlam1}}~
	\subfloat{\includegraphics[height=5cm]{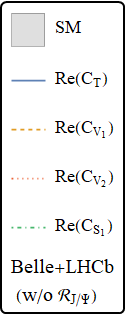}\label{fig:BELLHCbcorrpltLegend1}}
	\caption{\small Correlation plots among different observables for all one-operator scenarios listed in the first column of Table \ref{tab:beL2}. Blue (solid), Orange (dashed), Red(dotted), and Green (dot-dashed)  contours correspond to the scenarios with $\mathcal{R}e(C_{T})$, $\mathcal{R}e(C_{V_1})$, $\mathcal{R}e(C_{V_2})$, and $\mathcal{R}e(C_{S_1})$ respectively. For each of these scenarios, $1\sigma$ (filled region) and $3\sigma$ contours are given.}
	\label{fig:BELLHCbcorrplt1}
\end{figure*}

\begin{figure*}[!htb]
	\centering
	\subfloat[]{\includegraphics[height=4.5cm]{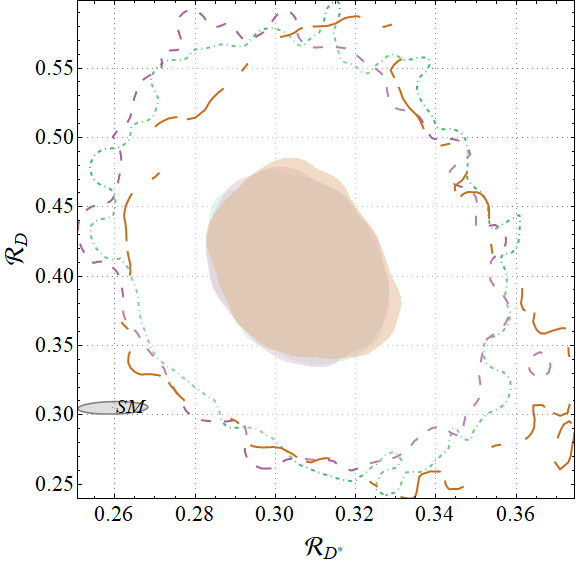}\label{fig:alldatcorrpltRDstRD2a}}~
	\subfloat[]{\includegraphics[height=4.5cm]{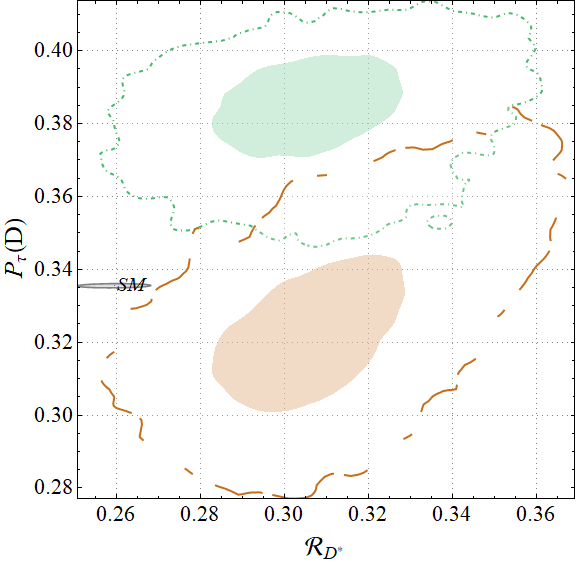}\label{fig:alldatcorrpltRDstPtauD2a}}~
	\subfloat[]{\includegraphics[height=4.5cm]{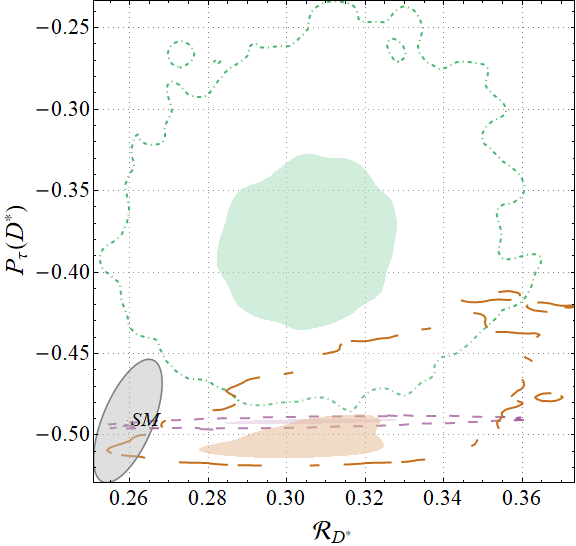}\label{fig:alldatcorrpltRDstPtauDst2a}}\\
	\subfloat[]{\includegraphics[height=4.5cm]{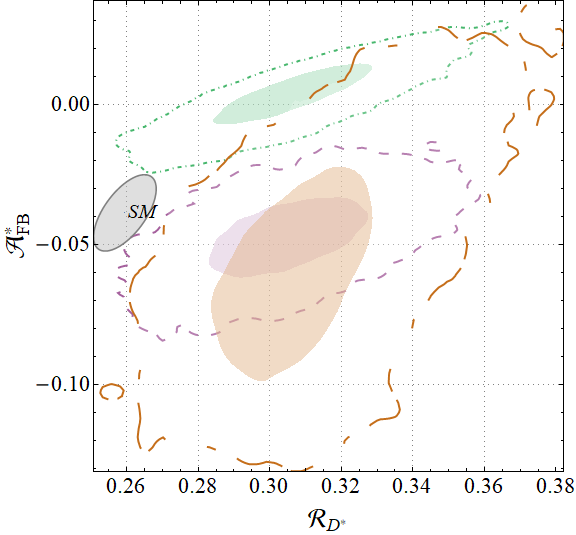}\label{fig:alldatcorrpltRDstAFBDst2a}}~
	\subfloat[]{\includegraphics[height=4.5cm]{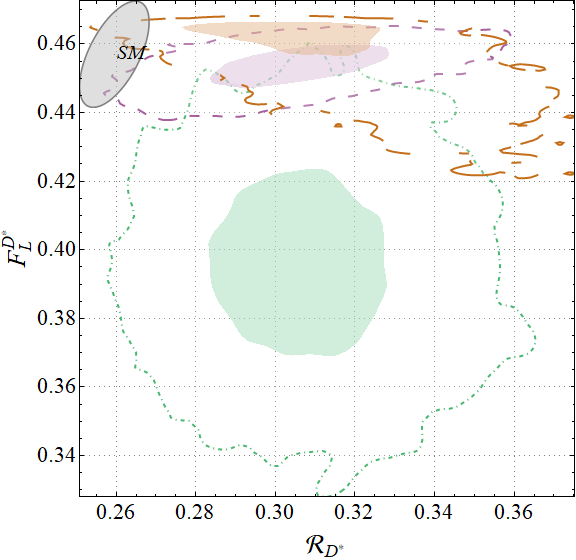}\label{fig:alldatcorrpltRDstFLDst2a}}~
	\subfloat[]{\includegraphics[height=4.5cm]{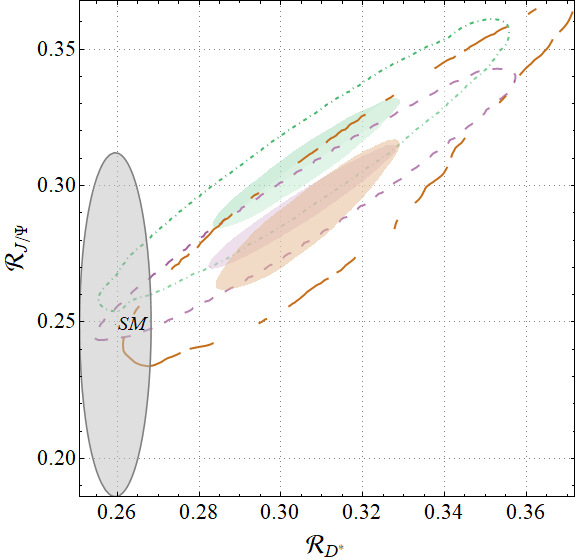}\label{fig:alldatcorrpltRDstRJ2a}}\\
	\subfloat[]{\includegraphics[height=4.5cm]{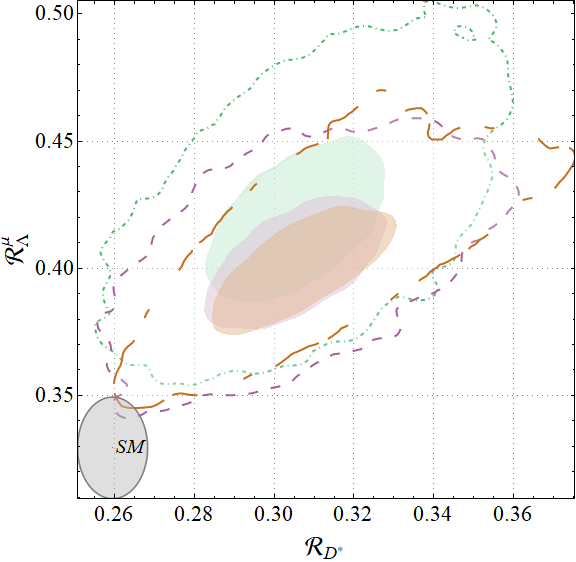}\label{fig:alldatcorrpltRDstRlamMu2a}}~~
	\subfloat[]{\includegraphics[height=4.5cm]{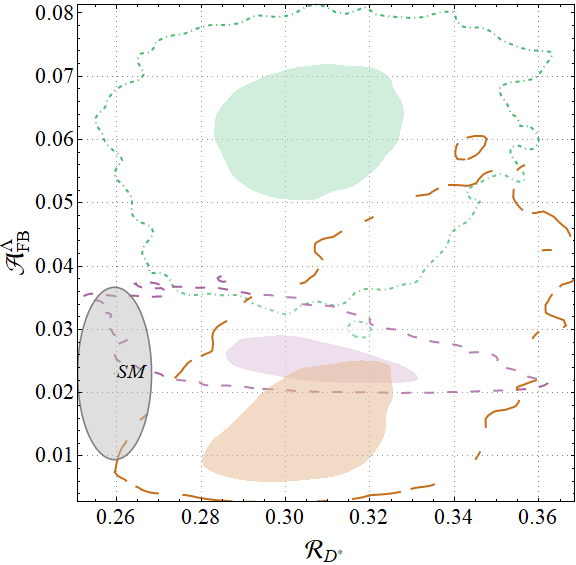}\label{fig:alldatcorrpltRDstAFBlam2a}}~
	\subfloat{\includegraphics[height=5cm]{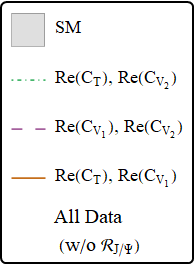}\label{fig:alldatcorrpltLegend2b}}
	\caption{\small Correlation plots among different observables for some of the two-operator scenarios listed in the first column of Table \ref{tab:alldat2}. Green (dot-dashed), Magenta (dashed) and Brown (non-uniform dot-dashed) contours correspond to the scenarios with $\left[\mathcal{R}e(C_{T}),\mathcal{R}e(C_{V_2})\right]$, $\left[\mathcal{R}e(C_{V_1}),\mathcal{R}e(C_{V_2})\right]$ and $\left[\mathcal{R}e(C_{T}),\mathcal{R}e(C_{V_1})\right]$ respectively. For each of these scenarios, $1\sigma$ (filled region) and $3\sigma$ contours are given. }
	\label{fig:alldatcorrplt2a}
\end{figure*}

\begin{figure*}[!htb]
	\centering
	\subfloat[]{\includegraphics[height=4.5cm]{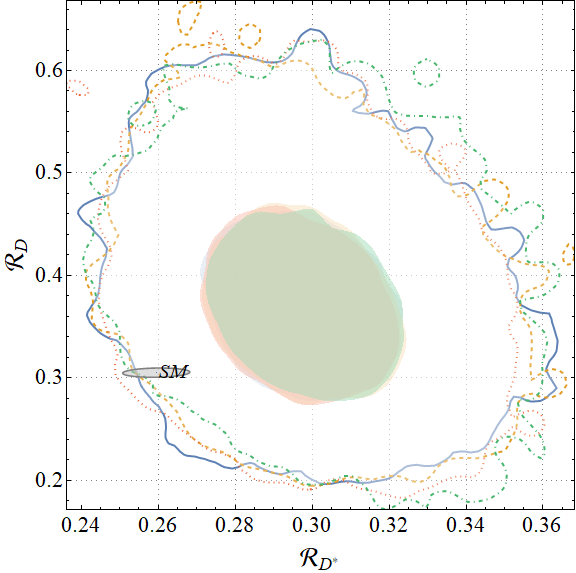}\label{fig:BELLHCBcorrpltRDstRD2}}~
	\subfloat[]{\includegraphics[height=4.5cm]{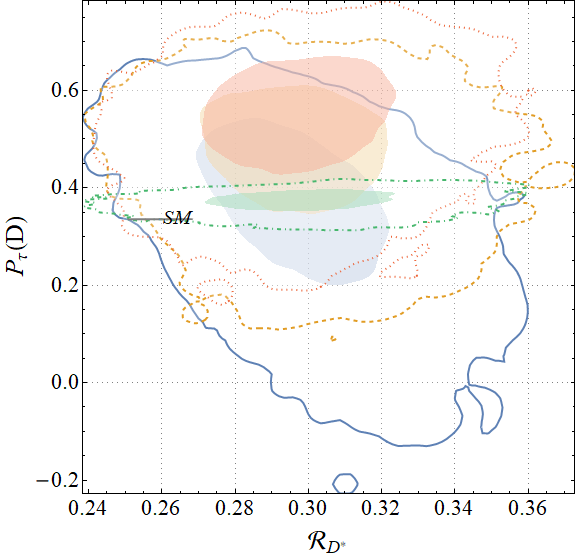}\label{fig:BELLHCBcorrpltRDstPtauD2}}~
	\subfloat[]{\includegraphics[height=4.5cm]{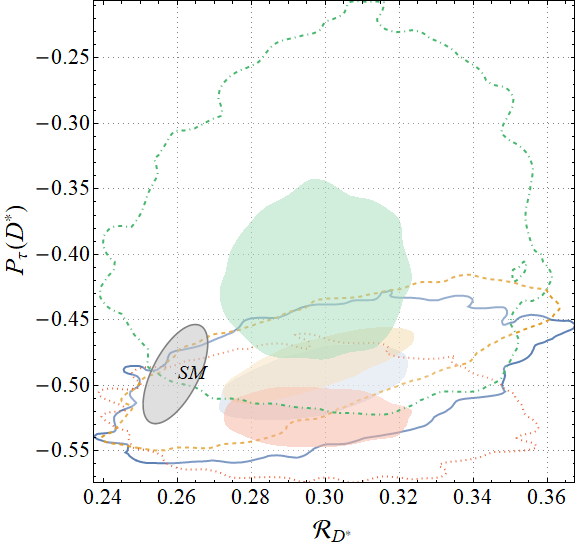}\label{fig:BELLHCBcorrpltRDstPtauDst2}}\\
	\subfloat[]{\includegraphics[height=4.5cm]{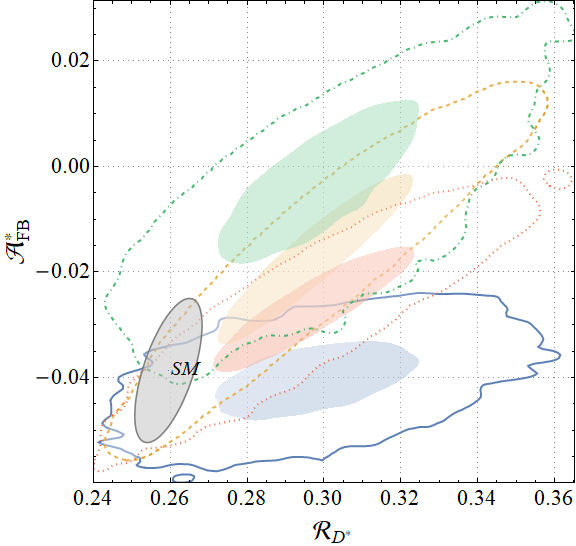}\label{fig:BELLHCBcorrpltRDstAFBDst2}}~
	\subfloat[]{\includegraphics[height=4.5cm]{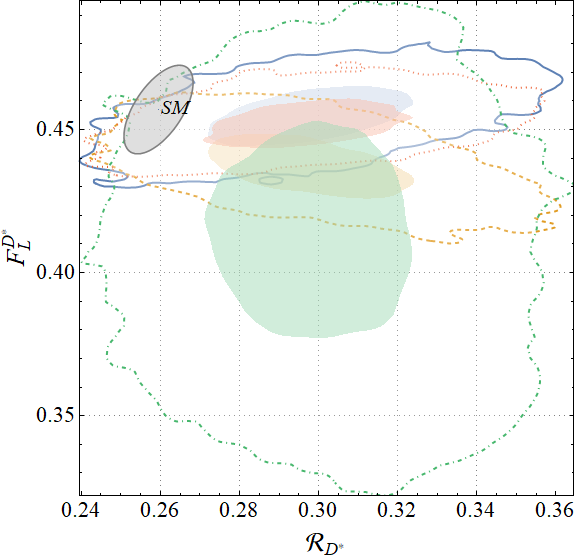}\label{fig:BELLHCBcorrpltRDstFLDst2}}~
	\subfloat[]{\includegraphics[height=4.5cm]{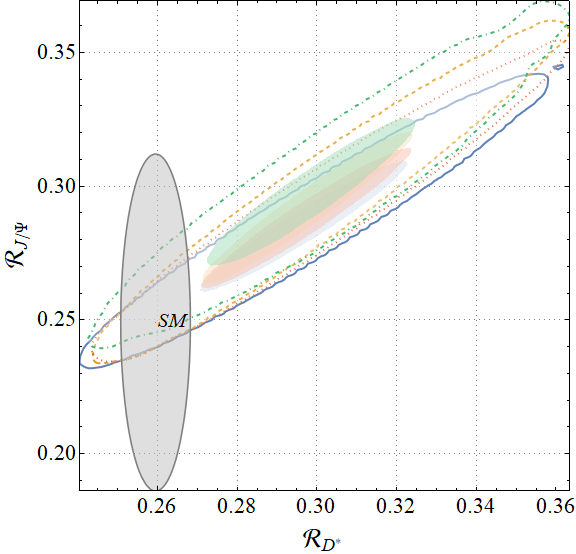}\label{fig:BELLHCBcorrpltRDstRJ2}}\\
	\subfloat[]{\includegraphics[height=4.5cm]{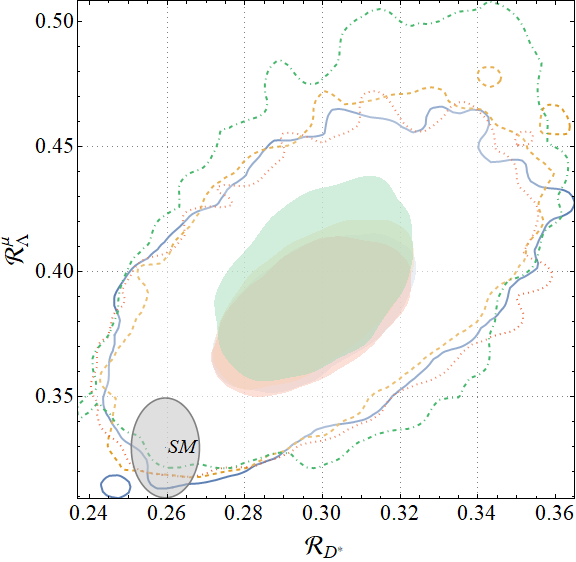}\label{fig:BELLHCBcorrpltRDstRlamMu2}}~
	\subfloat[]{\includegraphics[height=4.5cm]{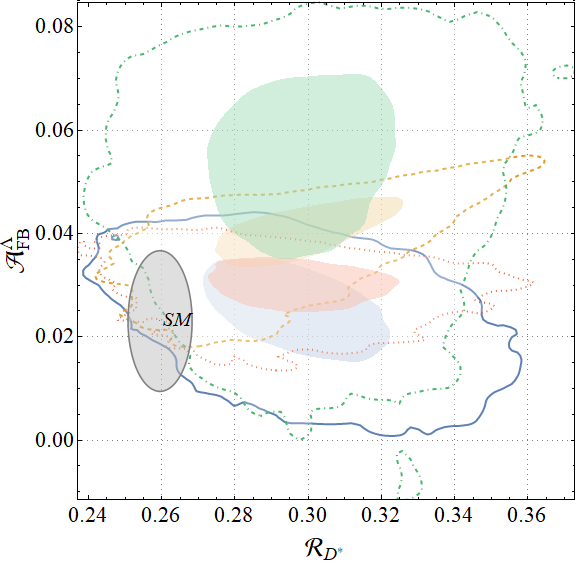}\label{fig:BELLHCBcorrpltRDstAFBlam2}}~
	\subfloat{\includegraphics[height=5cm]{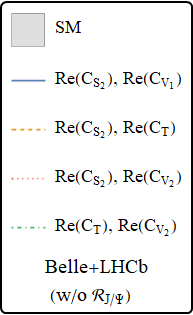}\label{fig:BELLHCbcorrpltLegend2}}
	\caption{\small Correlation plots among different observables for two-operator scenarios listed in the first column of Table \ref{tab:beL2}. Blue (solid), 		Orange (dashed), Red(dotted) and Green (dot-dashed) contours correspond to the scenarios with $\left[\mathcal{R}e(C_{S_2}),\mathcal{R}e(C_{V_1})\right]$, $\left[\mathcal{R}e(C_{S_2}),\mathcal{R}e(C_{T})\right]$, $\left[\mathcal{R}e(C_{S_2}),\mathcal{R}e(C_{V_2})\right]$ and $\left[\mathcal{R}e(C_{T}),\mathcal{R}e(C_{V_2})\right]$ respectively. For each of these scenarios, $1\sigma$ (filled region) and $3\sigma$ contours are given.}
	\label{fig:BELLHCbcorrplt2}
\end{figure*}

\begin{table*}[!htb]
	\centering
%	\begin{ruledtabular}
		\caption{\small Predictions for $\mathcal{R}_{X_c}$, in different NP scenarios (same as table \ref{tab:pred1}). Here, the mentioned accuracy levels are for the SM predictions. Two sets of calculations are done with $m_c$ in either $\overline{MS}$ or kinetic scheme. The corresponding SM predictions are given in table \ref{tab:SMres2}. Quoted uncertainties are for NP only. The SM uncertainties are to be added in quadrature to respective cases.}
		\label{tab:pred3}
%		\colorbox{yellow}{
			\begin{tabular}{ccccc}
				\hline
				\multicolumn{5}{c}{$\mathcal{R}_{X_c}$ (SM + NP) for Data without $\mathcal{R}_{J/\Psi}$}\\
				\hline
				& \multicolumn{4}{c}{$m_c$ in scheme:} \\
				\cline{2-5}
				Scenario & \multicolumn{2}{c}{$\overline{MS}$} & \multicolumn{2}{c}{Kinetic} \\
				\cline{2-5}
				& NNLO + $\mathcal{O}(1/m_b^2)$ & NNLO + $\mathcal{O}(1/m_b^3)$ & NNLO + $\mathcal{O}(1/m_b^2)$ & NNLO + $\mathcal{O}(1/m_b^3)$ \\
				\hline
				1 & \text{0.410(19)} & \text{0.387(19)} & \text{0.395(18)} & \text{0.372(18)} \\
				3 & \text{0.281(10)} & \text{0.258(10)} & \text{0.2741(98)} & \text{0.2511(98)} \\
				7 & \text{0.283(10)} & \text{0.260(10)} & \text{0.276(10)} & \text{0.253(10)} \\
				8 & \text{0.298(14)} & \text{0.275(14)} & \text{0.291(14)} & \text{0.268(14)} \\
				9 & \text{0.287(11)} & \text{0.264(11)} & \text{0.280(11)} & \text{0.257(11)} \\
				10 & \text{0.317(19)} & \text{0.294(19)} & \text{0.309(19)} & \text{0.286(19)} \\
				11 & \text{0.296(14)} & \text{0.273(14)} & \text{0.289(13)} & \text{0.266(13)} \\
				12 & \text{0.288(11)} & \text{0.265(11)} & \text{0.281(11)} & \text{0.258(11)} \\
				13 & \text{0.283(10)} & \text{0.260(10)} & \text{0.276(10)} & \text{0.253(10)} \\
				14 & \text{0.281(10)} & \text{0.258(10)} & \text{0.2743(99)} & \text{0.2513(99)} \\
				15 & \text{0.301(15)} & \text{0.278(15)} & \text{0.293(14)} & \text{0.270(14)} \\
				16 & \text{0.275(11)} & \text{0.252(11)} & \text{0.268(11)} & \text{0.245(11)} \\
				17 & \text{0.392(48)} & \text{0.369(48))} & \text{0.378(45)} & \text{0.356(45)} \\
				\hline
				%		      17 & \text{0.2832(95)} & \text{0.2597(95)} & \text{0.2691(87)} & \text{0.2461(87)} \\
		\end{tabular}%}
		
%	\end{ruledtabular}
\end{table*}

\begin{figure*}[!htb]
	\centering
	\subfloat[]{\includegraphics[width=0.45\linewidth]{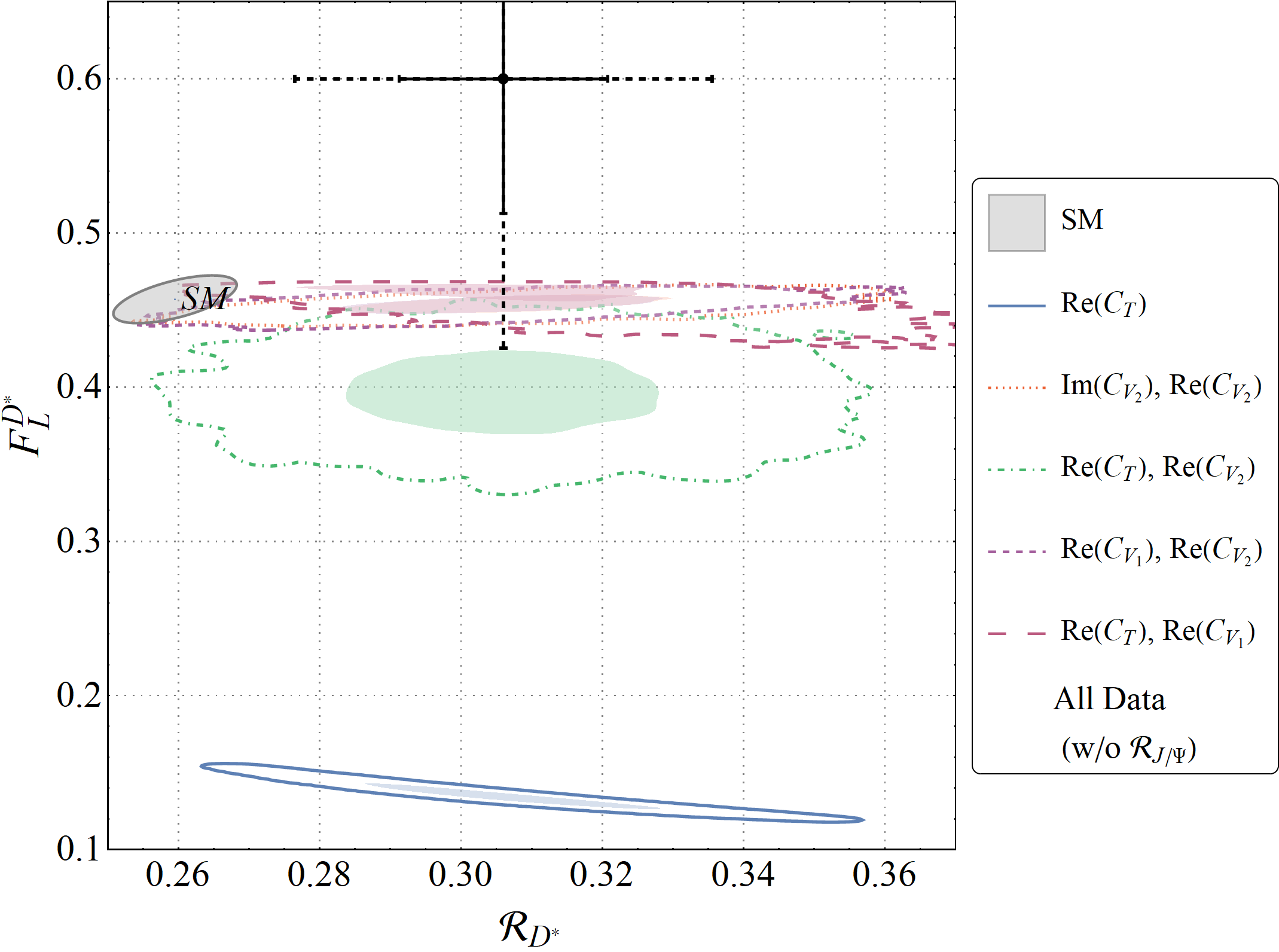}\label{fig:alldatFLcorrExp}}~
	\subfloat[]{\includegraphics[width=0.45\linewidth]{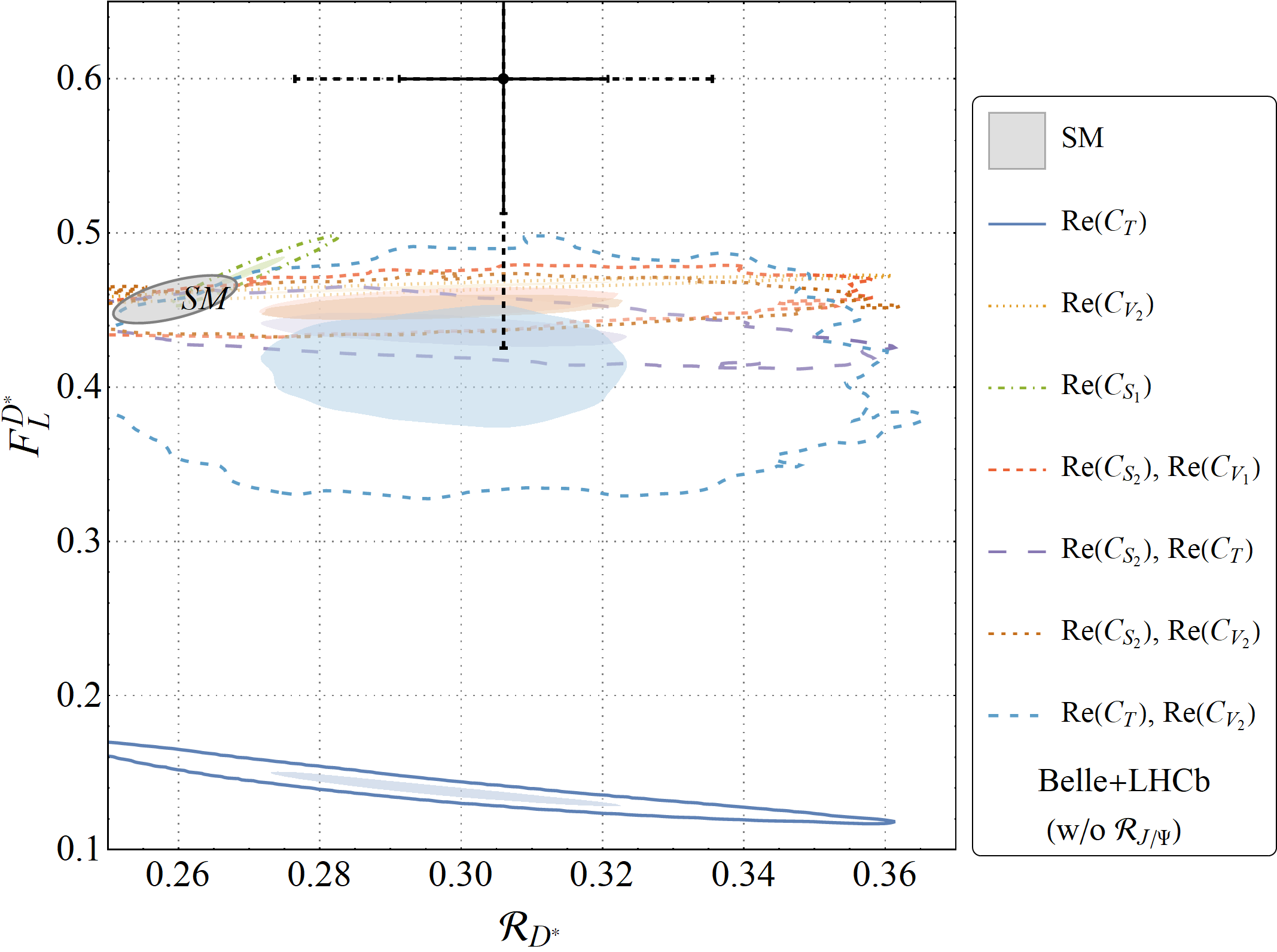}\label{BellLHCbFLcorrExp}}
	\caption{\small Correlation plot of $\mathcal{R}_{D^*}$ vs. $F^{D^*}_L$ for two-operator scenarios previously showed in figures \ref{fig:alldatcorrplt1}, \ref{fig:BELLHCbcorrplt1}, \ref{fig:alldatcorrplt2a}, and \ref{fig:BELLHCbcorrplt2}. We overlap these plots with the experimental results of $\mathcal{R}_{D^*}$ (eq. \ref{eq:expRDRDst}) and $F^{D^*}_L$ (eq. \ref{eq:FLDstExp}), upto $1\sigma$ (solid black) and $2\sigma$ (dashed black) ranges. The scenario with $\mathcal{R}e(C_T)$ is discarded in presence of the new result.}
	\label{fig:FLcorrExp}
\end{figure*}

After ensuring that we are dealing only with the scenarios allowed by $\Delta$AIC$_c$, as well as constraints, we estimate the values of the parameters 
	along with their uncertainties. Ideally, these would be obtained from the projections of the $\Delta\chi^2 = 1$ regions on the parameter line 
	\footnote{For an illustration with the one parameter case, check figure \ref{fig:alldat3JPQ}, where there are two minima. Only one of them is the global 
	minimum and the end points of the red projection region on the parameter line gives the $1\sigma$ uncertainty. There is another region allowed by $3\sigma$ 
	around the other local minimum, coloured in blue.}. For simplicity, and to avoid asymmetric uncertainties, we consider a parabolic approximation around the 
	chosen minimum and not only obtain the uncertainties of all parameters for each case, but also the correlation between them in the 2 parameter cases. These 
	results are tabulated in tables \ref{tab:alldat2}, \ref{tab:noptau2}, \ref{tab:beL2} and \ref{tab:allrdst2}. For some scenarios, instead of the results, the 
	reader is asked to check the corresponding plots. In these scenarios, the best fit, instead of being an isolated point, is actually a contour in the 
	parameter-space. Figure \ref{fig:alldat16JPQ} is such an example. We note that we do not need large values of the $C_W$s ($< 1$) to explain the observed 
	discrepancies in general. Among the best possible scenarios, the data is more sensitive to the model with operator ${\cal O}^{\ell}_{V_1}$ (with real $C_W$) or  
	${\cal O}^{\ell}_{V_2}$ (with complex $C_W$) than the one with operator ${\cal O}^{\ell}_T$. From the best fit values we note that  
	$\mathcal{R}e(C_{V_2}) < \mathcal{R}e(C_{V_1}) << \mathcal{R}e(C_T) < 1$.  
		
	These results could be used by model builders to effectively put bounds on the parameter space of their lepton-flavour universality violating model, satisfying $b\to c\ell \nu$ transitions. 
	
%%%%%%%%%%%%%%%%%%%%%%%%%%%%%%%%%%%%%%%%%%%%%%%%%%%%%%%%%%
\subsection{Prediction of observables and correlations amongst them}\label{sec:res3}
%%%%%%%%%%%%%%%%%%%%%%%%%%%%%%%%%%%%%%%%%%%%%%%%%%%%%%%%%%

Using these NP results, we have predicted the values of the observables listed in table \ref{tab:SMres}. Our predictions for all pertinent scenarios for the dataset without $\mathcal{R}_{J/\Psi}$ are listed in tables \ref{tab:pred1} and \ref{tab:pred2}. Predictions for the inclusive ratio $\mathcal{R}_{X_c}$ are given in a separate table (table \ref{tab:pred3}).
	
	All of the predicted values for NP show deviations from their respective SM predictions. Moreover, neither all observables are equally deviated for a particular type of NP scenario, nor a single observable has similar deviations for different types of NP scenarios. Therefore, in trying to explain the deviations in $\mathcal{R}_{D^{(*)}}$ for a specific type of NP, we get information about the expected deviations in other associated observables. The obtained patterns then can be compared with the future measurements of these observables for a consistency check of the SM and to look for the types of NP. Any result, inconsistent with SM, but consistent with a future prediction of some observable, could be an indirect evidence in support for that specific scenario. In this regard, the correlations between the observables will play an important role. In figs. \ref{fig:alldatcorrplt1}, \ref{fig:BELLHCbcorrplt1}, \ref{fig:alldatcorrplt2a}, %\ref{fig:alldatcorrplt2b},
	and \ref{fig:BELLHCbcorrplt2} we have shown the correlations between the various observables in different NP scenarios which are allowed by our model selection criteria. Following points illuminate our findings after scrutinising these plots: 
\begin{enumerate}
	\item Let us first note the very important correlations between $\mathcal{R}_{D}$ and $\mathcal{R}_{D^*}$ in the scenarios which are allowed by our model selection criteria. In figs. \ref{fig:alldatcorrpltRDstRD1} and \ref{fig:BELLHCBcorrpltRDstRD1}, we plot these correlations for the scenarios with one operator at a time, such as ${\cal O}^{\ell}_T$, ${\cal O}^{\ell}_{V_1}$, ${\cal O}^{\ell}_{V_2}$, and ${\cal O}^{\ell}_{S_1}$.  
	We note that in all the scenarios, except the one with ${\cal O}^{\ell}_{V_2}$, the above two observables are positively correlated but the slopes are very different. By looking at the correlations, one can distinguish between the contributions from different NP operators. In the presence of either $\mathcal{R}e(C_{V_1})$, $\mathcal{R}e(C_{V_2})$, or $\mathcal{R}e(C_{S_1})$, if one of the observables is consistent with the SM, then so must be the other. However, in the case of $\mathcal{R}e(C_T)$, there are regions in which $\mathcal{R}_{D^*}$ is consistent with the SM, whereas $\mathcal{R}_{D}$ is largely deviated from its SM prediction. Therefore, if future data shows that $\mathcal{R}_{D^*}$ is within the SM ballpark but $\mathcal{R}_{D}$ has a large value above its SM prediction, then any scenario with either $\mathcal{R}e(C_{V_1})$, $\mathcal{R}e(C_{V_2})$ or $\mathcal{R}e(C_{S_1})$ has less chance to explain the data, but one with the operator ${\cal O}^{\ell}_T$ will still be able to explain it. 
	The situation is completely opposite in the case of $\mathcal{R}e(C_{V_2})$, where the enhancement in $\mathcal{R}_{D^*}$ over its SM prediction is associated with a decrease in $\mathcal{R}_{D}$ from its SM value. On the other hand, the contributions from ${\cal O}^{\ell}_{V_2}$ with complex $C_{V_2}$ will show deviations in both the observables.
	
	\item All the asymmetric and angular observables are insensitive to the operator ${\cal O}_{V_1}$, as its effect gets cancelled in the ratios. Thus, if future measurements show deviations only in $\mathcal{R}_{D}$ and $\mathcal{R}_{D^*}$, and all the other asymmetric and angular observables in $\bdtau$ decays are consistent with their SM values, then the presence of ${\cal O}_{V_1}$ can be singled out. 
	
	\item On top of this, $P_{\tau}(D)$ is insensitive to the operator ${\cal O}_{V_2}$; detailed correlations can be seen in figs. \ref{fig:alldatcorrpltRDstPtauD1} and \ref{fig:BELLHCBcorrpltRDstPtauD1}. Therefore, large deviations in $P_{\tau}(D)$ in future experiments could point to either tensor or scalar operators. Also, a measured value, well above its SM prediction, can only be explained by the scalar operator. On the other hand, if the value is below its SM prediction, then depending on whether or not there are deviations in $\mathcal{R}_{D^*}$, tensor or scalar operators are favoured. Hence, if future measurements do not see large deviations in $\mathcal{R}_{D}$ and $P_{\tau}(D)$, then the presence of a scalar or tensor operator can be ruled out.
	
	\item As explained earlier, the effects of ${\cal O}_{V_2}$ can be distinguished from those of the other operators in the $\mathcal{R}_{D}-\mathcal{R}_{D^*}$ correlations. Further observations about ${\cal O}_{V_2}$, which can be seen in figs. \ref{fig:alldatcorrplt1} and \ref{fig:BELLHCbcorrplt1}, are as follows: 
	\begin{itemize}
	 \item Large deviations in both $\mathcal{R}_{D}$ and $\mathcal{R}_{D^*}$.
	 \item $P_{\tau}(D)$ will be consistent with its SM value.
	 \item Measured values of $P_{\tau}(D^*)$ and $F_L(D^*)$ are consistent with their respective SM predictions. 
	 \item The measured value of ${\cal A}_{FB}^*$ will be above its SM prediction.  
	\end{itemize}
	
	\item Let us accentuate a few other important points here. In figs. \ref{fig:alldatcorrpltRDstPtauD1} and \ref{fig:alldatcorrpltRDstPtauDst1}, we have shown the correlations between $P_{\tau}(D^{(*)})$ and $\mathcal{R}_{D^*}$. In the presence of a tensor operator ${\cal O}_T$, the $\mathcal{R}_{D^*}$ is negatively and positively correlated with $P_{\tau}(D)$ and $P_{\tau}(D^*)$,
	respectively. However, when $\mathcal{R}_{D^*}$ is consistent with the SM, the $\tau$ polarization asymmetries will not be consistent with their respective SM predictions. In the presence of $C_T$, the values of $P_{\tau}(D)$ and $P_{\tau}(D^*)$ will be below and above their respective SM predictions, respectively. Also, $P_{\tau}(D^*)$ can be positive, whereas the SM predicted value is negative. In the same scenario, the correlations of $\mathcal{R}_{D^*}$ with ${\cal A}_{FB}^*$ and $F_L(D^*)$ are similar to those obtained for $P_{\tau}(D^*)$ and  $P_{\tau}(D)$, respectively, for instances see figs. \ref{fig:alldatcorrpltRDstAFBDst1} and \ref{fig:alldatcorrpltRDstFLDst1}. Also, here the forward-backward and the $D^*$ polarisation asymmetries are largely deviated from their respective SM predictions even when $\mathcal{R}_{D^*}$ is consistent with the SM. On the other hand, we do not see such behaviour in the presence of ${\cal O}_{S_1}$. In this case, the $P_{\tau}(D^{(*)})$, ${\cal A}_{FB}^*$ and $F_L(D^*)$ are consistent with their SM predictions depending on whether or not $\mathcal{R}_{D^*}$ is consistent with its SM prediction (see figs. \ref{fig:BELLHCBcorrpltRDstPtauD1},
	 \ref{fig:BELLHCBcorrpltRDstPtauDst1}, \ref{fig:BELLHCBcorrpltRDstAFBDst1} and \ref{fig:BELLHCBcorrpltRDstFLDst1}). 
	
	\item In case of the dataset with only $\mathcal{R}_{J/\psi}$ dropped from the fit, the correlations of $\mathcal{R}_{D^*}$ with other observables like $\mathcal{R}_{J/\psi}$, $\mathcal{R}_{\Lambda}^{\mu}$, and $A_{FB}^{\Lambda}$ are shown in figs. \ref{fig:alldatcorrpltRDstRJ1}, \ref{fig:alldatcorrpltRDstRlamMu1},and \ref{fig:alldatcorrpltRDstAFBlam1}, respectively. Similar plots, which are obtained by dropping the \Babar~data, are given in \ref{fig:BELLHCBcorrpltRDstRJ1}, \ref{fig:BELLHCBcorrpltRDstRlamMu1}, and \ref{fig:BELLHCBcorrpltRDstAFBlam1}, respectively. In all the one-operator scenarios, the correlations are positive. Due to the large uncertainty in the SM prediction of $\mathcal{R}_{J/\psi}$, the predicted values of these observables are consistent with its SM prediction in all these cases. It is difficult to distinguish between the cases with either $\mathcal{R}e(C_{V_1})$, $\mathcal{R}e(C_{V_2})$ or $\mathcal{R}e(C_{S_1})$. A very large deviation in $\mathcal{R}_{D^*}$ may allow us to see a small deviation in $\mathcal{R}_{J/\psi}$. Also, the contribution from $\mathcal{R}e(C_T)$ can be distinguished from other new operators. In a high precision experiment, contributions of various above mentioned operators are separable from each other by observing the correlation between $\mathcal{R}_{\Lambda}^\mu$ and $\mathcal{R}_{D^*}$. The contribution from $\mathcal{R}e(C_T)$ may allow a large deviation in $\mathcal{R}_{\Lambda}^\mu$, with a sizeable effect in $\mathcal{R}_{D^*}$. Similar patterns are observed in the correlations of $A_{FB}^{\Lambda}$.
	
	\item Similar correlations in the allowed two-operator scenarios are shown in figs. \ref{fig:alldatcorrplt2a} and \ref{fig:BELLHCbcorrplt2}. We note that it will be hard to distinguish the allowed two-operator scenarios from each other just from the correlations of $\mathcal{R}_{D^*}$ with $\mathcal{R}_{D}$, $\mathcal{R}_{J/\psi}$, and $\mathcal{R}_{\Lambda}^{\mu}$, as all the scenarios have similar correlations. However, the shape of the confidence regions the two-operator scenarios are different from those of the one-operator ones. 
	
	For the two operator scenarios containing ${\cal O}_{V_1}$, the NP-predicted values of the angular and asymmetric observables are consistent with their SM values in general. Here too the $P_{\tau}(D)$ is insensitive to the operator-combination [${\cal O}_{V_1}, {\cal O}_{V_2}$]. We note that the contributions from [$\mathcal{R}e(C_T)$, $\mathcal{R}e(C_{V_2})$] in $P_{\tau}(D^*)$, ${\cal A}_{FB}^*$ and $F_L(D^*)$ can be identified. It is hard to distinguish the contributions of the rest of the operators with ${\cal O}_{S_2}$ and all the asymmetric and angular observables are consistent with their respective SM values for them. The correlation of $A_{FB}^{\Lambda}$ shows that except the contribution from  [$\mathcal{R}e(C_T)$, $\mathcal{R}e(C_{V_2})$], all other allowed two-operator scenarios are consistent with the SM even if there is a large deviation in $\mathcal{R}_{D^*}$. By looking at these correlations, one will be able to distinguish a two-operator scenario from the one-operator ones. 
	\end{enumerate}

\subsection{Recent measurement of $F^{D^*}_L$}
	
	Recently, after the publication of the first preprint of this work, a preliminary measurement of $F^{D^*}_L$ was announced \cite{AdamczykCKM18}. According to that talk, the recently measured value is 
	\begin{align}\label{eq:FLDstExp}
		F^{D^*}_L = 0.60 \pm 0.08 (stat.) \pm 0.035 (syst.)\,.
	\end{align}
	One immediate conclusion is that the operator ${\cal O}_{V_1}$ alone can not explain such a large value in $F_L(D^*)$, since its effect is cancelled in the ratio. With the hope that this result may somewhat help us discriminate between our selected models, we recreate the $F^{D^*}_L$ vs. $\mathcal{R}_{D^*}$ correlation plots of figures \ref{fig:alldatcorrplt1}, \ref{fig:BELLHCbcorrplt1}, \ref{fig:alldatcorrplt2a}, and \ref{fig:BELLHCbcorrplt2} in figure \ref{fig:FLcorrExp}. Keeping in mind the preliminary nature of this result, we show the corresponding experimental results up to $2 \sigma$. We note that the new tensor type operator with the Wilson coefficient $C_T$ cannot explain the observed result of $D^*$ polarisation asymmetry, though it is one of the best possible solutions for the explanations of the observed discrepancy in $R(D^*)$. However, the other operators like $ {\cal O}_{V_2}$ alone, and the combinations of operators like ${\cal O}_{V_1}$, 
	${\cal O}_{S_{1/2}}$ are amongst the most probable scenarios that can accommodate the present data.

%%%%%%%%%%%%%%%%%%%%%%%%%%%%%%%%%%%%%%%
\section{Summary }
%%%%%%%%%%%%%%%%%%%%%%%%%%%%%%%%%%%%%%%

	In this paper, we have predicted the SM values of the angular observables associated with the $\bdtau$ decays, following the results of an earlier up-to-date analysis on $\bdell$. Also, we have updated the SM prediction of $\mathcal{R}_{X_c}$ using the results of \cite {Alberti:2014yda} along with the proper correlations between the various non-perturbative parameters and masses. These predictions are based on two different schemes of the charm quark mass ($\overline{MS}$ and Kinetic). These include the NNLO perturbative corrections, and power-corrections up to order $1/{m_b}^3$. We have separately mentioned results with power-corrections up to $1/{m_b}^2$ order as well. Our best results are $\mathcal{R}_{X_c} = \text{0.214(4)}$ for $\overline{m_c} (3 {\it GeV}) = 0.987(13)$ {\it GeV}, and  $\mathcal{R}_{X_c} = \text{0.209(4)}$  when $m_c^{kin} = 1.091 (20)$ {\it GeV} where in both the cases we have taken $m_b^{kin} = 4.56(21)$ {\it GeV}.
	
	In the next part of our analysis, we have analysed the semitaunic $b\to c\tau\nu_{\tau}$ decays in a model independent framework with the 
	$\Delta B = \Delta C = 1$ semileptonic operators. We have included the complete set of vector, scalar and tensor operators, while assuming the neutrinos to be left handed. Different possible combinations of all the effective operators have been considered, and following AIC$_c$, the combinations, which are best suited for the available data, are considered for further studies. We have performed the analysis on several different prepared data sets. We note that for all of the data sets, the one-operator scenarios, with a real $C_W$, can best explain the available data. However, in most of them, the scalar operators are not allowed by the constraint $Br(B_c \to \tau\nu_{\tau}) \le 30\%$. The most favoured scenarios are the ones with tensor (${\cal O}_T$) or $(V-A)$ (${\cal O}_{V_1}$) type of operators. Also, the $(V+A)$ type of interactions, with a complex $C_W$, though less favoured, are allowed. In the absence of the \Babar~data on $\mathcal{R}_{D^{(*)}}$ from our analysis, one-operator scenarios like $(V \pm A)$, $S - P$, and tensor operators with real $C_W$ are the most favoured ones. These one operator scenarios are easily distinguishable from each other by studying the correlations of $\mathcal{R}_{D^*}$ with $\mathcal{R}_{D}$ and all the other asymmetric and angular observables. Also, the patterns of the future measurements of all such observables can easily discriminate the types of NP. Among all the possible combinations of $(V \pm A)$, tensor and $( S - P)$ operators, there are quite a few two-operator scenarios which pass all the selection criteria. In these cases, one cannot differentiate between the contributions from NP scenarios by looking at the correlations of $\mathcal{R}_{D^*}$ with $\mathcal{R}_{D}$, $\mathcal{R}_{J/\psi}$, and $\mathcal{R}_{\Lambda}^{\mu}$. However, the correlations of $\mathcal{R}_{D^*}$ with the various angular and asymmetric observables could be useful for such a discrimination. We have also predicted the numerical values of all the observables along with their errors, for the allowed scenarios.

	\section*{Acknowledgements}
	 We would like to thank Paolo Gambino for valuable inputs in the analysis and Zoltan Ligeti for his valuable comments. S.K.P. is supported by the grants IFA12-PH-34 and SERB/PHY/2016348.
	%\end{acknowledgments}

%\newpage

	\bibliography{ref_draft}
    \bibliographystyle{ieeetr}
% BibTeX users please use
% \bibliographystyle{}
% \bibliography{}
%\begin{thebibliography}{}
%Author, \textit{Book title} (Publisher, place year) page numbers
% etc
%\end{thebibliography}

\end{document}